\documentclass[twocolumn,aps,nofootinbib]{revtex4-1}
\usepackage{graphicx}
\usepackage{epstopdf}
\usepackage{latexsym}
\usepackage{amssymb}
\usepackage{amsmath}
\usepackage{color}
\usepackage{mathrsfs}
\usepackage{xparse}
\usepackage{float}
\usepackage{mathtools}
\usepackage[center]{subfigure}

\begin{document}
\renewcommand\arraystretch{2}
\newcommand{\bq}{\begin{equation}}
\newcommand{\eq}{\end{equation}}
\newcommand{\bqn}{\begin{eqnarray}}
\newcommand{\eqn}{\end{eqnarray}}
\newcommand{\nb}{\nonumber}
\newcommand{\lb}{\label}
\newcommand{\cb}{\color{blue}}
\newcommand{\cc}{\color{cyan}}
\newcommand{\cm}{\color{magenta}}
\newcommand{\rc}{\rho^{\scriptscriptstyle{\mathrm{I}}}_c}
\newcommand{\rd}{\rho^{\scriptscriptstyle{\mathrm{II}}}_c} 
\NewDocumentCommand{\evalat}{sO{\big}mm}{%
  \IfBooleanTF{#1}
   {\mleft. #3 \mright|_{#4}}
   {#3#2|_{#4}}%
}

\newcommand{\PRL}{Phys. Rev. Lett.}
\newcommand{\PL}{Phys. Lett.}
\newcommand{\PR}{Phys. Rev.}
\newcommand{\CQG}{Class. Quantum Grav.}
\newcommand{\parallelsum}{\mathbin{\!/\mkern-5mu/\!}}
\renewcommand\arraystretch{2}
 \newcommand{\subbq}{\begin{subequations}}
 \newcommand{\subeq}{\end{subequations}}
\newcommand{\rcone}{\rho^{\scriptscriptstyle{\mathrm{I}}}_c}
\newcommand{\La}{\Lambda}
\newcommand{\va}{\scriptscriptstyle}
\newcommand{\be}{\nopagebreak[3]\begin{equation}}
\newcommand{\ee}{\end{equation}}
\newcommand{\sign}{\text{sign}}

\newcommand{\ba}{\nopagebreak[3]\begin{eqnarray}}
\newcommand{\ea}{\end{eqnarray}}

\newcommand{\la}{\label}
\newcommand{\n}{\nonumber}
\newcommand{\su}{\mathfrak{su}}
\newcommand{\SU}{\mathrm{SU}}
\newcommand{\U}{\mathrm{U}}

\def\be{\nopagebreak[3]\begin{equation}}
\def\ee{\end{equation}}
\def\ba{\nopagebreak[3]\begin{eqnarray}}
\def\ea{\end{eqnarray}}
\newcommand{\f}{\frac}
\def\rmd{\rm d}
\def\lp{\ell_{\rm Pl}}
\def\d{{\rm d}}
\def\fe{\mathring{e}^{\,i}_a}
\def\fw{\mathring{\omega}^{\,a}_i}
\def\fq{\mathring{q}_{ab}}
\def\t{\tilde}

\def\db{\delta_b}
\def\dc{\delta_c}
\def\T{\mathcal{T}}
\def\GammaE{\Gamma_{\rm ext}}
\def\GammaEb{\bar\Gamma_{\rm ext}}
\def\GammaEh{\hat\Gamma_{\rm ext}}
\def\Hee{H_{\rm eff}^{\rm ext}}
\def\H{\mathcal{H}}

\newcommand{\R}{\mathbb{R}}


\title{Universal properties of the evolution of the Universe in modified loop quantum cosmology
}

\author{Jamal Saeed$^{a}$}
\email{jamal$\_$saeed1@baylor.edu}

\author{Rui Pan$^{a}$}
\email{rui$\_$pan1@baylor.edu}

\author{Christian Brown$^{b}$}
\email{Christian$\_$Brown4@baylor.edu}

\author{Gerald Cleaver$^{b}$}
\email{Gerald$\_$Cleaver@baylor.edu}

\author{Anzhong Wang$^{a}$ \footnote{The corresponding author}}
\email{anzhong$\_$wang@baylor.edu; the corresponding author}

\affiliation{ $^{a}$ GCAP-CASPER, Physics Department, Baylor University, Waco, TX 76798-7316, USA\\
$^{b}$ EUCOS-CASPER, Physics Department, Baylor University, Waco, TX 76798-7316, USA}

\date{\today}
\begin{abstract}
In this paper, we systematically study the evolution of the Universe in the framework of a modified loop quantum cosmological model (mLQC-I) with various inflationary potentials, including chaotic, Starobinsky, generalized Starobinsky, polynomials of the first and second kinds, generalized T-models and natural inflation. In all these models, the big bang singularity is replaced by a quantum bounce, and the evolution of the Universe both before and after the bounce is {\em universal and weakly depends on the inflationary potentials}, as long as the evolution is dominated by the kinetic energy of the inflaton at the bounce. In particular, the evolution in the pre-bounce region can be {\em universally} divided into three different phases: {\em pre-bouncing, pre-transition, and pre-de Sitter}. The pre-bouncing phase occurs immediately before the quantum bounce, during which the evolution of the Universe is dominated by the kinetic energy of the inflaton. Thus, the equation of state of the inflaton is about one, $w(\phi) \simeq 1$. Soon, the inflation potential takes over, so $w(\phi)$ rapidly falls from one to negative one. This pre-transition phase is very short and quickly turns into the pre-de Sitter phase, whereby the effective cosmological constant with a Planck size takes over and dominates the rest of the contracting phase. In the entire pre-bounce regime, the evolution of the expansion factor and the inflaton can be approximated by analytical solutions, which are universal and independent of the inflation potentials.

\end{abstract}

\maketitle

\section{
Introduction
}
\renewcommand{\theequation}{1.\arabic{equation}}
\setcounter{equation}{0}

Since its incarnation in 1980 \cite{1981PhRvD..23..347G}, the inflationary paradigm has achieved great success, resolving many long-standing problems of the standard big bang cosmology, and is consistent with all cosmological and astrophysical observations conducted so far \cite{Planck:2018jri}. However, the paradigm is past incomplete \cite{Borde:2001nh},
Due to the Big Bang singularity, where all physical quantities become infinite. Hence, the initial conditions of inflation are usually imposed at a moment sufficiently far from the singularity but early enough so that the wavelengths of all the observational modes are within the Hubble horizon.

Then, a natural question arises: What happened before inflation? This important question is expected to be answered by quantum gravity (QG), a theory that has not yet been established, despite enormous efforts in the past century, although many candidates have been proposed. Among them are string/M-theory \cite{Green_Schwarz_Witten_2012,Becker:2006dvp} and loop quantum gravity (LQG) \cite{Ashtekar:2004eh,Thiemann_2007,Bojowald_2010,Gambini:2011zz,Rovelli:2014ssa}. In particular, in the last two decades, LQG has been rigorously applied to understand singularity resolution in various cosmological models
(For reviews, see Refs. \cite{Bojowald:2005epg,Ashtekar:2011ni,Ashtekar:2015dja,Agullo:2016tjh,Wilson-Ewing:2016yan,ElizagaNavascues:2020uyf,Ashtekar:2021kfp,Li:2023dwy,Agullo:2023rqq}).  In all these models, a coherent picture of Planck scale physics has emerged: {\em the big bang singularity is replaced by a quantum bounce} \footnote{Bouncing universe inspired by string theory was first proposed in Ref. \cite{gasperini1993pre}, and later studied intensively by various authors, see, for example Refs. \cite{Gasperini:1996fu,Gasperini:2002bn,Haro:2013bea,article,Conzinu:2023fth} and references therein. Other bouncing models have been also studied extensively \cite{Khoury:2001wf,Brown:2004cs,Battefeld:2014uga,Brandenberger:2016vhg,Ijjas:2018qbo,Chandran:2024utf}. In this paper, we shall focus ourselves on the quantum bounce from LQG., purely due to quantum geometric effects}. This framework is often referred to as loop quantum cosmology (LQC).

The physical implications of LQC have  also been  studied using {\em the effective descriptions} of the quantum spacetime derived from coherent states \cite{Taveras:2008ke},
whose validity has been verified for various spacetimes both numerically \cite{Singh:2018rwa} and \cite{Corichi:2007am}, especially for states that sharply peaked on classical trajectories at late times \cite{Kaminski:2019qjn}.
Effective dynamics provide a definitive answer to the resolution of the big bang singularity, replaced by a quantum bounce when the energy density of matter reaches a maximum value determined purely by the underlying quantum geometry
\cite{Ashtekar:2011ni,ElizagaNavascues:2020uyf,Li:2023dwy,Agullo:2023rqq}.

Despite a wealth of results on the singularity resolution and phenomenology of the very early universe obtained in the framework of LQC, an important issue that has remained open is its connection to LQG (see, for example, \cite{Beetle:2017qle,Bojowald:2021kzv} for discussions). The starting point of LQC is to first classically reduce the Hamiltonian from infinitely many to a few gravitational degrees of freedom by imposing homogeneity and then to quantize the classically reduced Hamiltonian using the techniques of LQG. However, in LQG, the processes of symmetry reduction and quantization do not commute in general, and it is important to understand how well the physics of the full LQG is captured by LQC.
In addition, in LQG, the Hamiltonian usually consists of two parts, the Euclidean and Lorentz parts, and one follows different processes to quantize each part \cite{Ashtekar:2004eh,Thiemann_2007,Bojowald_2010,Gambini:2011zz,Rovelli:2014ssa}. However, in LQC, only the quantization of the Euclidean part was considered by taking advantage of the properties of the classical Hamiltonian, in which the Euclidean and Lorentz parts are proportional to each other for the flat Friedmann-Lema\'itre-Robertson-Walker (FLRW) universe \cite{Ashtekar:2011ni}.

In the past decade, the above issue has been extensively studied by both bottom-up and top-down approaches, from which an important conclusion is emerging: {\em LQC and its major predictions are robust}. In particular, the big bang singularity is resolved in the models studied so far. However, dramatic changes in the evolution of the universe in the pre-bounce phase are also found \cite{Li:2021mop}.

In the bottom-up approach \cite{Yang:2009fp}, symmetries are still imposed before quantization, but the Lorentzian term is treated independently by applying Thiemann's regularization from the full theory of LQG \cite{Thiemann:1996av,Thiemann:1996aw}. In doing so, it was found that the resultant wave function is now described by a fourth-order difference equation \cite{Yang:2009fp}, which is referred to as mLQC-I  \cite{Li:2021mop}. For a systematic derivation of the model, we refer readers to \cite{Assanioussi:2018hee,Assanioussi:2019iye}.  For sharply peaked states, the resulting quantum dynamics are well described by effective Friedman-Raychaudhuri (FR) equations \cite{Li:2018opr},  with which it was found that both the resolution of the big bang singularity \cite{Li:2018fco}  and the existence of a subsequent desired slow-roll inflation are generic  \cite{Li:2019ipm}.

In the top-down approach, using complexifier coherent states and treating the Euclidean and Lorentzian terms in the scalar constraint separately as in full LQG, Dapor and Liegener (DL) first obtained a modified LQC model \cite{Dapor:2017rwv,Dapor:2017gdk}, which is quite similar to mLQC-I but with the $\mu_0$-scheme, as DL used a fixed graph \cite{Dapor:2017rwv,Dapor:2017gdk}.  To address this shortcoming, the DL work was extended to allow for graph-changing dynamics in an approach based on the path-integral reformulation of LQG \cite{Han:2021cwb}, whereby a consistent mLQC-I model can be obtained.

In this paper, we shall systematically study the evolution of the Universe in the framework of mLQC-I. Note that such studies have already been conducted in the post-bounce regime ($t\ge t_B$) in \cite{Li:2018opr,Li:2018fco,deHaro:2018khb,Saini:2018tto,Li:2019ipm,Saini:2019tem,Li:2021mop,Li:2021fmu}, where $t_B$ denotes the bounce time. It was found that the evolution in this regime is universal and independent of the inflationary potential, as long as the evolution is dominated at the quantum bounce by the kinetic energy of the inflaton
\bq
\lb{eq1.1}
\frac{1}{2}\dot\phi_B^2 \gg V(\phi_B),
\eq where an over dot denotes the derivative with respect to the cosmic time, and $\phi_B$ is the value of the inflaton at the bounce. Initial conditions satisfying Eq.(\ref{eq1.1}) are important because they always lead to slow-roll inflation \cite{Li:2018opr,Li:2018fco,Li:2019ipm,Li:2021mop}, quick similar to that in LQC \cite{Bonga:2015xna,Zhu:2016dkn,Zhu:2017jew,Shahalam:2017wba,Shahalam:2018rby,Sharma:2018vnv,Sharma:2019okc,Shahalam:2019mpw}. Therefore, in this paper, we focus on the pre-bounce regime ($t\le t_B$), which has not been studied adequately so far and is important, especially when the initial conditions of cosmological perturbations are imposed in the contracting phase \cite{Ashtekar:2011ni,ElizagaNavascues:2020uyf,Li:2023dwy,Agullo:2023rqq}.

Specifically, the rest of the paper is organized as follows: In Sec.\ref{SecII} we briefly review the mLQC-I model and write down the dynamical Hamiltonian equations. To understand the major properties of mLQC-I, we also write down the corresponding FR equations.
In Sec.\ref{SecIII}, we study the evolution of the Universe with several inflationary potentials, including chaotic, Starobinsky, generalized Starobinsky, polynomials of the first and second kinds, generalized T-models and natural inflation. In the last case, we consider several choices of parameters involved in the model. Note that the chaotic and natural inflationary models have some tensions with current observations \cite{Planck:2018jri}.  The reasons that we still consider them here are two-folds: First, after quantum gravitational effects are taken into account, such tensions can be alleviated \cite{Zhu:2016srz}. Second, the universal properties of the evolution of the universe both before and after the quantum bounce are independent of the choice of inflationary potentials in  LQC  \cite{Zhu:2016dkn,Zhu:2017jew,Shahalam:2017wba,Shahalam:2018rby,Sharma:2018vnv,Sharma:2019okc,Shahalam:2019mpw,Levy:2024naz}. We find that this is also true in mLQC-I. In all these models, we find that the evolution of the Universe in the pre-bounce regime ($t \le t_B$) is universal and weakly depends on the inflationary potentials, as long as the kinetic-energy-dominated condition (\ref{eq1.1}) is satisfied at the bounce. In all these models, the evolution is clearly divided into three different epochs, {\em pre-bouncing, pre-transition, and pre-de Sitter}, as defined by Eq.(\ref{eq3.11}), in terms of the equation of state of the inflaton
\bq
\lb{eq1.2}
w(\phi) \equiv \frac{P(\phi)}{\rho(\phi)},
\eq 
where $P(\phi)$ and $\rho(\phi)$ are the pressure and energy density of the inflationary field, respectively, given by Eq.(\ref{eq2.8}). Physically, it can be understood as follows: In the pre-bouncing phase, the evolution of the universe is dominated by the kinetic energy as determined by the initial conditions imposed at the bounce, so it is expected that its evolution in this phase is independent of the inflationary potentials and $w(\phi) \simeq + 1$. However, an effective Planck-size cosmological constant exists in the pre-bounce regime [cf. Eq.(\ref{FRccA})], so it will soon dominate the evolution once apart from the bounce, during which we have $w(\phi) \simeq - 1$.  Clearly, during this phase the evolution of the universe will be also independent of the inflationary potentials. In between these two phases, a short pre-transition phase exists, during which  $w(\phi)$ drops rapidly  from $+ 1$ to $- 1$ and behaves like a step function\footnote{It should be noted that such universal properties in the pre-bounce regime may not be shared by the standard LQC \cite{Bonga:2015xna,Zhu:2016dkn,Zhu:2017jew,Shahalam:2017wba,Shahalam:2018rby,Sharma:2018vnv,Sharma:2019okc,Shahalam:2019mpw}, as in LQC such an effective Planck-size cosmological constant is absent in the pre-bounce regime \cite{Ashtekar:2011ni,ElizagaNavascues:2020uyf,Li:2023dwy,Agullo:2023rqq}.} As a result, in these epochs the expansion factor $a(t)$ and the inflaton $\phi(t)$ are universal,
as shown explicitly by Fig. \ref{fig31} of Sec. \ref{SecIV}. In this section, we also show that  $a(t)$ and  $\phi(t)$ can be well-approximated by Eqs.(\ref{eq3.12}) and (\ref{eq3.13}), respectively, where $d_n$ and $e_n$ are constants determined by fitting the analytical expressions with their numerical solutions, as given in Sec. \ref{SecIV}.  Therefore, {\em the evolution of the Universe in both of the pre-bounce regime ($t \le t_B$)  and the post-bounce regime ($t \ge t_B$) is universal in mLQC-I} \cite{Li:2018opr,Li:2018fco,Li:2019ipm,Li:2021mop}. This is important and will significantly facilitate the studies of cosmological perturbations \cite{Li:2019qzr,Li:2020mfi,Li:2024xxz,ElizagaNavascues:2020uyf,Li:2022evi}, as already shown in LQC \cite{Zhu:2017jew}. The paper  ends in Sec.\ref{SecV}, in which our main results and concluding remarks are presented.


\section{Effective Dynamics of Modified Loop Quantum Cosmology}
\renewcommand{\theequation}{2.\arabic{equation}}
\setcounter{equation}{0}
\lb{SecII}

In this section, we provide a summary of the modified Friedmann dynamics for mLQC-I \cite{Li:2021mop}. Its dynamics can be obtained directly from the effective Hamiltonian, given by \cite{Yang:2009fp}
\bq
\lb{Hamiltonian for mLQC-I}
\mathcal{H} = \frac{3v}{8\pi G \lambda^2 } \left\{ \sin^2 (\lambda b) -\frac{(\gamma^2+1)\sin^2 (2\lambda b)}{4\gamma^2}\right\}+\mathcal{H_M},
\eq 
where  $G$ is the Newtonian constant, $v \equiv v_0a^3$, and $v_0$ is the volume of a fiducial cell in the $\mathcal{R}^3$ spatial manifold, and $a$ is the expansion factor of the Universe. The variable $b$ denotes the momentum conjugate of $v$ and satisfies the canonical relation
\bq\lb{canonical relation}
\{b,v\} = 4 \pi G \gamma,
\eq where  $\gamma$ is known as the Barbero-Immirzi parameter whose value is set to $\gamma \approx 0.2375$ using black hole thermodynamics in LQG \cite{Meissner:2004ju}. The parameter $\lambda$ is defined as $\lambda^2 \equiv \Delta = 4 \sqrt{3}\pi\gamma\ell^2_{Pl}$, where $\Delta$ 
denotes the minimal area gap of the area operator in LQG \cite{Rovelli:1994ge,Ashtekar:1996eg,Thiemann_2007,Ashtekar:2011ni}.

In this paper, we consider only the case in which the matter is characterized by a single scalar field $\phi$ with potential $V(\phi)$, for which $\mathcal{H}_M$ is given by

\bqn
\lb{MatterH}
{\mathcal{H}}_M = \frac{1}{2} v\left(\frac{p^2_\phi}{v^2}+2V(\phi)\right),
\eqn where $p_{\phi}$ is the momentum of $\phi$. Then, the corresponding energy density and pressure of the scalar field are given by
\bqn
\lb{eq2.8}
\rho &\equiv& \frac{\mathcal{H_M}}{v}  = \frac{1}{2}\dot\phi^2 + V(\phi), \nb\\
P &\equiv& -\frac{\partial\mathcal{H}_M}{\partial v} = \frac{1}{2}\dot{\phi}^2 - V(\phi).
\eqn
\vfill\null

The basic variables $b$,  $v$, $\phi$ and $p_{\phi}$ satisfy the Hamiltonian equations
\bqn
\lb{eqA}
\dot v &=& \{v,\mathcal{H}\} \nb\\
&=& \frac{3v\sin{(2\lambda b)}}{2\gamma \lambda}\left\{(\gamma^2+1)\cos{(2\lambda b)}-\gamma^2\right\}, \\
\lb{eqB}
\dot b &=& \{b,\mathcal{H}\}
= \frac{3\sin^2{(\lambda b)}}{2\gamma\lambda^2}\left\{\gamma^2\sin^2{(\lambda b )}-\cos^2{(\lambda b )}\right\} \nb\\
&& ~~~~~~~~~~~~~ - 4\pi G \gamma P, \\
\lb{eqC}
\dot \phi &=& \{\phi,\mathcal{H}\} = \frac{p_{\phi}}{v}, \\
\lb{eqD}
\dot{p}_{\phi} &=& \{p_{\phi},\mathcal{H}\} = - v V_{,\phi},
\eqn where $V_{,\phi} \equiv dV(\phi)/d\phi$.
Eqs.(\ref{eqA}) - (\ref{eqD}) are the first-order ordinary differential equations for the four canonical variables ($v, b; \phi, p_{\phi}$). Once the initial conditions are specified at a given moment, for example, $t = t_B$, they uniquely determine the trajectory of the evolution of the Universe. Such initial conditions are often imposed at the quantum bounce \cite{Ashtekar:2011ni,Li:2021mop}, at which the expansion factor reaches its minimal value and the energy density reaches its maximum.

To see the above clearly, it is very suggestive to write the above four dynamical equations in the form of the modified FR equations \cite{Li:2021mop}
\begin{widetext}
\bqn
\lb{FRa}
H^2 &=& \frac{8\pi G \rho}{3} \left(1-\frac{\rho}{\rcone} \right) \left(1+\frac{\gamma^2 \rho/\rcone}{\left(\gamma^2+1\right)\left(1+\sqrt{1-\rho/\rcone}\right)^2}\right), \;\;\; (t \ge t_B), \\
\lb{FRb}
\frac{\ddot{a}}{a} &=& -\frac{4 \pi G}{3}\left(\rho+3P\right) + \frac{4\pi G \rho^2}{3 \rcone}\left(\frac{\left(7\gamma^2+8\right)-4\rho/\rcone\left(5\gamma^2+8\right)\sqrt{1-\rho/\rcone}}{\left(\gamma^2+1\right)\left(1+\sqrt{1-\rho/\rcone}\right)^2}\right)  \nb\\
&& +4\pi G P\left(\frac{3\gamma^2+2+2\sqrt{1-\rho/\rcone}}{\left(\gamma^2+1\right)\left(1+\sqrt{1-\rho/\rcone}\right)}\right)\frac{\rho}{\rcone}, \;\;\; (t \ge t_B),
\eqn
\end{widetext}
where $H \equiv \dot{v}/(3v) = \dot{a}/a$, and
\bqn
\lb{eq_rho}
\rcone \equiv \frac{\rho_c}{4(1+\gamma^2)}, \quad
\rho_c \equiv \frac{3}{8\pi \lambda^2 \gamma^2 G}.
\eqn
It is evident from these two equations that the energy conservation law $\dot\rho+3H(\rho+P) = 0$ holds.
Substituting  Eq.(\ref{eq2.8}) into this equation, we find
\bq\lb{eq2.12}
\quad \ddot{\phi} + 3H\dot{\phi} + V_{,\phi} = 0,
\eq while in terms of $\rho$ and $P$, we also have
\bq\lb{eq2.13}
\dot b = -4\pi G\gamma(\rho+P) = -4\pi G \gamma \dot\phi^2.
\eq

It should be noted that,   Eqs.(\ref{FRa}) and (\ref{FRb}) hold only after the quantum bounce ($t \ge t_B$), as already indicated in these equations, at which we have $\rho(t_B) = \rcone$ and $H(t_B) = 0$, so the expansion factor reaches its minimal value $a_B \equiv a(t_B)$. When $t \gg t_B$ (or equivalently, $\rho/\rcone \ll 1$), Eqs.(\ref{FRa}) and (\ref{FRb}) reduce to their relativistic limits
\bqn
\lb{FRc1}
H^2 &\simeq& \frac{8\pi G}{3} \rho,\; (t \gg t_B), \\
\lb{FRc2}
\frac{\ddot{a}}{a} &\simeq&  - \frac{4\pi G}{3}\left(\rho + 3P\right), \; (t \gg t_B).
\eqn
In particular, it is interesting to note that $\rho/\rcone \simeq 10^{-12}$ at the onset of inflation \cite{Ashtekar:2011ni,Li:2021mop}.
Therefore, during the inflationary phase, the modified FR equations are well approximated by its classical limits (\ref{FRc1}) and (\ref{FRc2}).

In the pre-bounce phase ($t \le t_B$), the modified FR equations take the form \cite{Li:2021mop}
\begin{widetext}
\bqn
\lb{FRaa}
H^2 &=& \frac{8\pi G_{\alpha} \rho_\Lambda}{3}\left(1-\frac{\rho}{\rcone}\right) \left(1+\frac{\rho\left(1-2\gamma^2+\sqrt{1-\rho/\rcone}\right)}{4\gamma^2\rcone\left(1+\sqrt{1-\rho/\rcone}\right)}\right), \;\;\; (t \le t_B), \\
\lb{FRbb}
\frac{\ddot{a}}{a} &=& -\frac{4\pi G_{\alpha}}{3}\left(\rho+3P-2\rho_\Lambda\right)
+4\pi G_{\alpha} P \left(\frac{2-3\gamma^2+2\sqrt{1-\rho/\rcone}}{\left(1-5\gamma^2\right)\left(1+\sqrt{1-\rho/\rcone}\right)}\right)\frac{\rho}{\rcone} \nb\\
&&
-\frac{4\pi G_{\alpha} \rho^2\left(2\gamma^2 + 5\gamma^2\left(1+ \sqrt{1-\rho/\rcone}\right)  - 4 \left(1+\sqrt{1-\rho/\rcone}\right)^2\right)}{3\rcone\left(1-5\gamma^2\right)\left(1+\sqrt{1-\rho/\rcone}\right)^2}, \;\;\; (t \le t_B),
\eqn
\end{widetext}
where $G_{\alpha} \equiv \alpha G$, and
\bq\lb{eq2.18}
\alpha \equiv \frac{1-5\gamma^2}{\gamma^2+1}, \quad
\rho_\Lambda \equiv \frac{3}{8\pi G\alpha\lambda^2(1+\gamma^2)^2}.
\eq
From Eqs.(\ref{FRaa}) and (\ref{FRbb}) we can see that at the bounce $\rho(t_B) = \rcone$, the universe contracts to its minimal volume $v = v_0 a_B^3$ at $t = t_B$.
Afterward, it smoothly passes to the expansion phase, but is now described by Eqs.(\ref{FRa}) and (\ref{FRb}). The smoothness is shown explicitly in \cite{Li:2018opr,Li:2018fco,Li:2019ipm}, and can be also seen from  Eqs.(\ref{eq2.12}) and (\ref{eq2.13}), which hold across the bounce.

When $t \ll t_B$ (or $\rho/\rcone \ll 1$),  Eqs.(\ref{FRaa}) and (\ref{FRbb}) reduce to
\bqn
\lb{FRccA}
H^2 &\simeq& \frac{8\pi G_{\alpha}}{3} \rho_{\Lambda} \left(1 - \frac{\rho}{\rho_{\Lambda}}\right),
\; (t \ll t_B),\\
\lb{FRccB}
\frac{\ddot{a}}{a} &\simeq&   \frac{8\pi G_{\alpha}}{3}\rho_{\Lambda}\left( 1 - \frac{\rho + 3P}{2\rho_{\Lambda}}\right), \; (t \ll t_B), ~~~~
\eqn which are quite different from Eqs.(\ref{FRc1}) and (\ref{FRc2}). In particular, the effective Planck-scale cosmological constant $\rho_{\Lambda}$ soon dominates the evolution of the pre-bounce phase, whereby a de Sitter spacetime is obtained in the pre-bounce phase but with a Planck-scale cosmological constant $\rho_{\Lambda} \simeq {\cal{O}}(\rho_{\text{pl}})$. In addition, the Newtonian constant $G$ is replaced by $G_{\alpha} (= \alpha G)$, where $\alpha$
is defined by Eq.(\ref{eq2.18}). More remarkably, this Planck-scale cosmological constant is filtered out by the quantum bounce and disappears miraculously after the bounce, whereby the classical FR equations are obtained, as shown explicitly by Eqs.(\ref{FRc1}) and (\ref{FRc2}). This is significantly different from LQC \cite{Ashtekar:2011ni}, in which the evolution of the Universe is symmetric with respect to the bounce \footnote{More precisely, it is symmetric for kinetic energy-dominated initial conditions $\dot{\phi}_B^2 \gg 2V(\phi_B)$ \cite{Ashtekar:2011ni,Li:2021mop}.}.

\begin{figure}[h!]
\includegraphics[width=0.85\linewidth]{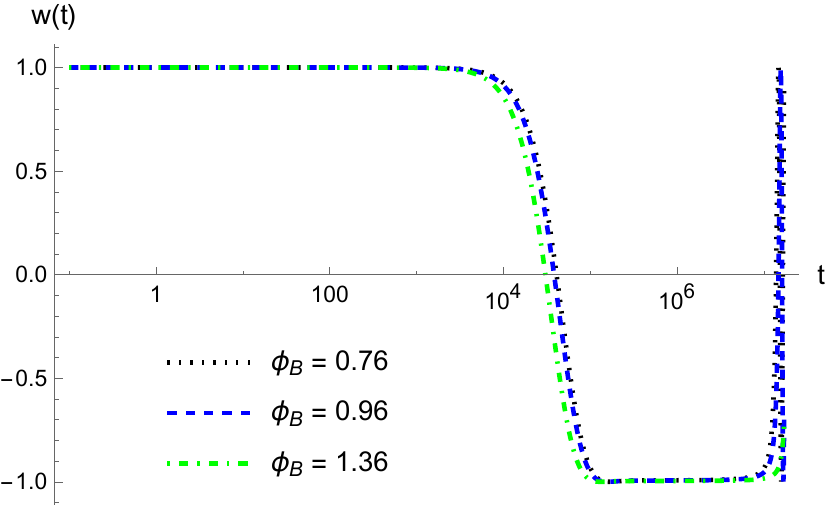}
\caption{The evolution $w(t)$ of the equation of state in the post-bounce phase ($t \ge t_B$) for the chaotic potential with different initial conditions $\phi_B$ for $\dot\phi_B > 0$. In this plot, $t_B$ is set to zero. It should be noted that the equation of state, $w(t)$, behaves the same not only for  $\dot\phi_B < 0$ with the same  chaotic potential, but also for other potentials, as long as the evolution of the universe at the bounce is kinetic-energy dominated \cite{Li:2018opr,Li:2018fco,Li:2019ipm,Li:2021mop}.}
\label{fig1}
\end{figure}

The background evolution of the universe in the post-bounce regime ($t \geq t_B$) has been studied extensively  in \cite{Li:2018opr,Li:2018fco,Li:2019ipm,Li:2021mop}, with various potentials, including chaotic, fractional monodromy, Starobinsky, and non-minimally coupling Higgs potentials. Among several remarkable features, it was found that {\em the evolution of the universe is universal for any given potential, as long as the  initial conditions imposed at the bounce are kinetic-energy dominated}, that is
\bq
\lb{eq3.1}
\frac{1}{2}\dot\phi_B^2 \gg V(\phi_B),
\eq where $\phi_B (\equiv \phi(t_B))$ is the value of the scalar field at the bounce. 
Moreover, evolution can be universally divided into three phases:
\bqn
\lb{eq3.1b}
&& (i) \; \text{bouncing} \; (w(t) \simeq 1),\nb\\
&& (ii)\; \text{transition}\;  (-1 < w(t) < 1), \nb\\
&& (iii) \; \text{inflationary},\; (w(t) \simeq -1),
\eqn as shown in Fig. \ref{fig1}. Here, the equation of state of the scalar field  $w(t)$ is defined as follows:
\begin{equation}
\label{eq3.2}
w(t) \equiv \frac{P(t)}{\rho(t)} = \frac{\frac12 \dot\phi^2-V}{\frac12 \dot\phi^2+V}.
\end{equation}

It should be noted that in Fig. \ref{fig1}, only the case of the chaotic potential was considered. However, the general behavior of the evolution of the universe is universal and independent of the choice of potential, as long as the initial conditions are kinetic energy dominated. These remarkable features are also shared by LQC, first found in \cite{Zhu:2017jew} and later confirmed in various cases \cite{Shahalam:2017wba,Shahalam:2018rby,Sharma:2018vnv,Sharma:2019okc,Shahalam:2019mpw}.

\begin{figure}[h]
\graphicspath{ {./Plots/} }
\includegraphics[width=7cm]{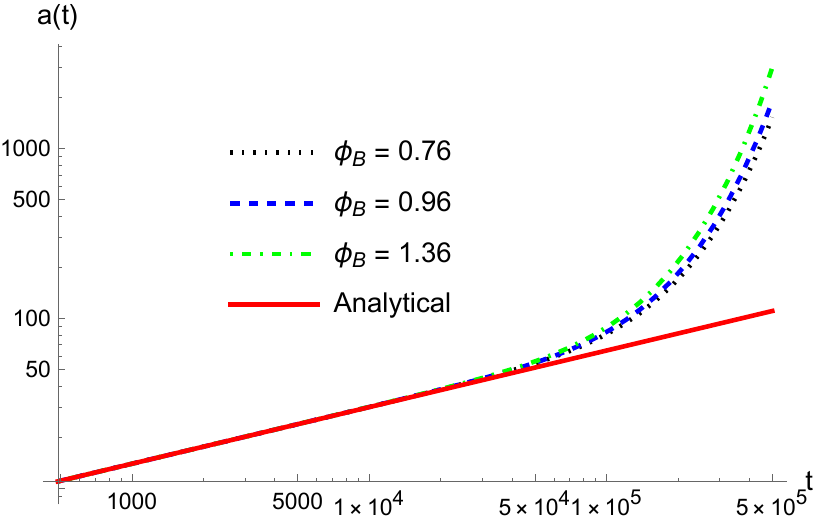}\\
\vspace{.5cm}
\includegraphics[width=7cm]{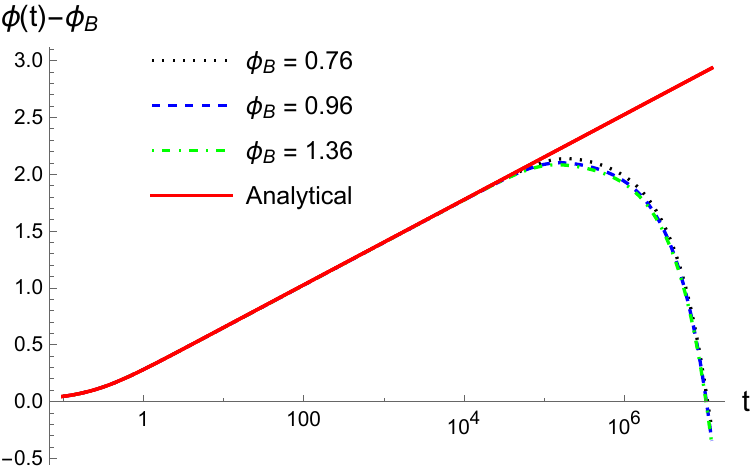}
\caption{The numerical and analytical solutions of $a\left(t\right)$ and $\phi(t)$ for different initial values of $\phi_B$ that result in at least 50 e-folds for the chaotic potential with $\dot\phi_B > 0$.
This universal behavior is true for any given inflation potential $V(\phi)$ as long as the condition (\ref{eq3.1}) holds at the bounce.}
\label{fig2}
\end{figure}

Moreover, during the bouncing phase, the expansion factor $a(t)$ and the scalar field $\phi(t)$ can be well described by the analytical solution \cite{Li:2019ipm}
\bqn
\lb{eq3.3}
a(t) &=& \left(1+24\pi\rho^I_c\left( 1+\frac{A_0\gamma}{1+B_0 t} \right)t^2   \right)^\frac{1}{6},\nb\\
\phi(t) &=& \phi_B +\text{sgn}\left(\dot\phi_B\right) \frac{\text{sinh}^{-1}{\sqrt{24\pi\rho^I_c\left(1+\frac{C_0\gamma^2}{1+D_0t}\right)t}}}{\sqrt{12\pi\left(1+\frac{C_0\gamma^2}{1+D_0t}\right)}},\nb\\
\eqn where $A_0 = 1.2$, $B_0 = 6$, $C_0 = 1.2$, and $D_0 = 2$ with relative errors no larger than $0.3\%$,
As shown in Fig. \ref{fig2}. Again, the analytical solutions are universal and independent of the choice of inflationary potential $V(\phi)$ as long as the initial kinetic-energy dominated condition (\ref{eq1.1}) holds.

In addition, it was shown that the probability of the occurrence of a desired slow-roll inflation is generic. In particular, it was found that for such a desired slow-roll inflation to not occur, the probability is \cite{Li:2019ipm}
\bq\lb{eq3.4}
P_{\;\text{mLQC-I}}(\text{not realized})\lesssim  1.12\times 10^{-5},
\eq
where ``desired" means the one that is consistent with current observations. 
This is comparable to
$P_{\text{LQC}}(\text{not realized})\lesssim2.74\times 10^{-6}$, obtained in
LQC \cite{Ashtekar:2011rm,Ashtekar:2009mm}.

Therefore, in this paper, we shall focus on the evolution of the universe in the pre-bounce regime ($t \leq t_B$) and pay particular attention to the universal properties of evolution.

\section{Numerical solutions of the Evolution of the Universe}
\renewcommand{\theequation}{3.\arabic{equation}}
\setcounter{equation}{0}
\lb{SecIII}

Since the kinetic-energy dominated initial conditions imposed at the bounce always lead to slow-roll inflation for any given potential, in the rest of this paper, we shall consider only such initial conditions with several well-studied inflationary potentials \cite{Planck:2018jri,Kallosh:2022feu}, including {\em chaotic, Starobinsky, $\alpha$-attractive, and natural inflation potentials}.

The advantage of imposing the initial conditions at the bounce is that $\dot\phi_B$ is determined uniquely up to a sign for any given initial data $\phi_B$ via the relation $\rho(t_B) = \rho^{\text{I}}_c$, which yields
\bq\lb{Initial condition relation}
\dot\phi_B = \pm\sqrt{2(\rcone-V(\phi_B))}.
\eq
On the other hand, from Eqs.(\ref{FRa})-(\ref{FRb}) and  (\ref{FRaa})-(\ref{FRbb}) we can see that these equations are scaling-invariant with respect to the expansion factor $a \rightarrow a/L_o$. Therefore, without loss of generality, we can always set $a_B = 1$. Then, the initial conditions are reduced to the choice of
\bq
\lb{eq3.3b}
\left(\phi_B, \text{sgn}\left(\dot\phi_B\right)\right).
\eq
In addition, using the translation invariance $t \rightarrow t + t_0$, in the rest of this paper, we shall set $t_B = 0$.

To process further, following \cite{Li:2018vzr,Li:2018fco,Li:2019ipm,Li:2021mop}, let us first introduce the following physical quantities:

\begin{itemize}

\item The first-order Hubble rate and potential slow-roll parameters
\subbq\lb{Hubble rate and potential slow-roll parameters}
\begin{align}
\epsilon_H &= -\frac{\dot H}{H^2}, &\eta_H = - \frac{\ddot H}{2H \dot H}, \\
\epsilon_V &= \frac{1}{16\pi G}\left( \frac{V_{,\phi}}{V}\right)^2,  &\eta_V = \frac{V_{,\phi\phi}}{8\pi G V}.
\end{align}
\subeq
These sets of slow-roll parameters are typically used for different purposes \cite{Baumann:2009ds}. In particular, the slow-roll parameters with the subscript ``$V$" can be used to determine which part of the potential can successfully drive inflation. On the other hand, slow-roll parameters with subscript ``$H$" are used for numerical simulations to define when slow-roll inflation begins and ends. In the classical regime, the scale factor acceleration equation satisfies the following relation:
\begin{equation}\label{a and H relation}
\ddot a = aH^2(1-\epsilon_H).
\end{equation}
The Universe experiences an accelerated expansion when $\epsilon_H < 1$, whereas slow-roll inflation occurs only when $\epsilon_H \ll 1$ and $\eta_H \ll 1$  \cite{Baumann:2009ds}. For the sake of concreteness, we define the onset of inflation as the time $t_i$ when $|\eta_H| = 0.03$ for the first time in the transition phase. The end of the slow-roll inflation is defined at the time $t_{\text{end}}$ when $|\epsilon_H| = 1$ for the first time after $t_i$.

\item The e-fold $N$ during the inflationary phase: This number is usually defined as follows:
\begin{equation}\label{efolds definition}
N  = \ln\left({\frac{a(t_{\text{end}})}{a(t_i)}}\right).
\end{equation}
To have a successful slow-roll inflation, the inflation potential has to be very flat, so that the Universe can expand large enough \cite{Baumann:2009ds}. All the cosmological problems can be resolved if the Universe expands about 60 e-fold during the inflationary phase, although its exact value depends on the inflationary models \cite{Planck:2018jri}. Therefore, 
in the following one will see that the minimal e-fold $N_{\text{min}}$ will be different in different models.

\end{itemize}

With the above in mind, we are now ready to solve the four dynamical equations (\ref{eqA}) - (\ref{eqD}) for any given initial conditions (\ref{eq3.3b}) \footnote{It is found that numerically it is more convenient to solve the dynamical equations (\ref{eqA}) - (\ref{eqD}) than the ones of Eqs.(\ref{FRa})-(\ref{FRb}) and (\ref{FRaa})-(\ref{FRbb}), although they should give the same results. However, in the latter, the integration needs to be carried out in the pre- and post-bounce separately, and then connect them smoothly across the bounce.}.  In the following, we shall study the chaotic, Staribinsky, $\alpha$-attractor and natural inflation potentials, separately.

\begin{figure}[H]
\includegraphics[width=0.95\linewidth]{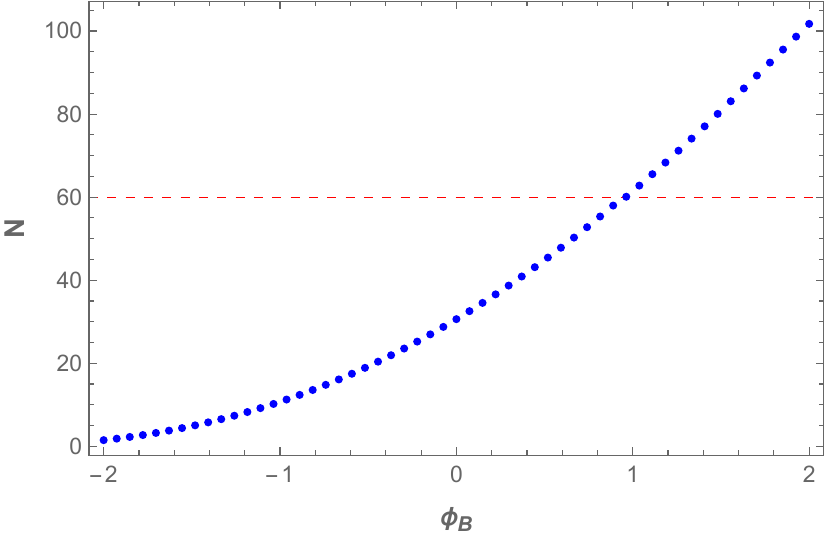}
\caption{The e-folds $N$ during the inflationary period for the chaotic potential (\ref{Chaotic Potential}) with $m = 1.26 \times 10^{-6}\; m_{pl}$ for $\dot \phi_B > 0$. The case $\dot\phi_B < 0$ can be obtained by the symmetry $(\phi, \dot\phi) \rightarrow (-\phi, -\dot\phi)$, a particular property owned by this model.}
\lb{fig3}
\end{figure}

\subsection{Chaotic Potential}

The chaotic inflationary potential is given by \cite{Linde:1983gd}
\begin{equation}
\label{Chaotic Potential}
V(\phi) = \frac12 m^2\phi^2,
\end{equation}
where the mass is set to $m = 1.26 \times 10^{-6}\; m_{pl}$ \cite{Planck:2018jri}. Since it has already been shown that the occurrence of a desired slow-roll inflation is generic  \cite{Li:2019ipm}, it is sufficient for us to find the initial conditions that lead to   $N \ge 45$ e-folds.

In this model, the modified Friedmann and Klein-Gordon equations are invariant under the replacement
\bq
\lb{symmetry}
(\phi, \dot\phi) \rightarrow (-\phi, -\dot\phi),
\eq rendering it sufficient to consider only half of the parameter space with $\dot \phi_B > 0$. Here, $\phi_B$ can assume any value in the range $|\phi_B| \le \phi^I_{\text{max}} = \sqrt{2\rho^I_c}/m$, where $\phi^I_{\text{max}}$ is determined by $V(\phi^I_{\text{max}}) = \rho^I_c$ \cite{Li:2019ipm}.

\begin{figure}[H]
\includegraphics[width=7.7cm]{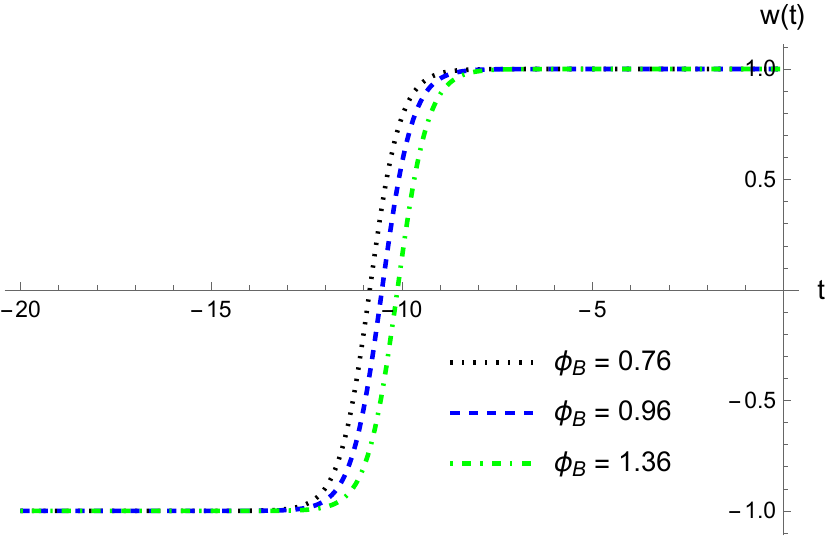}\\
\centerline{(a)}\\
\includegraphics[width=7.7cm]{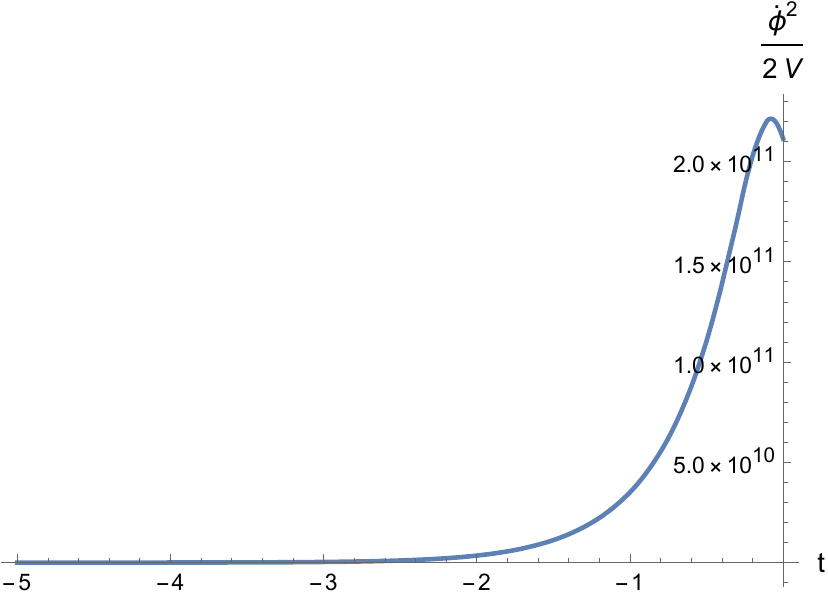}\\
\centerline{(b)}\\
\includegraphics[width=7.7cm]{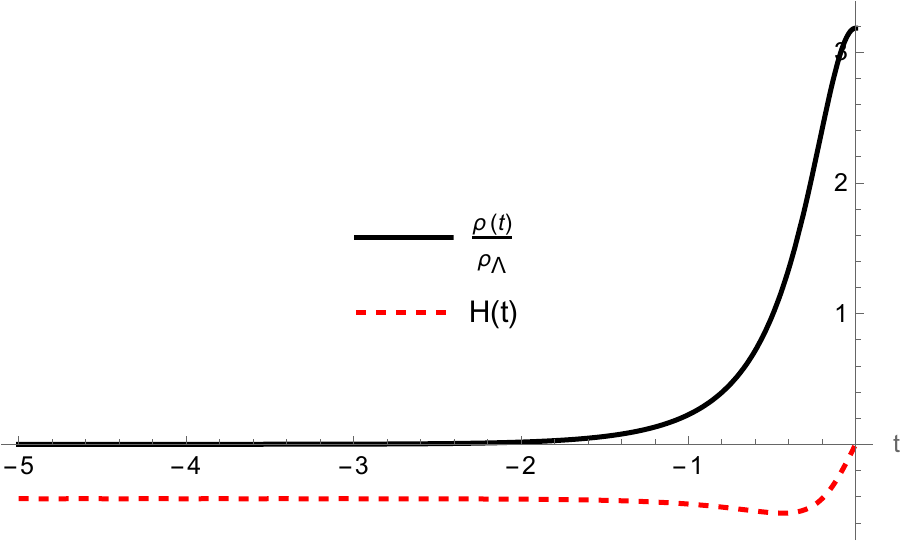}\\
\centerline{(c)}\\
\caption{(a) The evolution of the equation of state in the pre-bounce phase for the chaotic potential with various choices of
$\phi_B$ for $\dot\phi_B > 0$. (b) Ratio between the kinetic energy and potential of the scalar field. (c) The ratio $\rho(t)/\rho_{\Lambda}$ between the energy density of the scalar field and the effective cosmological constant, together with the Hubble parameter $H(t)$.}
\label{fig4}
\end{figure}

\begin{figure}[H]
\includegraphics[width=7.7cm]{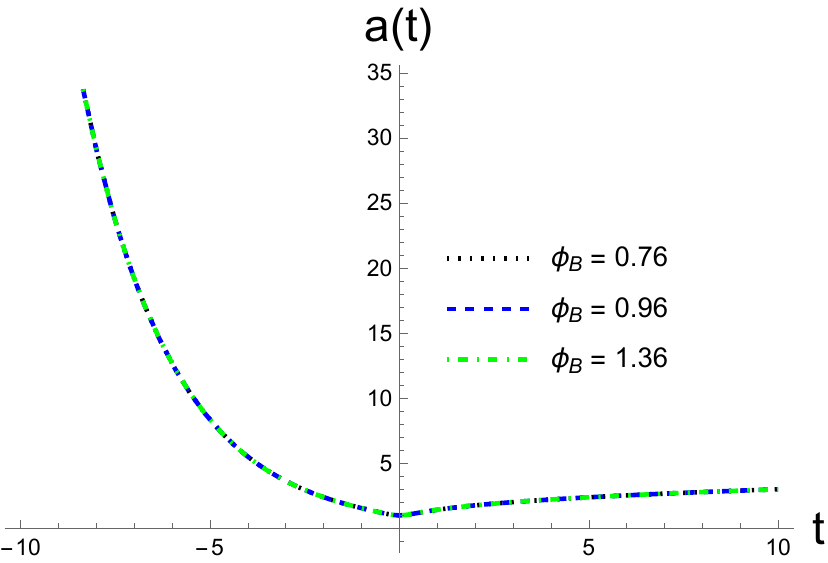}\\
\includegraphics[width=7.7cm]{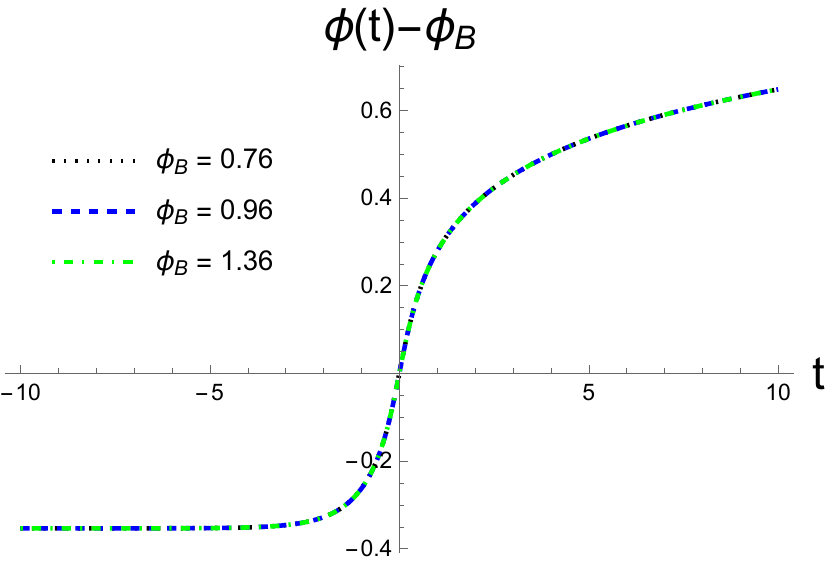}
\caption{The numerical solutions of $a(t)$ and $\phi(t)$  for the chaotic potential with $\dot\phi_B > 0$. In the figures, the numerical curves with different initial conditions are indistinguishable.}
\label{fig5}
\end{figure}

\begin{figure*}[htbp]
\resizebox{\linewidth}{!}
{\begin{tabular}{cc}
\includegraphics[height=4.cm,width=7cm]{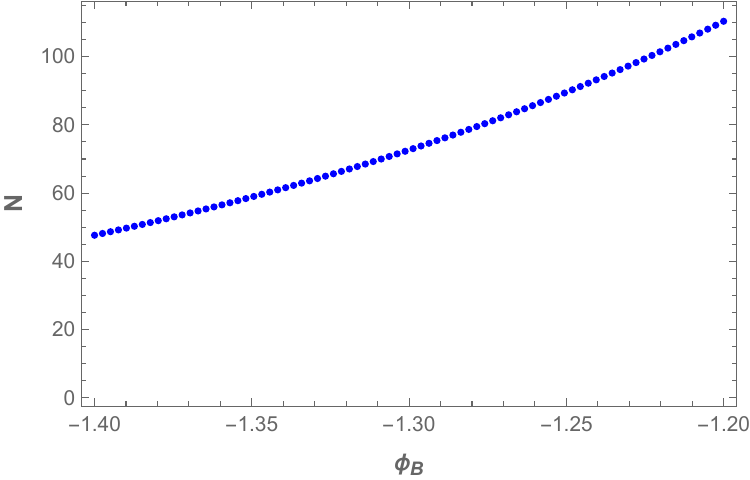}
\includegraphics[height=4.cm,width=7cm]{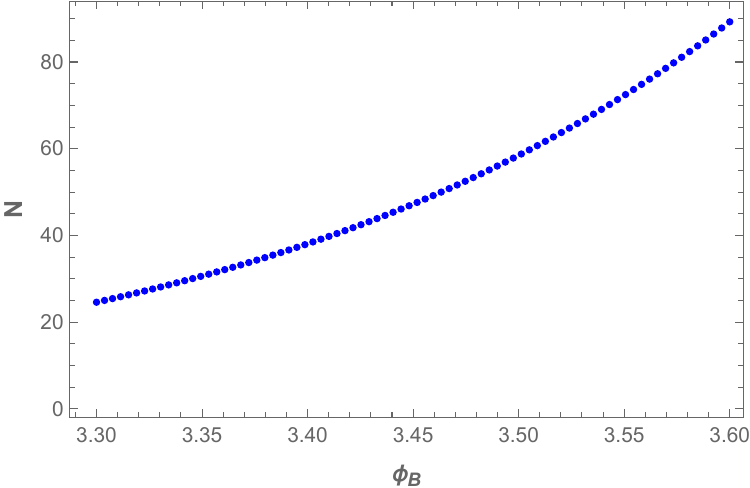}\\
\end{tabular}}
\caption{E-folds produced during slow inflation for different values of $\phi_B$ for the Starobinsky potential (\ref{eq30}). {\bf Left-panel}: $\dot\phi_B >0$. {\bf Right-panel}: $\dot\phi_B < 0$. }
\label{fig6}
\end{figure*}

For any chosen initial condition $\phi_B$ with $\dot\phi_B > 0$, which yields approximately 60 e-folds [cf. Fig. \ref{fig3}],
we numerically solve the dynamical equations (\ref{eqA}) - (\ref{eqD}), and find that the evolution of the universe in the post-bounce regime ($t \ge 0$) can always be divided into three different phases, {\em bouncing, transition and inflation}, as shown in Fig. \ref{fig1}. In Fig. \ref{fig2}, three numerical solutions are shown together with the analytical solution given by Eq.(\ref{eq3.3}) during the bouncing phase, $t \in (0, 10^{4})\; t_{\text{pl}}$. Note that the universal behavior of the evolution of the universe during the bouncing phase does not depend on the specific choice of $\phi_B$, as long as the kinetic-energy dominated condition
(\ref{eq1.1}) is satisfied at the bounce \cite{Li:2019ipm}.

With the same initial conditions imposed at the bounce, we numerically solve the dynamic equations (\ref{eqA})- (\ref{eqD}) for $t \le 0$ and obtain the solutions in the pre-bounce regime. In Fig. \ref{fig4} (a), the corresponding equation of state $w(t)$ defined by Eq.(\ref{eq3.2}) is plotted. From this figure, we can see that the evolution of the universe in the pre-bounce phase can be also {\em universally} divided into three different phases:
\bqn
\lb{eq3.11}
&& (i) \; {\text pre-bouncing} \; (w(t) \simeq 1), \nb\\
&& (ii)\; {\text pre-transition}\;  (- 1 < w(t) < 1), \nb\\
&& (iii) \; {\text pre-de \; Sitter} \; (w(t) \simeq - 1).
\eqn

The transition occurs at $t \simeq -10 t_{\text{pl}}$ and is very rapid. From Fig. \ref{fig4} (b), we can see that this is because the kinetic energy of the scalar field decreases dramatically as time becomes more and more negative and is soon dominated by its potential energy. In addition, the effective cosmological constant quickly dominates the evolution of the universe $\rho(t)/\rho_{\Lambda} \ll 1$, as shown in Fig. \ref{fig4} (c), whereby the Hubble parameter becomes a (negative) constant, and spacetime quickly approaches de Sitter. As a result, the pre-bouncing phase is also very short compared with the pre-de Sitter phase.

In Fig. \ref{fig5}, the numerical  solutions of $a(t)$ and $\phi(t)$  are plotted for three different initial conditions, which all satisfy (\ref{eq1.1}). Form this figure it can be seen that the numerical solutions are indistinguishable. In fact, they are not only indistinguishable with different initial conditions, but also indistinguishable with different potentials, as long as the condition (\ref{eq1.1}) holds at the bounce, as to be shown below.

\subsection{Starobinsky Potential}

The Starobinsky potential results from adding an $R^2$ term to the Einstein-Hilbert action \cite{starobinskii1979spectrum}. After a conformal transformation, in the Einstein frame, the theory is equivalent to a scalar field with the potential \cite{PhysRevD.37.858}
\begin{equation}\label{eq30}
V(\phi) = \frac{3m^2}{32\pi G}\left(1-\exp\left(-\sqrt{\frac{16\pi G}{3}}\phi\right)\right)^2,
\end{equation}
where the mass is set to $2.49 \times 10^6 m_{\text{pl}}$  \cite{Planck:2018jri}.
Unlike the case with chaotic potential,  now the slow-roll inflation can take place only when   $\phi > \phi_{\text{end}} \simeq 0.188 m_{\text{pl}}$ \cite{Li:2019ipm}.\\

\begin{figure*}[htbp]
\resizebox{\linewidth}{!}
{\begin{tabular}{cc}
\includegraphics[height=4.cm,width=7cm]{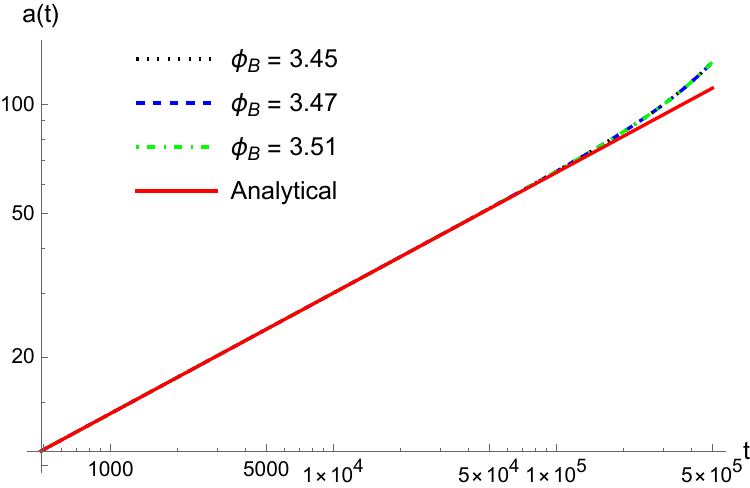}
\includegraphics[height=4.cm,width=7cm]{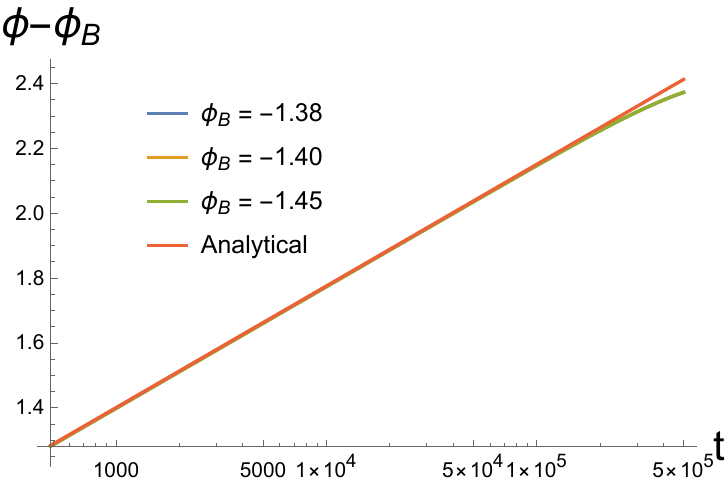}\\
(a) \\[6pt]
\vspace{.5cm}
\includegraphics[height=4.cm,width=7cm]{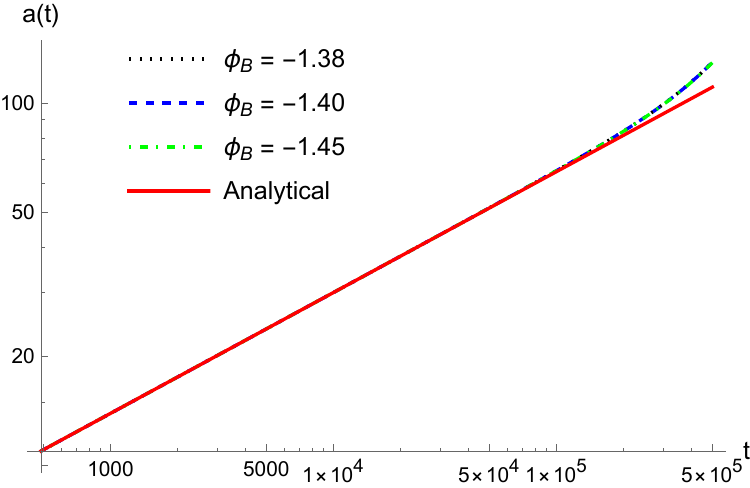}
\includegraphics[height=4.cm,width=7cm]{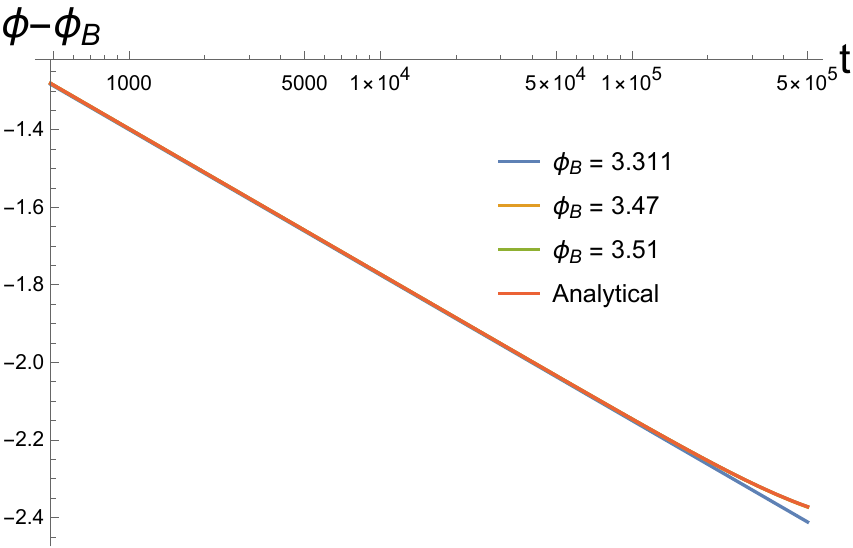}\\
(b)\\
\end{tabular}}
\caption{The numerical solutions $a(t)$ and $\phi(t)$ with indicated initial conditions for the Starobinsky potential (\ref{eq30}) in the post-bounce regime ($t > 0$), where (a) For $\dot\phi_B > 0$, and
(b) for $\dot\phi_B < 0$. To compare with the analytical solutions given by
Eq.(\ref{eq3.3}), we also plot them out, denoted by the red solid lines.}
\label{fig7}
\end{figure*}

\begin{figure*}[htbp]
\resizebox{\linewidth}{!}
{\begin{tabular}{cc}
\includegraphics[height=3cm,width=6cm]{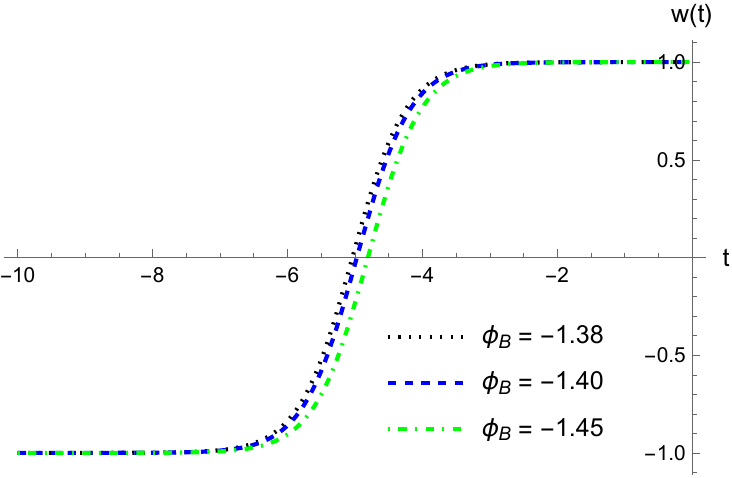}
\includegraphics[height=3cm,width=6cm]{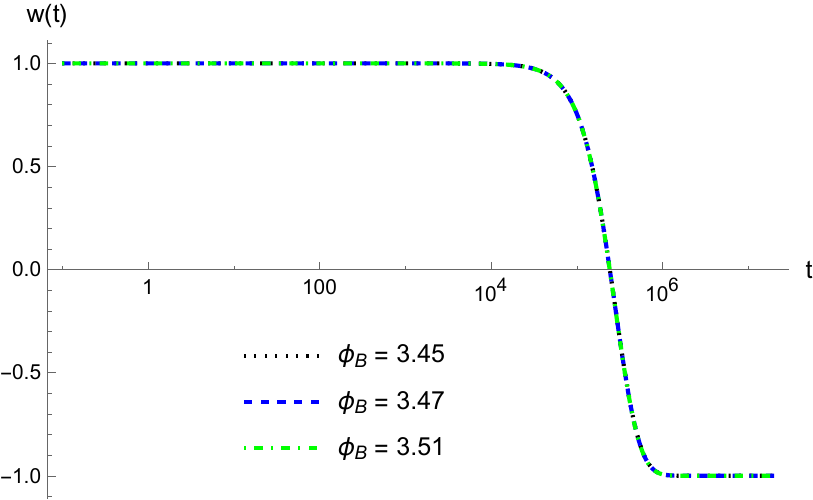}\\
(a) \\
\vspace{.5cm}
\includegraphics[height=3cm,width=6cm]{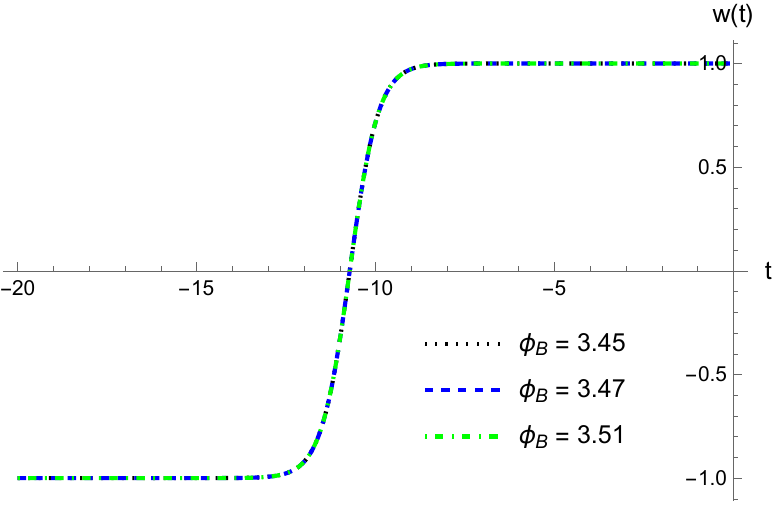}
\includegraphics[height=3cm,width=6cm]{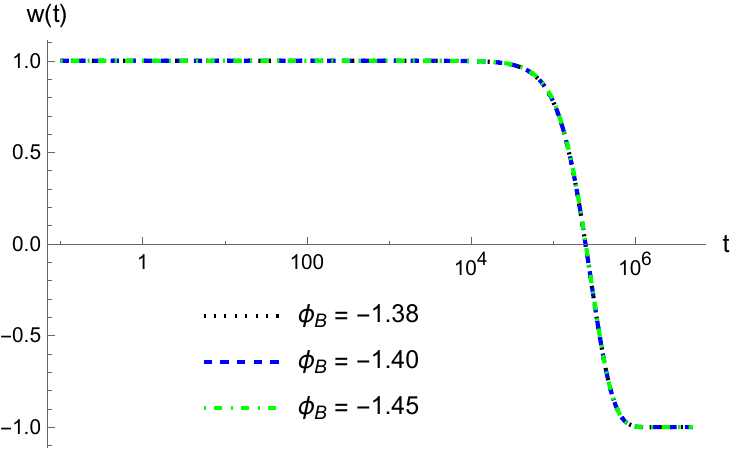}\\
(b)\\
\end{tabular}}
\caption{The equation of state for the Starobinsky potential (\ref{eq30}) for different values of $\phi_B$ that result in greater than 50 e-Folds for (a)  $\dot\phi_B > 0$ and (b) $\dot \phi_B < 0$.}
\label{fig8}
\end{figure*}

\begin{figure*}[htbp]
\resizebox{\linewidth}{!}
{\begin{tabular}{cc}
\includegraphics[height=4.cm,width=7cm]{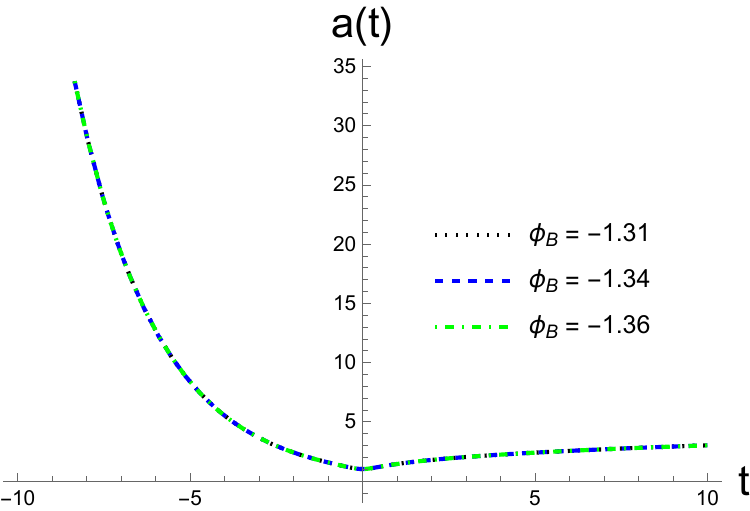}
\includegraphics[height=4.cm,width=7cm]{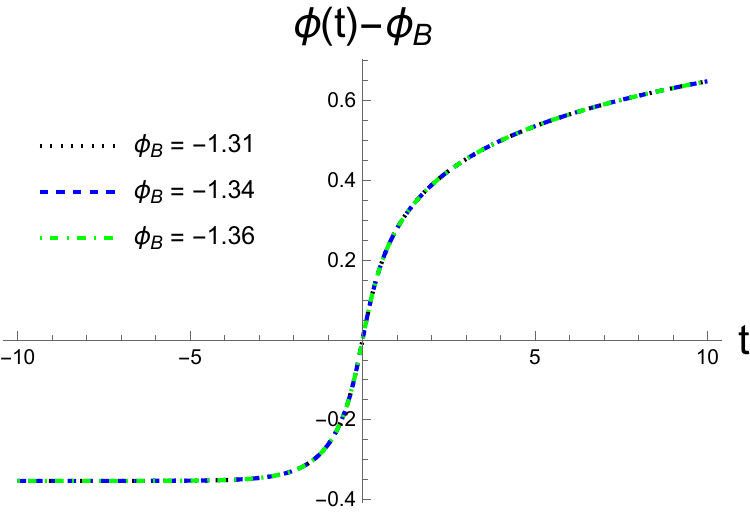}\\
(a)\\
\vspace{.1cm}\\
\includegraphics[height=4.cm,width=7cm]{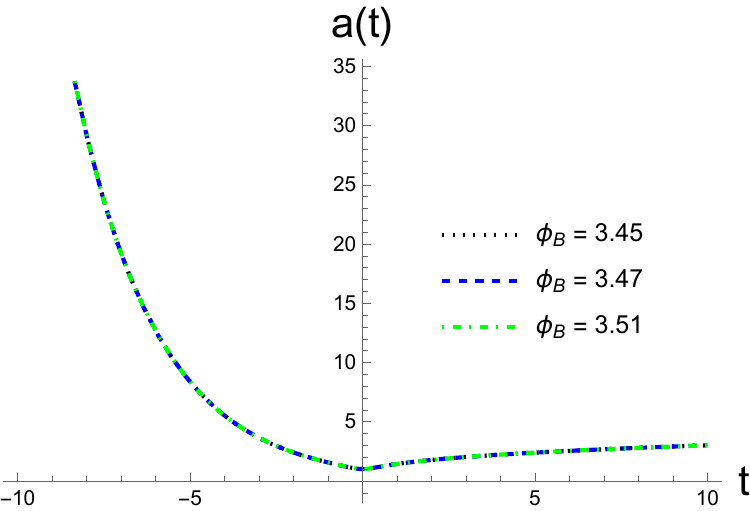}
\includegraphics[height=4.cm,width=7cm]{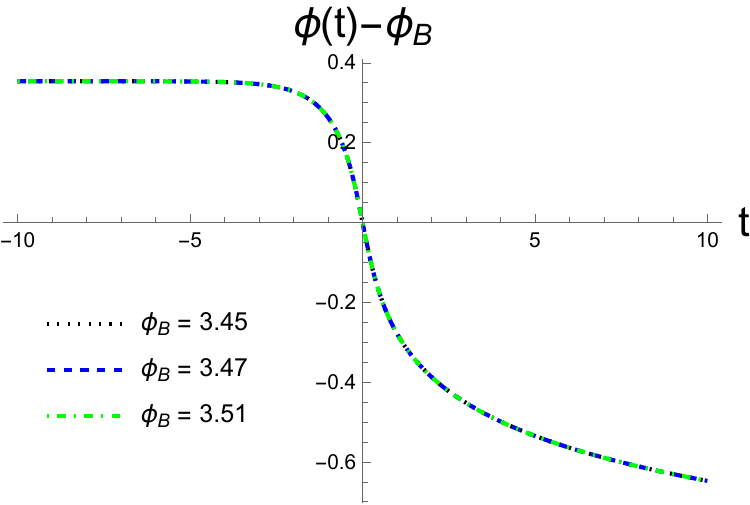}\\
(b)
\end{tabular}}
\caption{The numerical  solutions of $a(t)$ and $\phi(t)$ with various initial values of $\phi_B$ for  the Starobinsky potential (\ref{eq30}), where (a) for $\dot\phi_B > 0$ and (b)
For $\dot\phi_B < 0$.}
\label{fig9}
\end{figure*}

In addition, the Starobinsky potential does not share the symmetry of the chaotic one, given by Eq.(\ref{symmetry}). Therefore, in the following, we consider both $\dot\phi_B > 0$ and $\dot \phi_B < 0$ separately.
In particular, in Fig. \ref{fig6} we plot the e-fold $N$ of inflation defined by Eq.(\ref{efolds definition}) vs different initial conditions $\phi_B$ for
$\dot\phi_B > 0$ and $\dot\phi_B < 0$, respectively.

To confirm the universality of the evolution of the universe during the post-bounce regime ($t \ge 0$), in Fig. \ref{fig7} we  plot out $a(t)$ and $\phi(t)$ with the initial conditions
\bqn
\lb{eq3.19a}
\phi_B = \begin{cases}
1.38,\; -1.40,\; -1.45, & \dot\phi_B > 0, \cr
3.45,\;\;\;\;\; 3.47,\;\;\;\;\; 3.51, & \dot\phi_B < 0, \cr
\end{cases}
\eqn so that during slow-roll inflation, about 60 e-folds can be produced. 

On the other hand, with the same initial conditions as those chosen in Fig. \ref{fig7}, we plot the equation of state $w(t)$ in the post-bounce ($t \ge 0$) and the pre-bounce ($t \le 0$) regimes in Fig. \ref{fig8}. 
From these figures, it can be clearly seen that in the post-bounce regime, three different phases defined  by Eq.(\ref{eq3.1b}) exist. The only difference from the chaotic potential case is that the transition phase now occurs at approximately $t \simeq 10^{5}\; t_{\text{pl}}$, instead of $t \simeq 10^{4}\; t_{\text{pl}}$ as shown in Fig. \ref{fig1}.

Equally remarkable, the evolution of the universe in the pre-bounce regime ($t \le 0$) is also universally divided into three phases defined by Eq.(\ref{eq3.11}). However, the time $t_m$ defined by $\dot{H}(t_m) \simeq 0$
in the pre-transition phase weakly depends on the initial values of $\phi_B$ but can be significantly different with different signs of $\dot\phi_B$, as shown in Fig. \ref{fig8}, where $t_m \simeq -5 t_{\text{pl}}$
for $\dot\phi_B > 0$, and $t_m \simeq -10.5 t_{\text{pl}}$
for $\dot\phi_B < 0$. \\

In Figs. \ref{fig9},  the numerical solutions of $a(t), \; \phi(t)$
for several different choices of initial conditions $\phi_B$ with
$\dot\phi_B > 0$ and $\dot\phi_B < 0$ are plotted out, respectively. From the figures we can see that
the scalar field $\phi(t)$ quickly approaches a constant, which is negative  for $\dot\phi_B > 0$ and positive for $\dot\phi_B < 0$.
\newpage
\subsection{$\alpha$- attractor Potentials}

\begin{figure*}[htbp]
\resizebox{\linewidth}{!}
{\begin{tabular}{cc}
\includegraphics[height=4.cm,width=7cm]{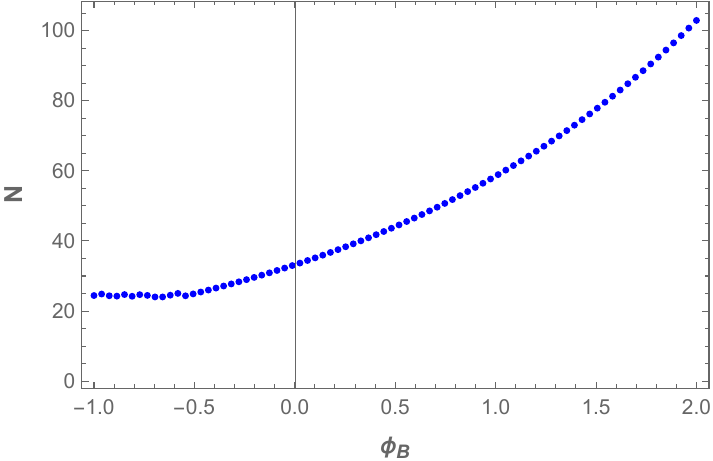}
\includegraphics[height=4.cm,width=7cm]{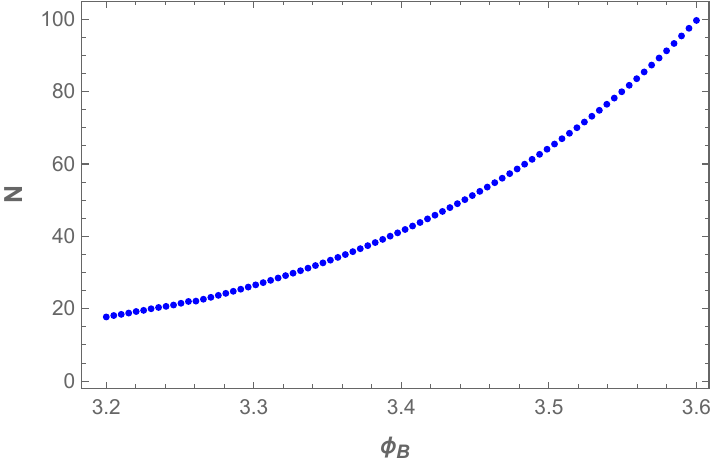}
\end{tabular}}
\caption{E-folds for different values of $\phi_B$ for the generalized Starobinsky potentials (\ref{eq3.20}) with $\alpha  = 0.0962$ and  $V_0 = 10^{-12}$ GeV.}
\label{fig10}
\end{figure*}

\begin{figure*}[htbp]
\resizebox{\linewidth}{!}
{\begin{tabular}{cc}
\includegraphics[height=4.cm,width=6cm]{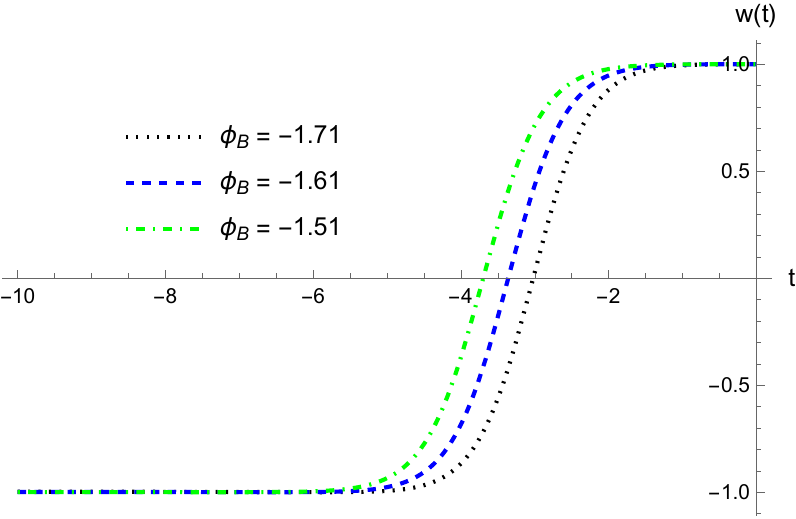}
\includegraphics[height=4.cm,width=6cm]{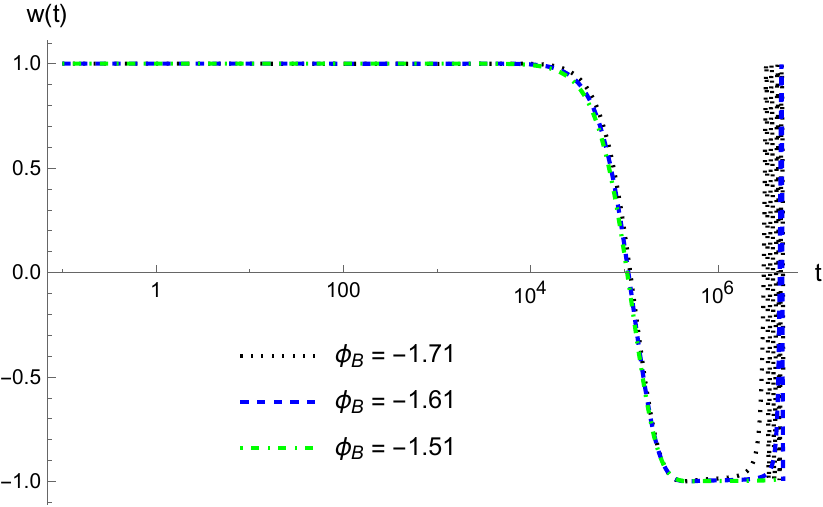}\\
(a) \\
\vspace{.5cm}
\includegraphics[height=4.cm,width=6cm]{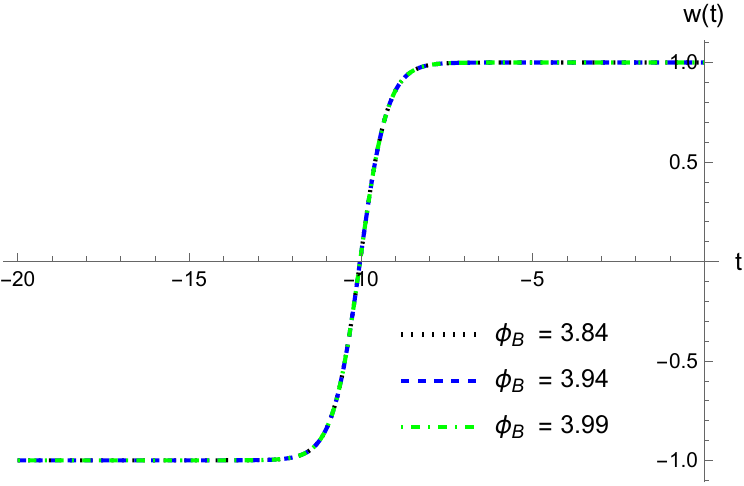}
\includegraphics[height=4.cm,width=6cm]{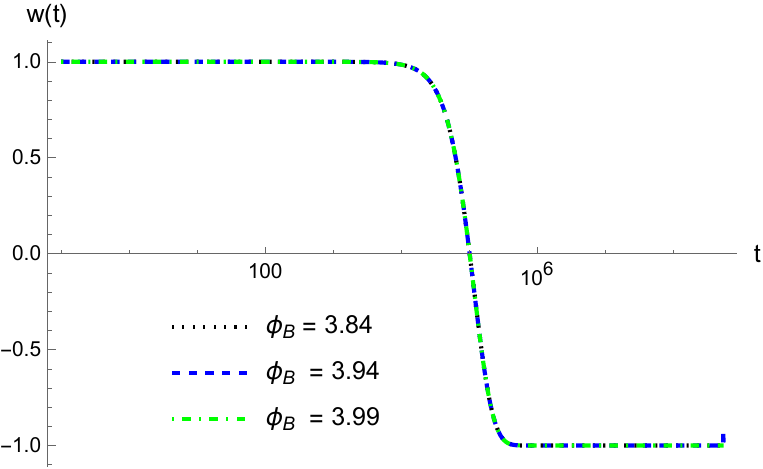}\\
(b)\\
\end{tabular}}
\caption{The equation of state for the generalized Starobinsky potentials (\ref{eq3.20}) for different values of $\phi_B$ that result in at least 50 e-Folds   with $\alpha  = 0.0962$ and  $V_0 = 10^{-12}$ GeV, where  (a)  for $\dot\phi_B > 0$ and (b) for $\dot \phi_B < 0$.}
\label{fig11}
\end{figure*}

\begin{figure*}[htbp]
\resizebox{\linewidth}{!}
{\begin{tabular}{cc}
\includegraphics[height=4.cm,width=7cm]{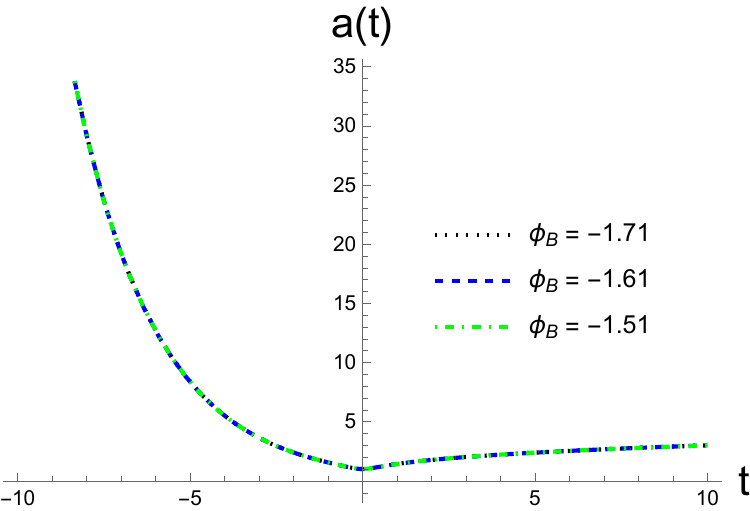}
\includegraphics[height=4.cm,width=7cm]{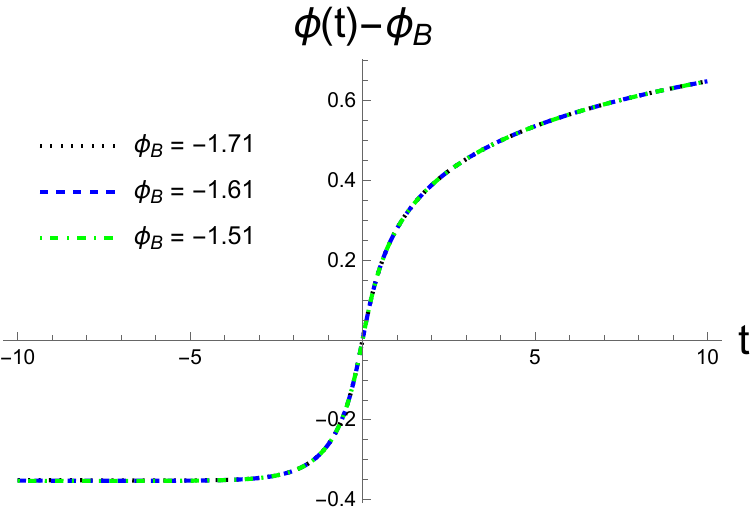}\\
(a)\\
\vspace{.1cm}\\
\includegraphics[height=4.cm,width=7cm]{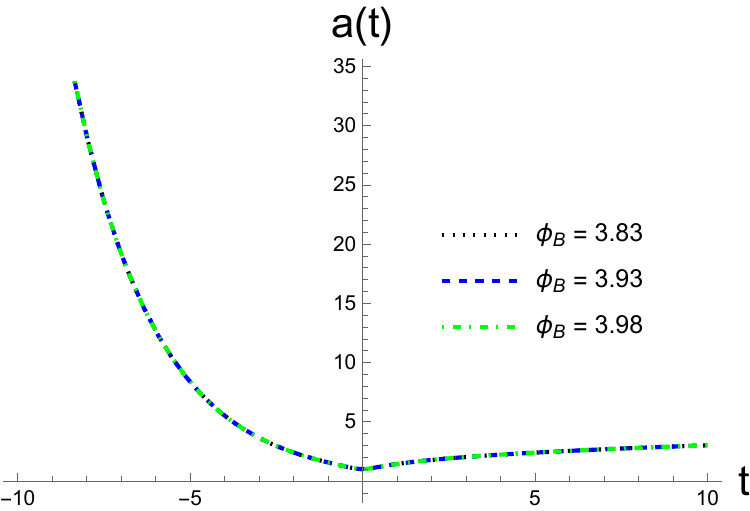}
\includegraphics[height=4.cm,width=7cm]{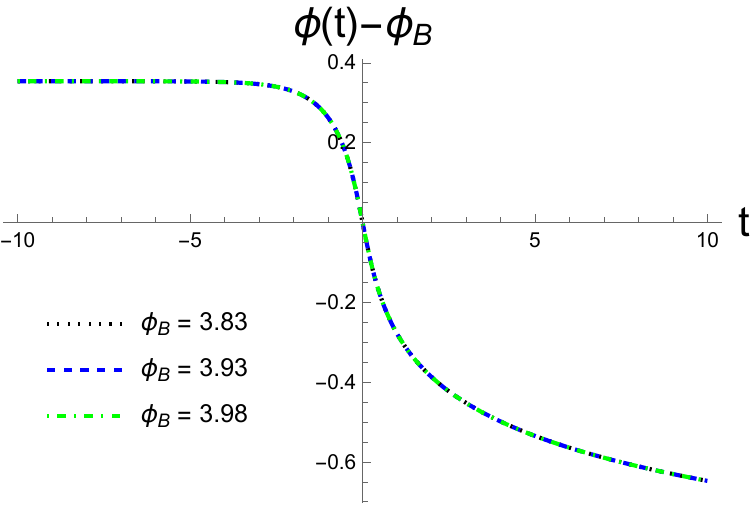}\\
(b)
\end{tabular}}
\caption{The numerical   solutions of   $a(t)$
and $\phi(t)$ for  the generalized Starobinsky potentials (\ref{eq3.20})  with $\alpha = 0.0962$ and $V_0 = 10^{-12}$ GeV. (a) For $\dot\phi_B >0$, and (b) for $\dot\phi_B < 0$.}
\label{fig12}
\end{figure*}

\begin{figure*}[htbp]
\resizebox{\linewidth}{!}
{\begin{tabular}{cc}
\includegraphics[width=7cm]{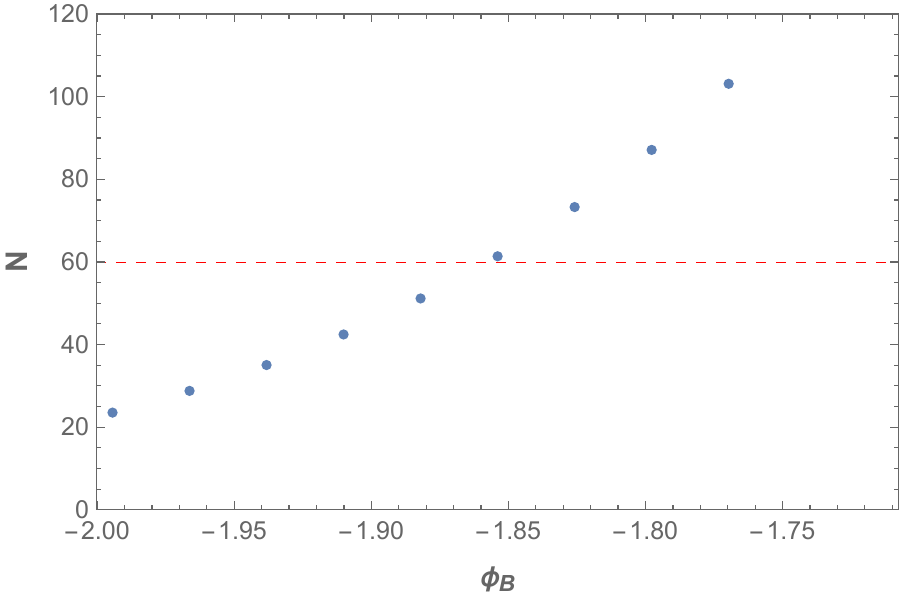}
\end{tabular}}
\caption{E-fold $N$ for $\dot\phi_B >0$ with different values of $\phi_B$ for the polynomial potential of the first kind given by Eq.(\ref{eq3.23}) where  $V_0=3.3787*10^{-15}$ GeV and $\mu=0.31075$. The case with  $\dot\phi_B<0$ can be obtained by the symmetry (\ref{symmetry}).}
\label{fig13}
\end{figure*}

\begin{figure*}[htbp]
\resizebox{\linewidth}{!}
{\begin{tabular}{cc}
\includegraphics[height=4.cm,width=7cm]{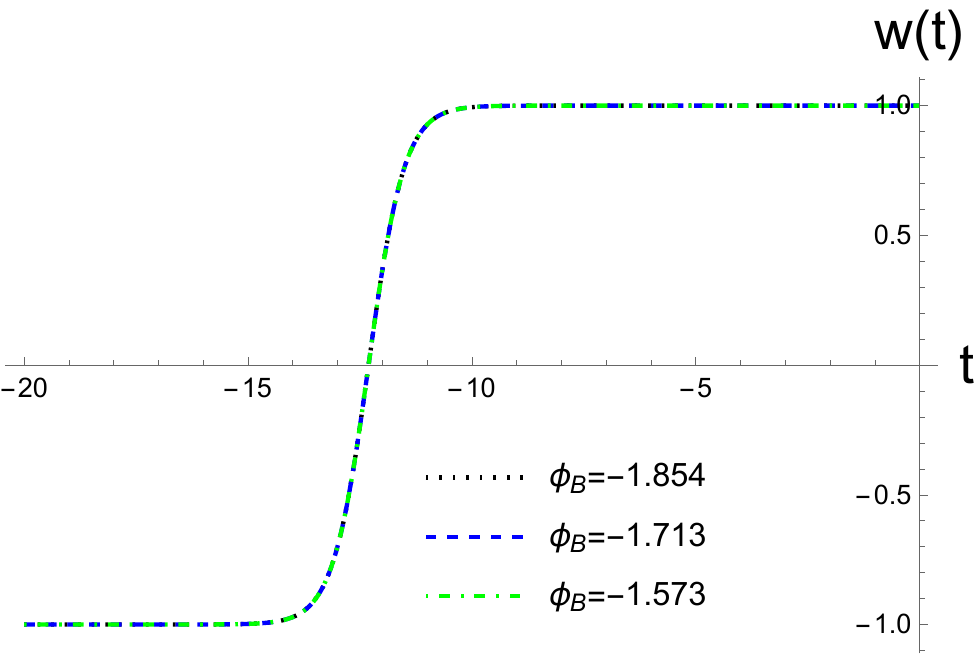}
\includegraphics[height=4.cm,width=7cm]{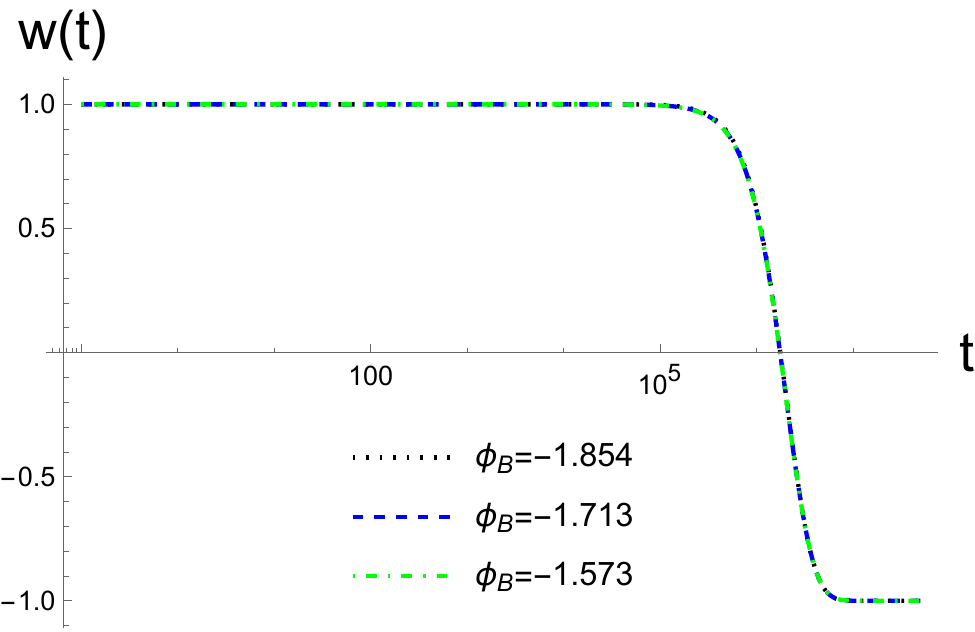}\\
\end{tabular}}
\caption{The equation of state $w(t)$  for the polynomial potential of the first kind given by Eq.(\ref{eq3.23}) with $\dot\phi_B >0$ and  different values of $\phi_B$, where $V_0=3.3787\times 10^{-15}$ GeV and $\mu=0.31075$.}
\label{fig14}
\end{figure*}

\begin{figure*}[htbp]
\resizebox{\linewidth}{!}
{\begin{tabular}{cc}
\includegraphics[height=4.cm,width=7cm]{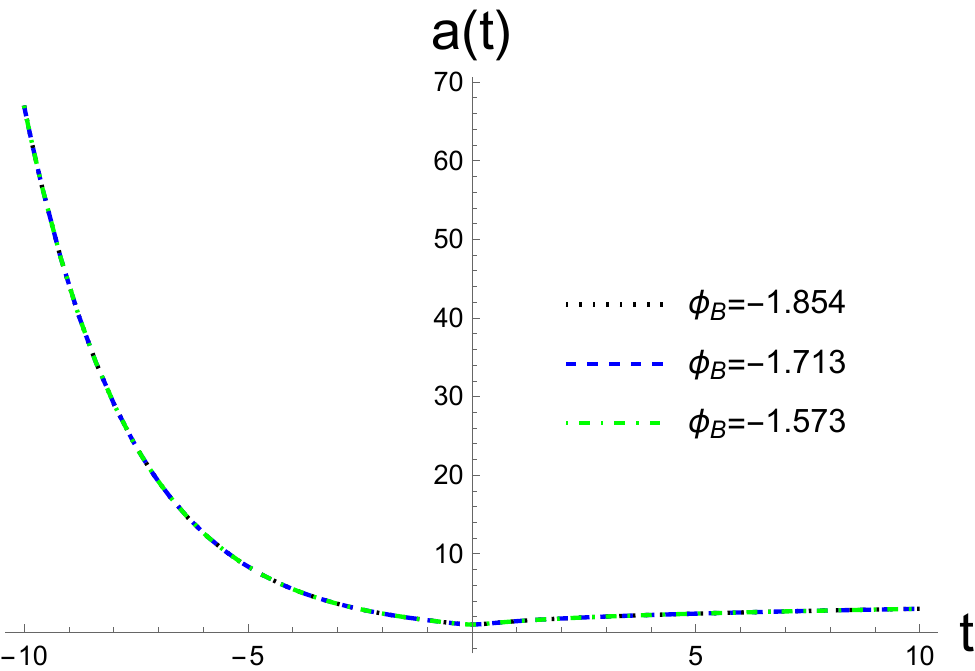}
\includegraphics[height=4.cm,width=7cm]{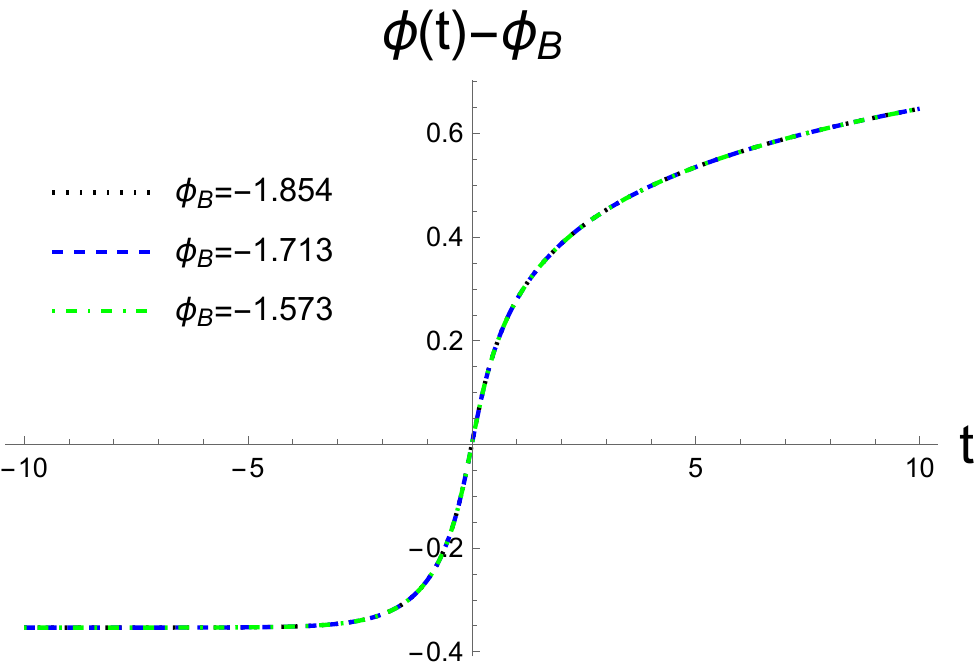}\\
\end{tabular}}
\caption{The numerical  solutions of $a(t)$ and $\phi(t)$ with different initial values of $\phi_B$ for   the polynomial potential of the first kind given by Eq.(\ref{eq3.23}) with $\dot\phi_B > 0$, where $V_0=3.3787\times 10^{-15}$ GeV and $\mu=0.31075$.}
\label{fig15}
\end{figure*}

\begin{figure*}[htbp]
\resizebox{\linewidth}{!}
{\begin{tabular}{cc}
\includegraphics[height=4.cm,width=7cm]{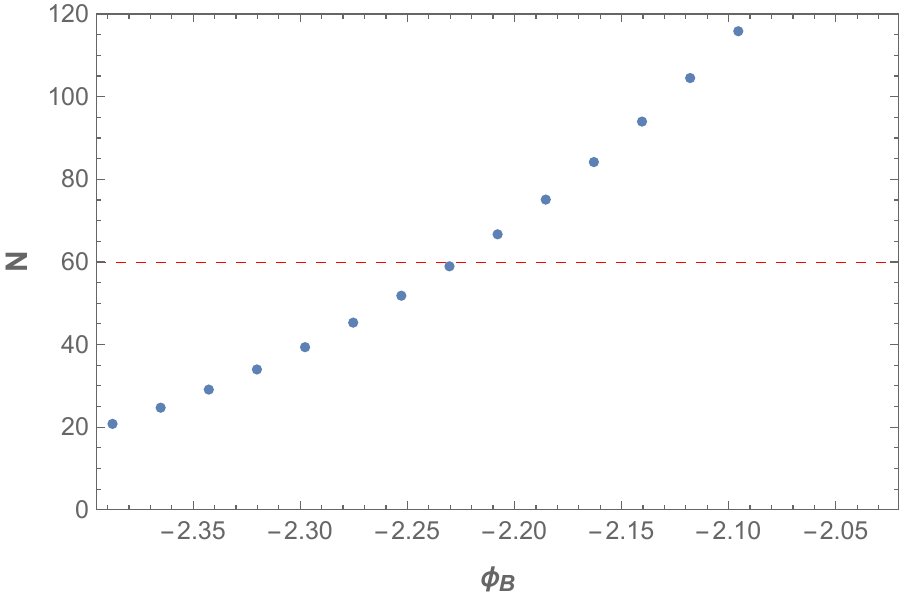}
\end{tabular}}
\caption{E-fold $N$ for different values of $\phi_B$ for the polynomial potential of the second kind given by Eq.(\ref{eq3.26}) with  $\dot\phi_B >0$,  $V_0=3.2599\times 10^{-15}$ GeV and $\mu=0.01043$.}
\label{fig16}
\end{figure*}

\begin{figure*}[htbp]
\resizebox{\linewidth}{!}
{\begin{tabular}{cc}
\includegraphics[height=4.cm,width=7cm]{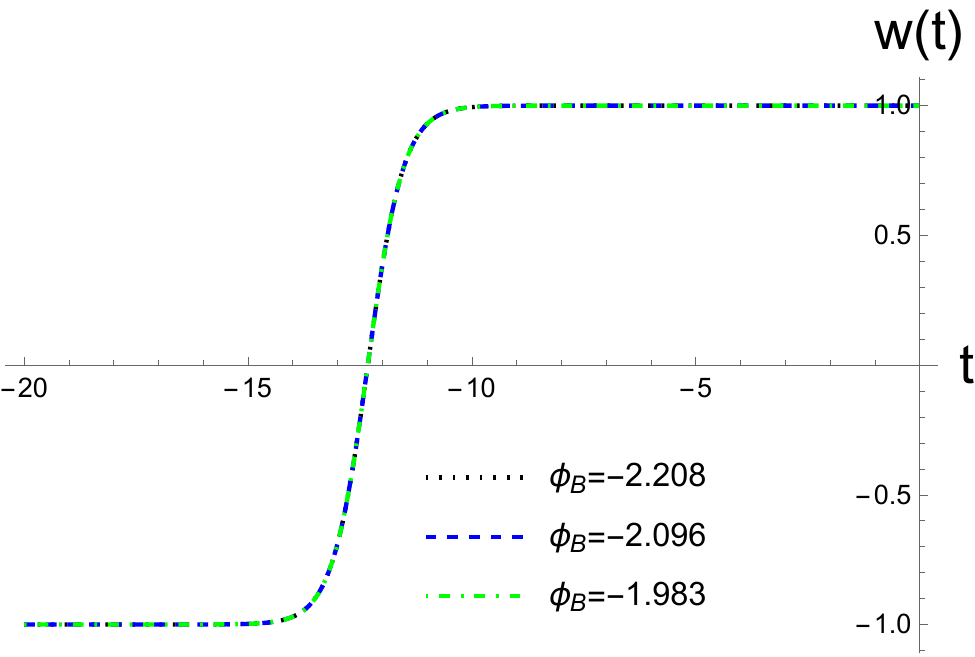}
\includegraphics[height=4.cm,width=7cm]{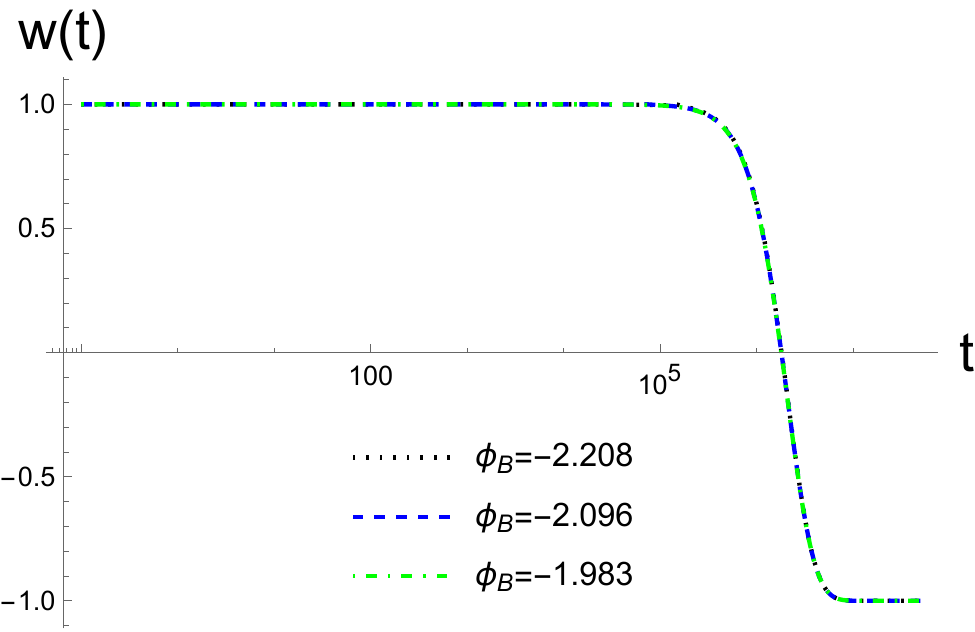}
\end{tabular}}
\caption{The equation of state $w(t)$ with different values of $\phi_B$ for for the polynomial potential of the second kind given by Eq.(\ref{eq3.26}) with $\dot\phi_B >0$,   $V_0=3.2599\times 10^{-15}$ GeV and $\mu=0.01043$.}
\label{fig17}
\end{figure*}

\begin{figure*}[htbp]
\resizebox{\linewidth}{!}
{\begin{tabular}{cc}
\includegraphics[height=4.cm,width=7cm]{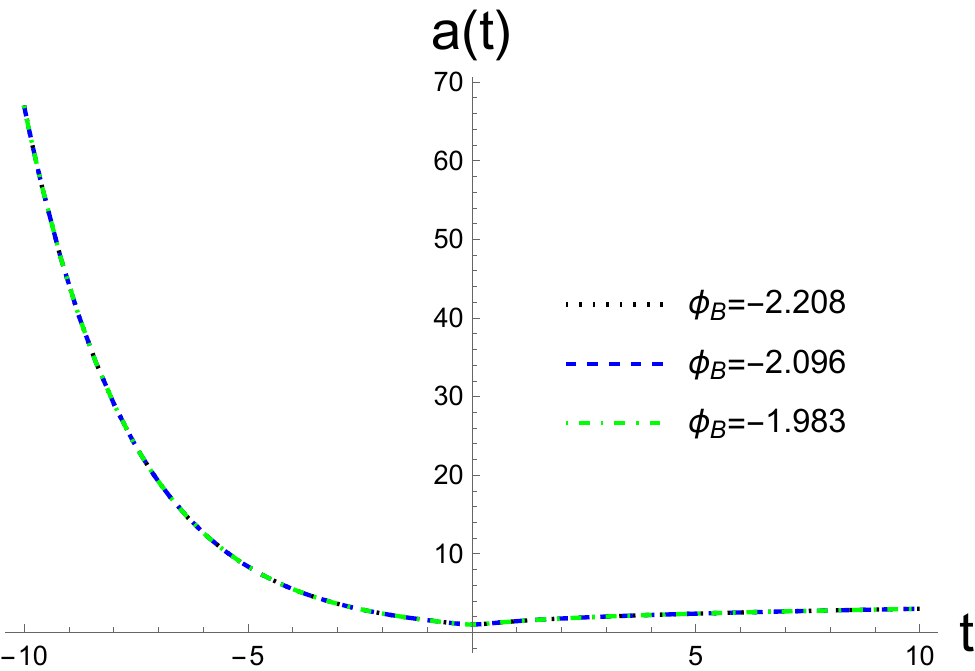}
\includegraphics[height=4.cm,width=7cm]{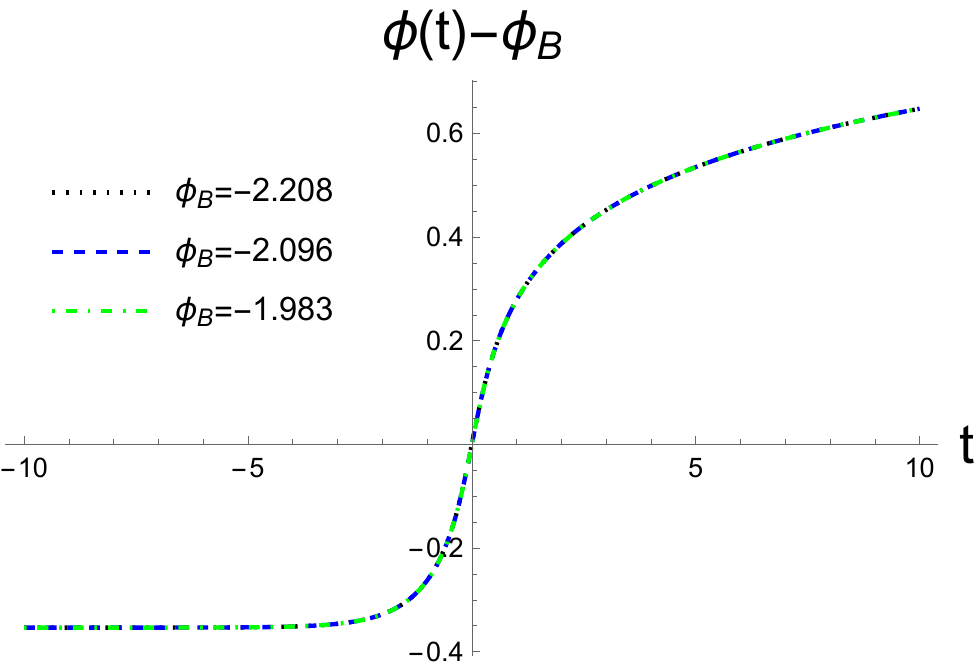}
\end{tabular}}
\caption{The numerical and analytical solutions of $a(t)$ and $\phi(t)$ with different initial values of $\phi_B$ for $\dot\phi_B > 0$ and the polynomial potential of the second kind given by Eq.(\ref{eq3.26}), where  $V_0=3.2599\times 10^{-15}$ GeV and $\mu=0.01043$.}
\label{fig18}
\end{figure*}

\begin{figure*}[htbp]
\resizebox{\linewidth}{!}
{\begin{tabular}{cc}
\includegraphics[height=4.cm,width=7cm]{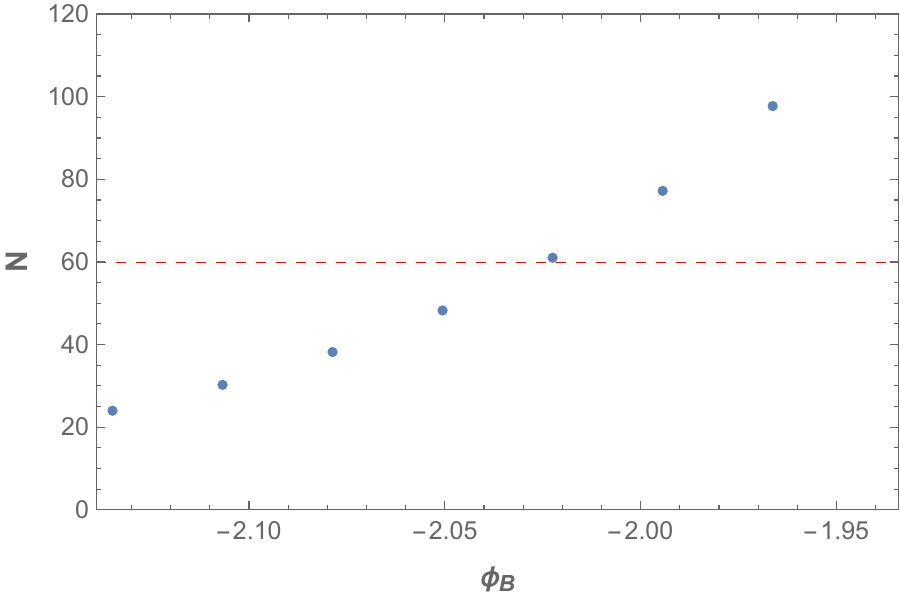}
\end{tabular}}
\caption{E-folds $N$ for the generalized T-Models given by (\ref{eq3.29}) with $\dot\phi_B >0$, $V_0=1.96\times 10^{-15}$ and $\alpha= 0.0962$.}
\label{fig19}
\end{figure*}

\begin{figure*}[htbp]
\resizebox{\linewidth}{!}
{\begin{tabular}{cc}
\includegraphics[height=4.cm,width=7cm]{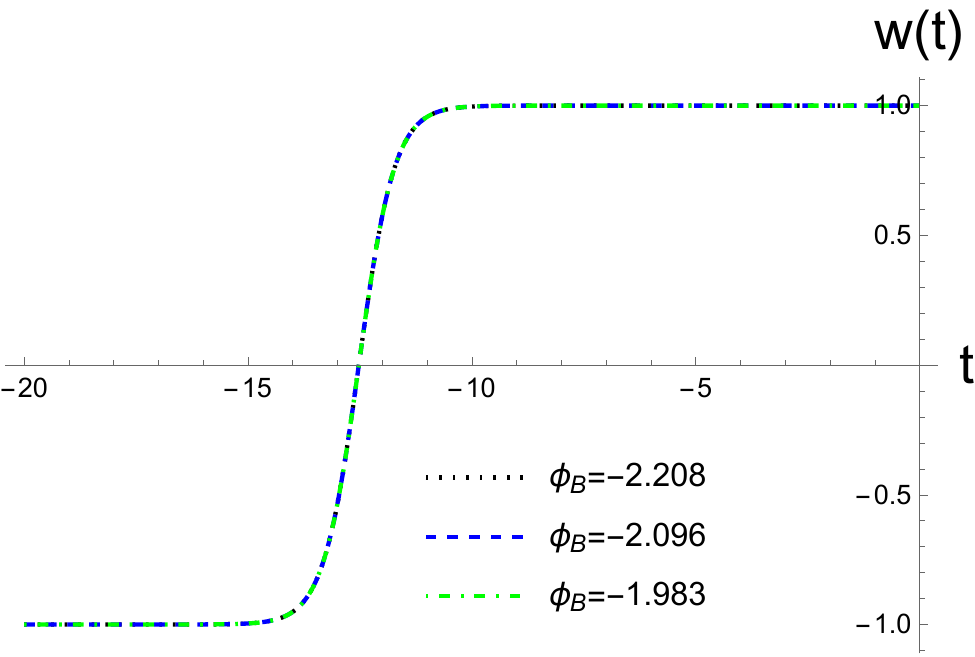}
\includegraphics[height=4.cm,width=7cm]{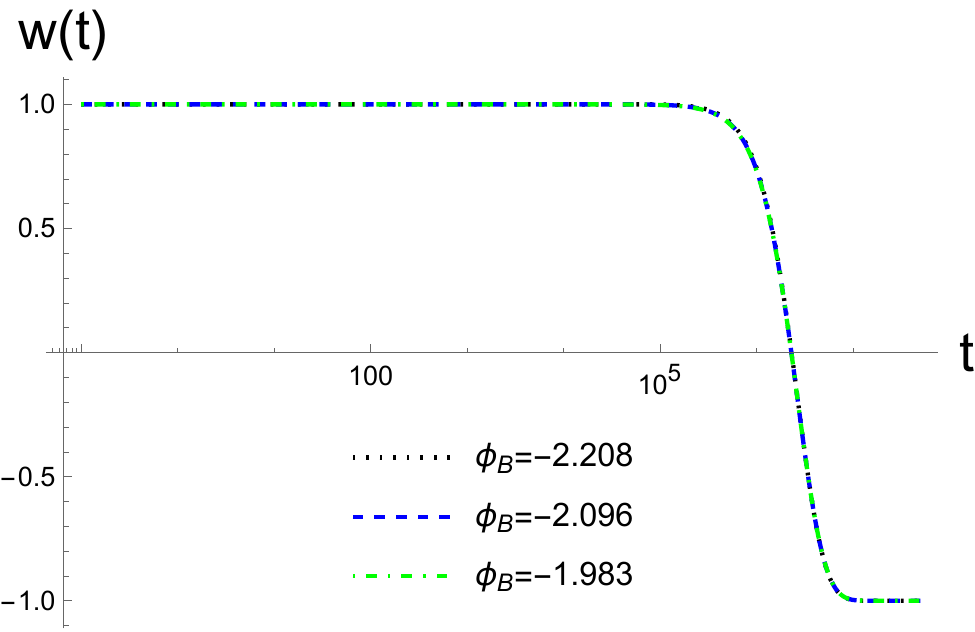}\\
\end{tabular}}
\caption{The equation of state $w(t)$ with different values of $\phi_B$ for the the generalized  T-Models given by (\ref{eq3.29}) with $\dot\phi_B >0$, $V_0=1.96\times 10^{-15}$ and $\alpha= 0.0962$.  }
\label{fig20}
\end{figure*}

\begin{figure*}[htbp]
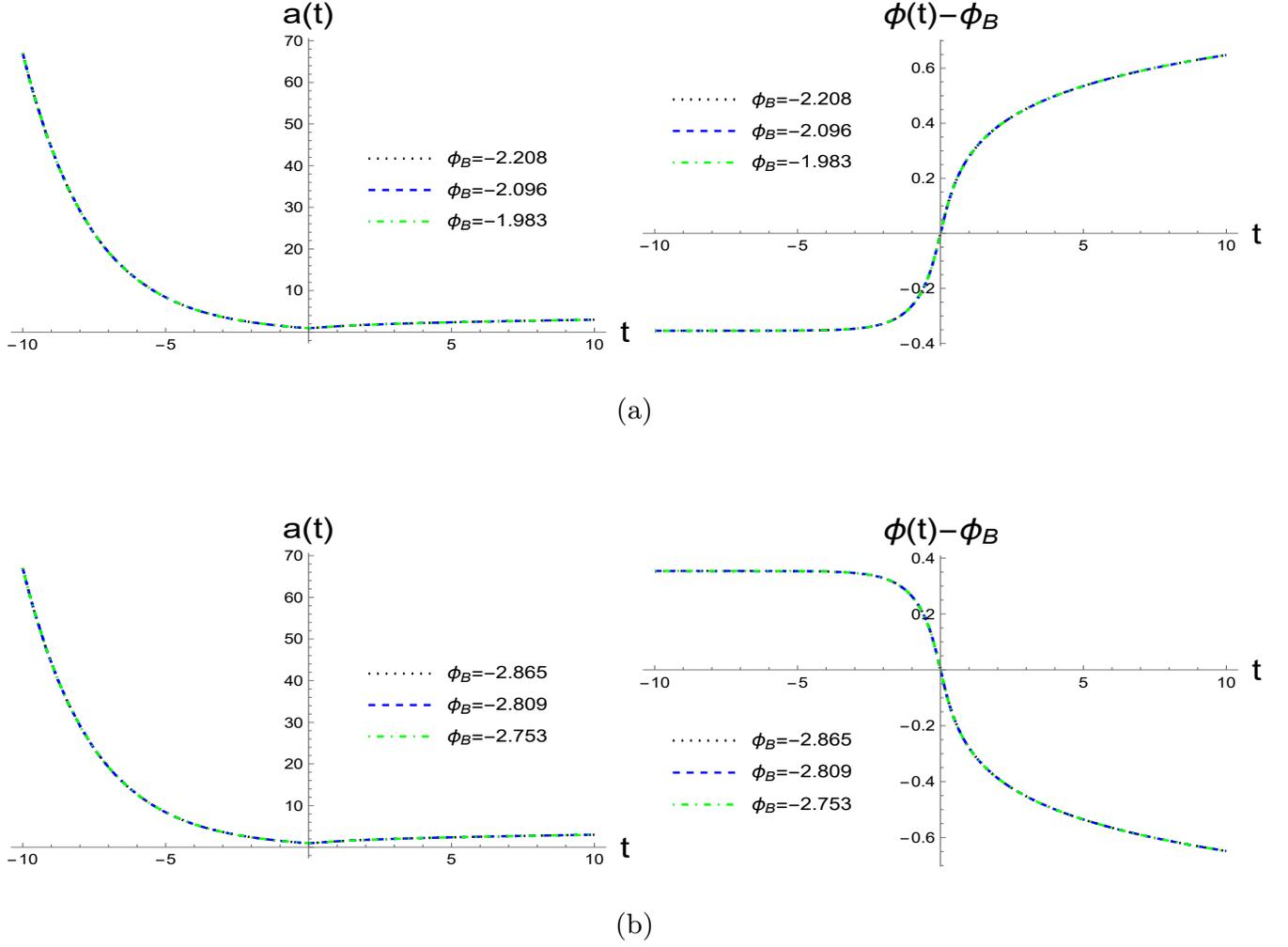

\resizebox{\linewidth}{!}
{\begin{tabular}{cc}
\includegraphics[height=4.cm,width=7cm]{Fig21a-T-model-a-fit-dphi-pos.pdf}
\includegraphics[height=4.cm,width=7cm]{Fig21b-T-model-phi-fit-dphi-pos.pdf}\\
\end{tabular}}
\caption{The numerical solutions of $a(t)$ and $\phi(t)$ for various initial conditions of $\phi_B$  for the generalized T-Models given by (\ref{eq3.29}) with $\dot\phi_B >0$, $V_0=1.96*10^{-15}$ and $\alpha= 0.0962$. }
\label{fig21}
\end{figure*}

The $\alpha$-attractor potentials usually take several different forms \cite{Planck:2018jri}.  In this subsection, we consider several representative cases.

\subsubsection{Generalized Starobinsky  Potentials}

One kind of the $\alpha$-attractor potentials from supergravity takes the form \cite{Ferrara:2013rsa}
\begin{equation}
\lb{eq3.20}
V(\phi) = V_0\left( 1-e^{-\frac{\phi}{\sqrt{6}\alpha}} \right)^2,
\end{equation}
which is often referred to as the generalized Starobinsky potentials or E-models \cite{Kallosh:2022feu}. For $\alpha = \sqrt{9/(8\pi G)}$, the above potentials are precisely reduced to the Starobinsky potential given by Eq.(\ref{eq30}).

In the following, we consider the case $\alpha  = 0.0962$ with  $V_0 = 10^{-12}$ GeV. Then, we find that the initial values of $\phi_B$ for both $\dot\phi_B > 0$
and $\dot\phi_B < 0$, with which the e-fold $N$ is always no less than 45, as shown in Fig.\ref{fig10}. With several such values of $\phi_B$, in Fig. \ref{fig11}    the corresponding equation of state $w(t)$ for both $\dot\phi_B > 0$
and $\dot\phi_B < 0$ are plotted out. From these figures, we can see that each of the post-bounce and pre-bounce regimes can be universally divided into three phases, as given in Eqs.(\ref{eq3.1b}) and (\ref{eq3.11}).

In Fig. \ref{fig12}, we plot the numerical solutions of
$a(t)$ and $\phi(t)$ for several initial values of $\phi_B$ for $\dot\phi_B > 0$
and $\dot\phi_B < 0$, respectively. 
From these figures, it can be seen that, similar to the previous cases, the numerical solutions are almost indistinguishable from each other.

\subsubsection{Polynomial of the first kind}

The polynomial potential of the first kind takes the form \cite{Kallosh:2022feu}
\begin{equation}
\lb{eq3.23}
V(\phi) = V_0\frac{\phi^4}{\phi^4+\mu^4},
\end{equation}
where $V_0$ and $\mu$ are two constants.
In this paper, we take the values of $V_0$ and $\mu$ as   $V_0=3.3787\times 10^{-15}$ GeV and $\mu=0.31075$ \cite{Bhattacharya:2022akq}.

Note that, similar to the chaotic potential, now the polynomial potential of the first kind has the symmetry 
\bq
\lb{Vsymmetry}
V(\phi) = V(-\phi).
\eq
As a results, the case with $\dot\phi_B < 0$ can be obtained from the case with $\dot\phi_B > 0$ via the symmetry (\ref{symmetry}). Clearly this is true for all the cases with the above symmetry, including the polynomial potential of the second kind and the generalized T-models to be considered below. Therefore, in all these cases, we shall consider only the cases with $\dot\phi_B > 0$ without furthering noticing it.

Then, in Figs. \ref{fig13} and \ref{fig14} we plot the e-fold $N$ and the equation of state $w(t)$ for $\dot\phi_B > 0$.
Similar to the previous cases, the evolution of the universe in each of the pre- and post-bounce regimes can be universally divided into three epochs, describing by Eqs.(\ref{eq3.1b}) and (\ref{eq3.11}), respectively.
In Fig. \ref{fig15}, we plot the functions $a(t)$ and $\phi(t)$ with various initial values of $\phi_B$ for $\dot\phi_B > 0$.

\subsubsection{Polynomial of the second kind}

The polynomial potential of the second kind is given by \cite{Kallosh:2022feu}
\begin{equation}
\lb{eq3.26}
V(\phi) = V_0\frac{\sqrt{\phi^2+\mu^2}-\mu}{\sqrt{\phi^2+\mu^2}+\mu}.
\end{equation}
In this paper, we adopt the best fitting values of $V_0$ and $\mu$ with current observations given by    $V_0=3.2599\times 10^{-15}$ and $\mu=0.01043$ \cite{Bhattacharya:2022akq}.

In Figs. \ref{fig16} and  \ref{fig17} we plot $N(\phi_B)$ and the corresponding equation of state $w(t)$ for several initial conditions $\phi_B$ for
$\dot\phi_B > 0$. These values of $\phi_B$ guarantee that the e-folds during inflation will be about 60 as shown in Fig. \ref{fig16}.
On the other hand, Fig. \ref{fig17} shows again that the evolution of the universe in each of the pre- and post-bounce regimes can be universally divided into three different epochs.

In Fig. \ref{fig18} we plot numerical  solutions of $a(t)$ and $\phi(t)$. Again, these numerical solutions are almost indistinguishable with different initial conditions as specified in the figures, which show again the universality of the evolution of the Universe.

\subsubsection{Generalized T-models}

The generalized T-model potentials of the superconformal inflationary models are given by \cite{Kallosh:2013yoa}
\begin{equation}
\lb{eq3.29}
V(\phi) = V_0\tanh^2\left({\frac{\phi}{\sqrt{6} \alpha}}\right),
\end{equation}
where the best fitting values of $V_0$ and $\alpha$ with current cosmological observations are given by
$V_0=1.96\times 10^{-15}$ and $\alpha=0.0962$ \cite{Bhattacharya:2022akq}.

In Figs. \ref{fig19} and  \ref{fig20} we plot $N(\phi_B)$ and the corresponding equation of state $w(t)$ for several initial conditions $\phi_B$ with
$\dot\phi_B > 0$. These values of $\phi_B$ also guarantee that the e-folds of the expansions of the universe during the inflationary period will be about 60, as shown in Fig. \ref{fig20}.
On the other hand, Fig. \ref{fig20} shows again that the evolution of the universe in each of the pre- and post-bounce regimes can be universally divided into three different epochs, as described by Eq.(\ref{eq3.1b}) in the post-bounce regime and by Eq.(\ref{eq3.11}) in the pre-bounce regime.

In Fig. \ref{fig21}, we plot the numerical  solutions of $a(t)$ and $\phi(t)$ with several initial conditions for $\dot\phi_B > 0$, which all satisfy the condition (\ref{eq1.1}) at the bounce. Again, they show the universality of the solutions.

\subsection{Natural Inflation}

The Natural inflationary potential, motivated from a pseudo-Nambu-Goldstone boson, takes the form \cite{Adams:1992bn},
\begin{equation}
\lb{eq3.32}
V\left(\phi\right) = \Lambda^4\left[1+\cos{\left(\frac{\phi}{\mu}\right)}\right],
\end{equation}
where $\mu \equiv f/{\cal{N}}$, $\Lambda \simeq m_{\text{GUT}} \simeq 10^{16}$ GeV, $f \simeq m_{\text{pl}}$, and ${\cal{N}}$ is a dimensionless constant.
Because of the periodic property of the potential, without loss of the generality, we can restrict ourselves only to the region 
\bq
\lb{phiBrange}
- \pi \le \frac{\phi}{\mu} \le \pi.
\eq
 In addition, the potential is also symmetric under the replacement $\phi \rightarrow - \phi$. So, in this case it is also sufficient to consider only the case $\dot\phi_B > 0$. In the following, we consider only two representative cases,   $\mu/m_{\text{pl}} = \left(1, \; 5\right)$.

\begin{figure*}[htbp]
\resizebox{\linewidth}{!}
{\begin{tabular}{cc}
\includegraphics[height=4.cm,width=7cm]{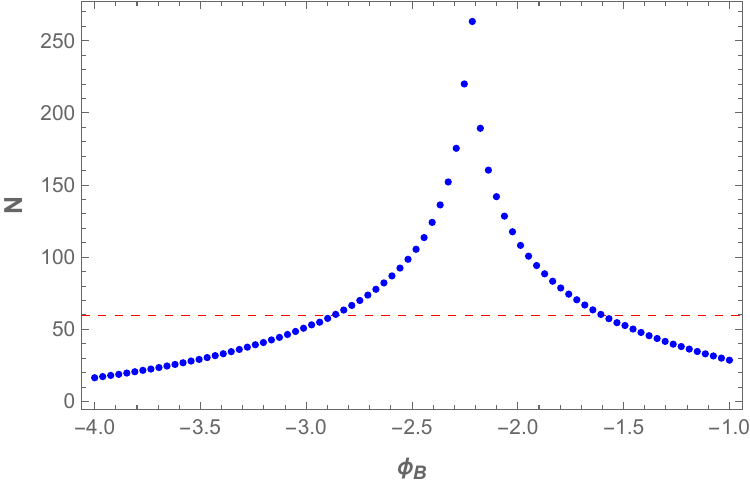}
\end{tabular}}
\caption{E-folds of natural inflation with the potential given by Eq.(\ref{eq3.32}), $\dot\phi_B >0$, $\Lambda = 10^{16}$ GeV, and $\mu = m_{\text{pl}}$ for different values of the initial conditions $\phi_B$. }
\label{fig25}
\end{figure*}

\begin{figure*}[htbp]
\resizebox{\linewidth}{!}
{\begin{tabular}{cc}
\includegraphics[height=4.cm,width=7cm]{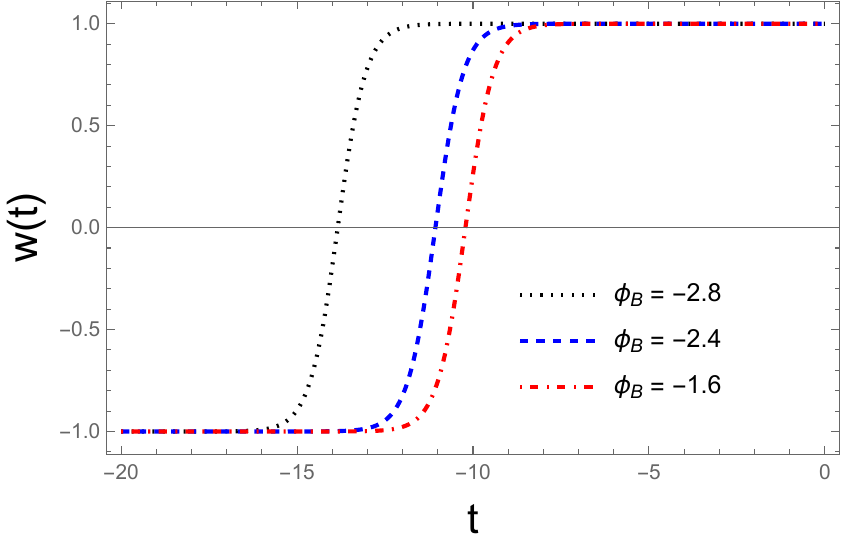}
\includegraphics[height=4.cm,width=7cm]{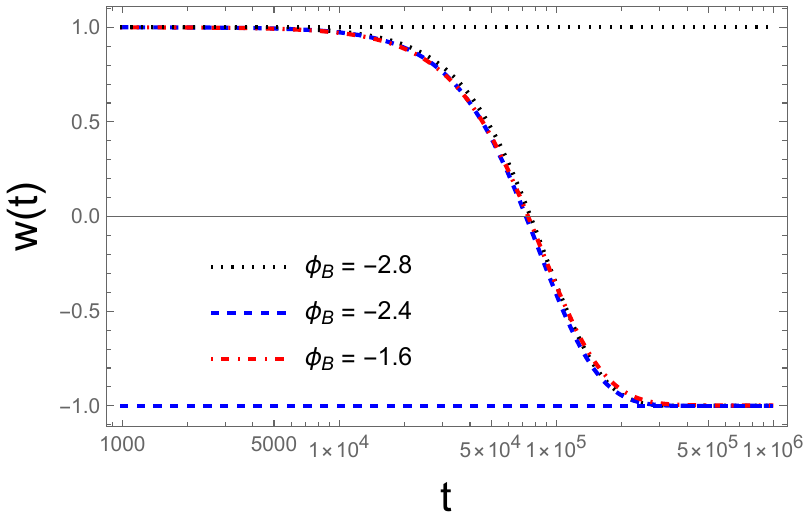}
\end{tabular}}
\caption{The equation of state $w(t)$ with different values of $\phi_B$ for for the  natural inflation with the potential given by Eq.(\ref{eq3.32}), $\dot\phi_B >0$, $\Lambda = 10^{16}$ GeV and $\mu = 1\; m_{\text{pl}}$.}
\label{fig26}
\end{figure*}

\begin{figure*}[htbp]
\resizebox{\linewidth}{!}
{\begin{tabular}{cc}
\includegraphics[height=4.cm,width=7cm]{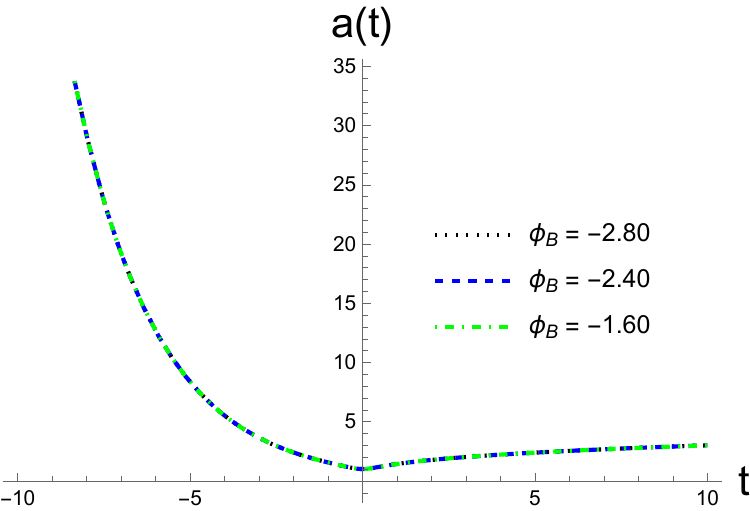}
\includegraphics[height=4.cm,width=7cm]{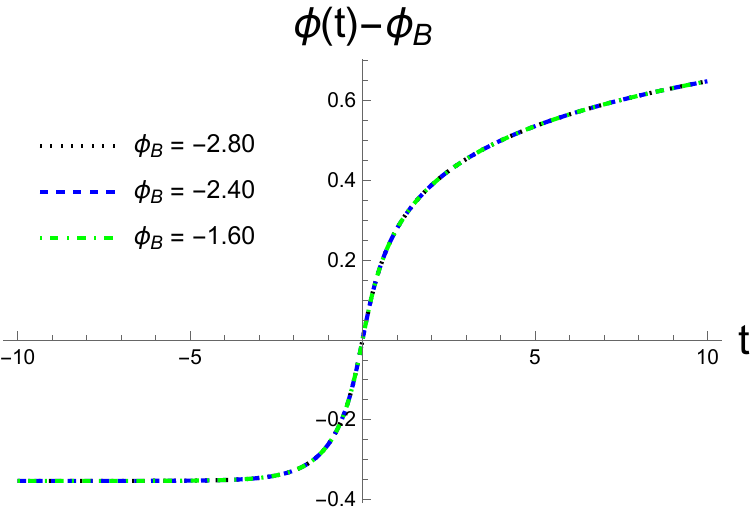}\\
\end{tabular}}
\caption{The numerical solutions of $a(t)$ and $\phi(t)$ for various initial conditions of $\phi_B$  for  the  natural inflation with the potential given by Eq.(\ref{eq3.32}), $\dot\phi_B >0$, $\Lambda = 10^{16}$ GeV and $\mu = 1 m_{\text{pl}}$. }
\label{fig27}
\end{figure*}

\subsubsection{$\mu = m_{\text{pl}}$}

In this subsection,  we first identify the initial conditions $\phi_B$ for both $\dot\phi_B > 0$ and $\dot\phi_B < 0$, with which the e-folds of inflation are approximately 60. In  Fig. \ref{fig25} we show two sets of such data. Note that the number of such sets is infinitely large because the potential is now a periodic function of $\phi$.
With some of the initial conditions presented in Fig. \ref{fig25}, we plot the corresponding equation of state $w(t)$ in Fig. \ref{fig26}, which again shows the universal properties described by Eqs.(\ref{eq3.1b}) and (\ref{eq3.11}).
The corresponding numerical solutions of $a(t)$ and $\phi(t)$ are given in Fig. \ref{fig27} for
$\dot\phi_B > 0$ and $\dot\phi_B <0$, respectively.

\begin{figure*}[htbp]
\resizebox{\linewidth}{!}
{\begin{tabular}{cc}
\includegraphics[height=4.cm,width=7cm]{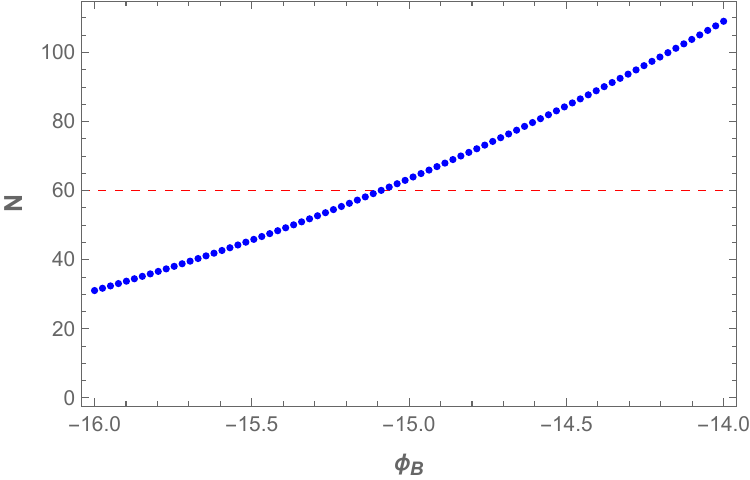}
\end{tabular}}
\caption{E-folds of natural inflation with the potential given by Eq.(\ref{eq3.32}), $\dot\phi_B >0$,  $\Lambda = 10^{16}$ GeV and $\mu = 5m_{\text{pl}}$ for different values of the initial conditions $\phi_B$.}
\label{fig28}
\end{figure*}

\begin{figure*}[htbp]
\resizebox{\linewidth}{!}
{\begin{tabular}{cc}
\includegraphics[width=7cm]{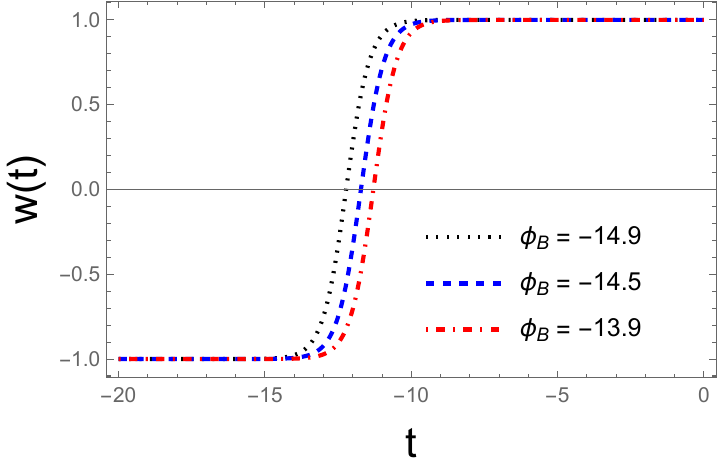}
\includegraphics[width=7cm]{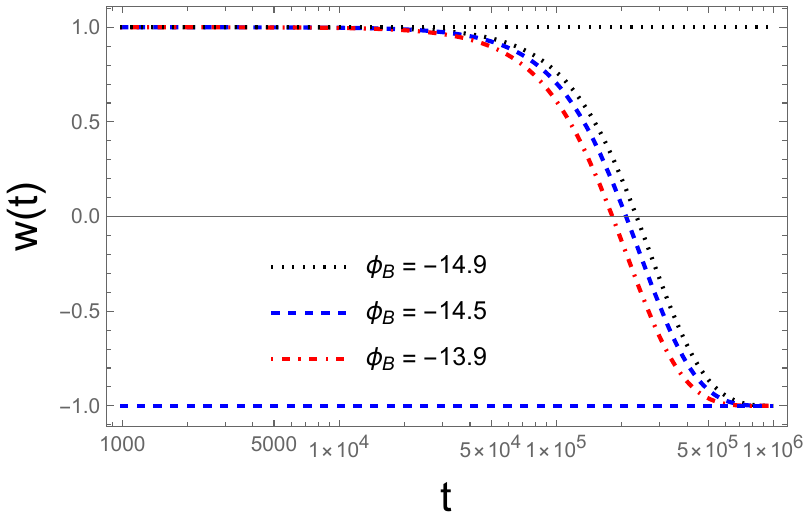}
\end{tabular}}
\caption{The equation of state $w(t)$ with different values of $\phi_B$ for for the  natural inflation with the potential given by Eq.(\ref{eq3.32}), $\dot\phi_B >0$, $\Lambda = 10^{16}$ GeV and $\mu = 5\; m_{\text{pl}}$.}
\label{fig29}
\end{figure*}

\begin{figure*}[htbp]
\resizebox{\linewidth}{!}
{\begin{tabular}{cc}
\includegraphics[height=4.cm,width=7cm]{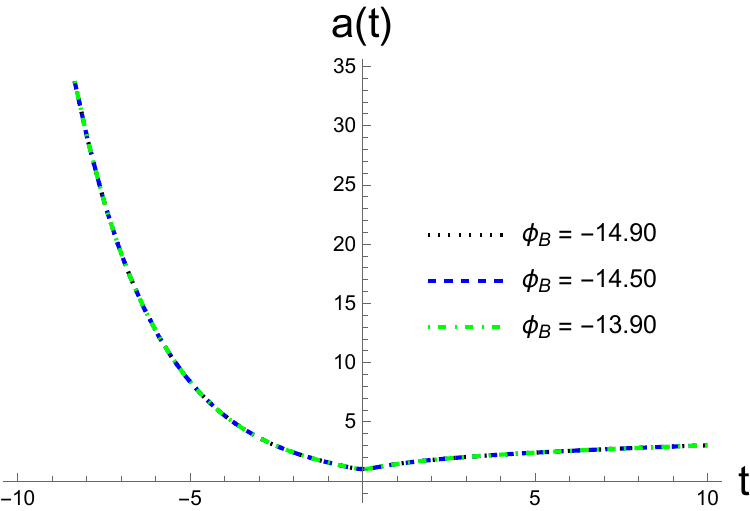}
\includegraphics[height=4.cm,width=7cm]{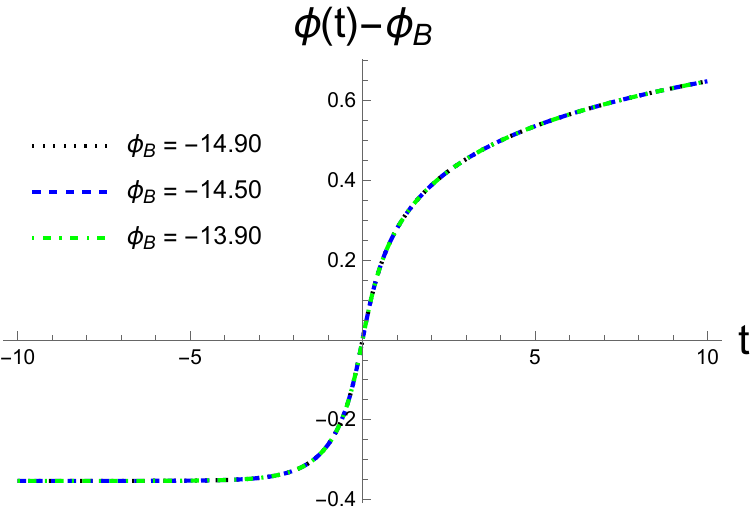}\\
\end{tabular}}
\caption{The numerical solutions of $a(t)$ and $\phi(t)$ for various initial conditions of $\phi_B$ with $\dot\phi_B > 0$ for natural inflation with the potential given by Eq.(\ref{eq3.32}), $\dot\phi_B >0$, $\Lambda = 10^{16}$ GeV and $\mu = 5 m_{\text{pl}}$. }
\label{fig30}
\end{figure*}

\subsubsection{$\mu = 5\; m_{\text{pl}}$}

In this case, in Fig. \ref{fig28} we plot the e-folds of the natural inflation for both $\dot\phi_B > 0$ and $\dot\phi_B < 0$, whereby we can read off the initial conditions of $\phi_B$ for which the inflation will last about 60 e-folds.
Then, with several choices of such initial values of $\phi_B$ presented in Fig. \ref{fig28}, the equation of state $w(t)$ is plotted in Fig. \ref{fig29}. Again, we can clearly identify the three distinguishable epochs in each of the pre- and post-bounce regimes.
The corresponding numerical solutions of $a(t)$ and $\phi(t)$ are plotted in Fig. \ref{fig30} for both $\dot\phi_B >0$ and  $\dot\phi_B < 0$.

\begin{figure*}[htbp]
\resizebox{\linewidth}{!}
{\begin{tabular}{cc}
\includegraphics[height=4.cm,width=7cm]{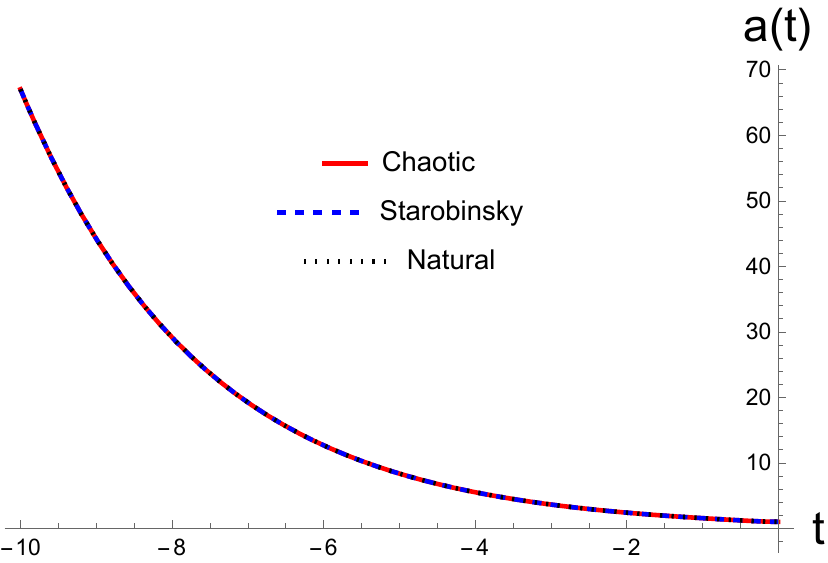}
\includegraphics[height=4.cm,width=7cm]{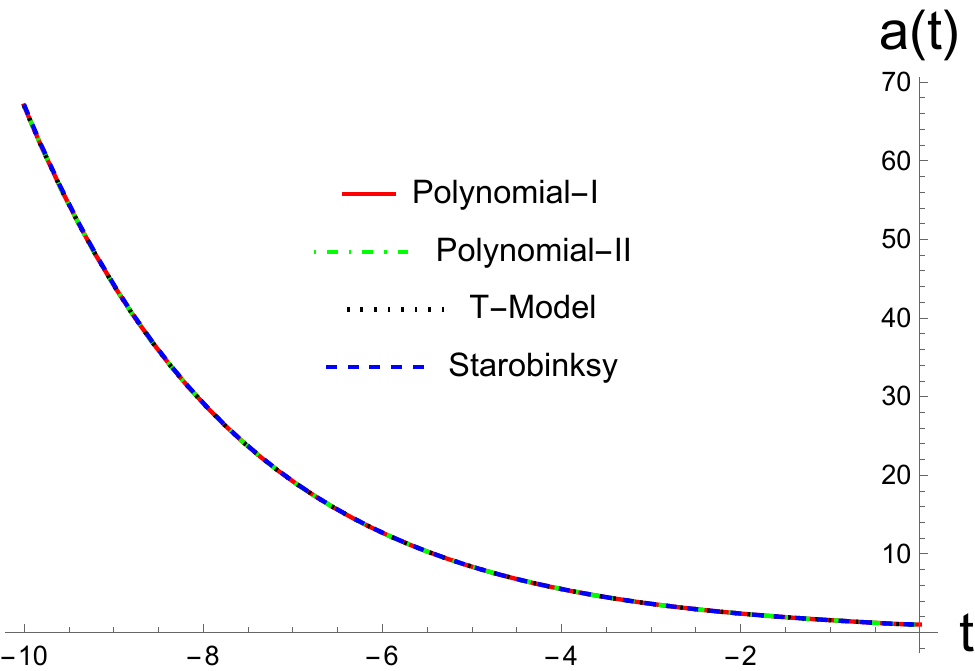}\\
(a)\\
\vspace{.1cm}\\
\includegraphics[height=4.cm,width=7cm]{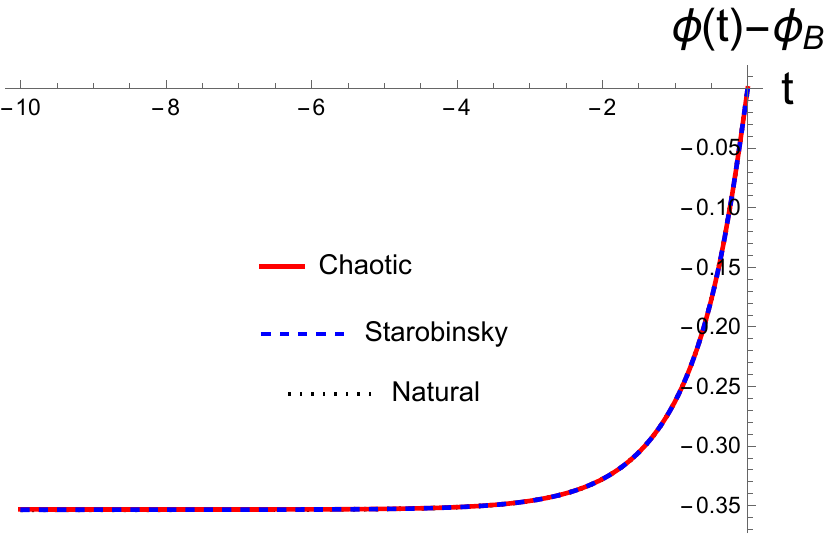}
\includegraphics[height=4.cm,width=7cm]{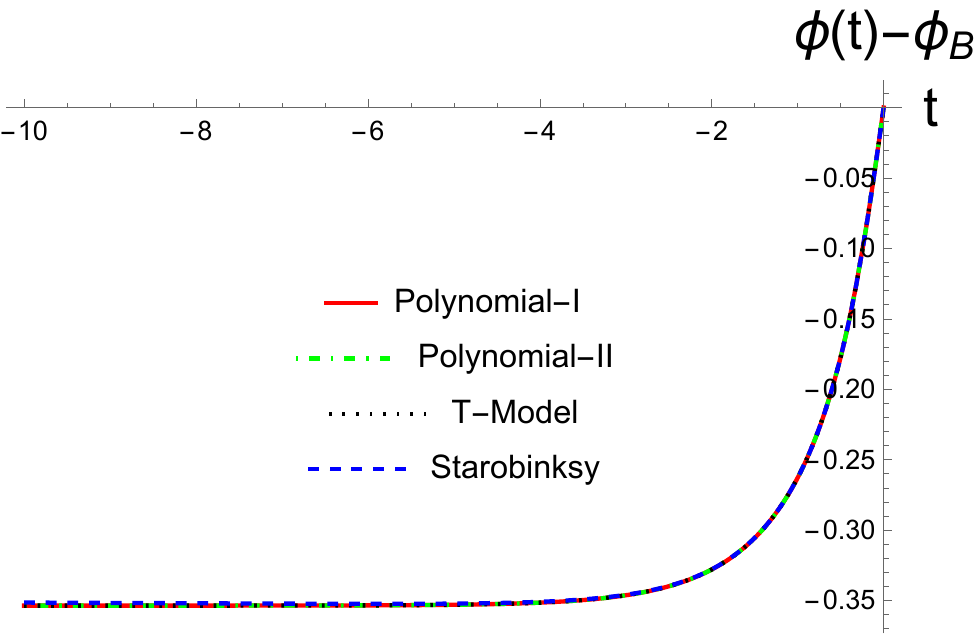}\\
(b)\\
\vspace{.1cm}\\
\includegraphics[height=4.cm,width=7cm]{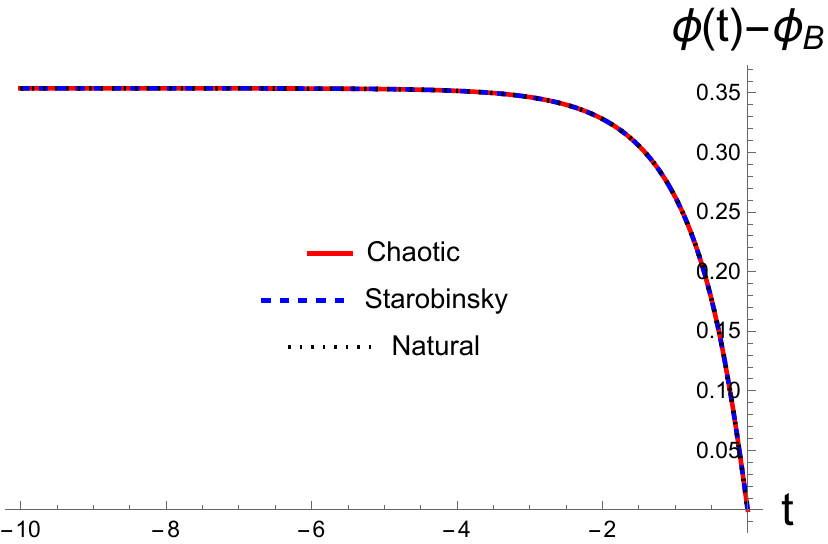}
\includegraphics[height=4.cm,width=7cm]{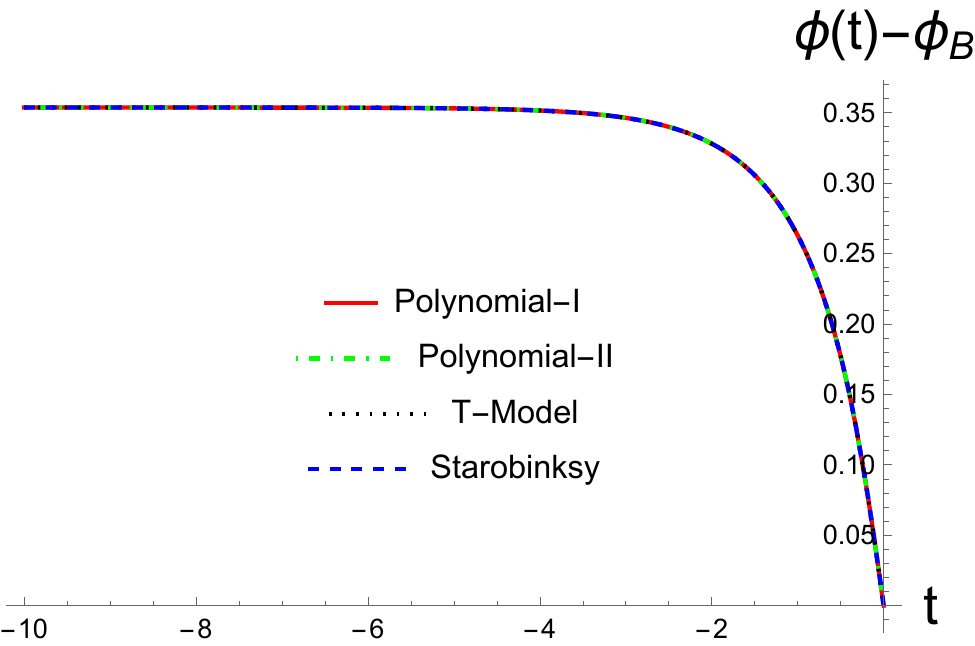}\\
(c)\\
\end{tabular}}
\caption{The numerical solutions of $a(t)$ and $\phi(t)$ for various potentials as specified in the figures. To see clearly the difference among the different models, we group the plot into two groups: {\bf Left Panels}: Models with chaotic, Starobinsky and natural inflationary potentials.
{\bf Right Panels}: Models with the polynomials of the first and second kinds and the generalized T-models, although they are almost indistinguishable among all these models.  (a) The evolution of the expansion factor $a(t)$ for both $\dot\phi_B > 0$ and $\dot\phi_B < 0$.
(b) The evolution of the scalar field $\phi(t)$ for $\dot\phi_B > 0$. (c) The evolution of the scalar field $\phi(t)$ for $\dot\phi_B < 0$.}
\label{fig31}
\end{figure*}

\section{Analytical Solutions for the evolution of the Universe}
\lb{SecIV}

As shown previously \cite{Li:2019ipm}, the evolution of the Universe in the post-bounce regime ($t \ge t_B$) is universal, and can be divided into three different epochs defined as in Eq.(\ref{eq3.1b}). In addition, during the bouncing phase, the solutions of $a(t)$ and $\phi(t)$ are given by the analytical solutions of Eq.(\ref{eq3.3}) [Also see Fig. \ref{fig2}].

In this section, we are going to show that this is also the case in the pre-bounce regime ($t \le t_B$). To show this, let us first plot the numerical solutions of $a(t)$ and $\phi(t)$ together for all the cases considered in the last section. Since for each potential the numerical solutions are indistinguishable for different initial values of $\phi_B$, we plot only one solution for each model. In addition, to see clearly the difference among the models, we also plot them in two groups, one for chaotic, Starobinsky and natural inflaitonary potentials and the other for the polynomials of the first and second kinds and the generalized T-models, although they are indistinguishable among all these models, as can be seen from Fig. \ref{fig31}, denoted by the right-hand and left-hand panels, respectively. The numerical solutions of $a(t)$ is independent of the signs of $\dot\phi_B$ and are given by Fig. \ref{fig31} (a), while The numerical solutions of $\phi(t)$ depend on the signs of $\dot\phi_B$ and are given respectively by Figs. \ref{fig31} (b) and (c).
From Fig. \ref{fig31} (a) we can see that the universe soon enters the pre-de Sitter phase, and $a(t)$ increases exponentially $a(t) \propto e^{-H_{\Lambda}t}$ as $t$ becomes more and more negative, where $H_{\Lambda} \equiv \sqrt{{8\pi \alpha G\rho_{\Lambda}/3}}$ [cf. Eq.(\ref{FRccA})]. 
On the other hand, the evolution of the scalar field indeed depends on the signs of $\dot\phi_B$,
as in the post-bounce regime, as cna be seen from Eq.(\ref{eq3.3}). In particular, it quickly approaches a negative constant for $\dot\phi_B > 0$ and a positive constant for $\dot\phi_B <0$, as shown explicitly by Figs. \ref{fig31} (b) and (c).

\begin{figure*}[htbp]
\resizebox{\linewidth}{!}
{\begin{tabular}{cc}
\includegraphics[height=4.cm,width=7cm]{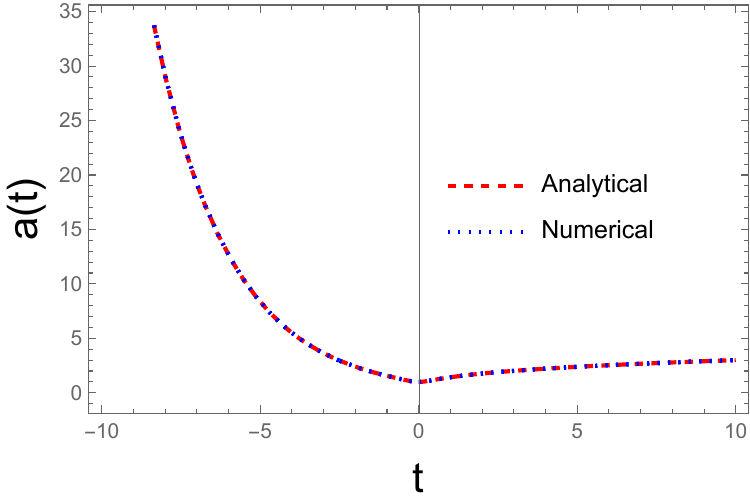}
\includegraphics[height=4.cm,width=7cm]{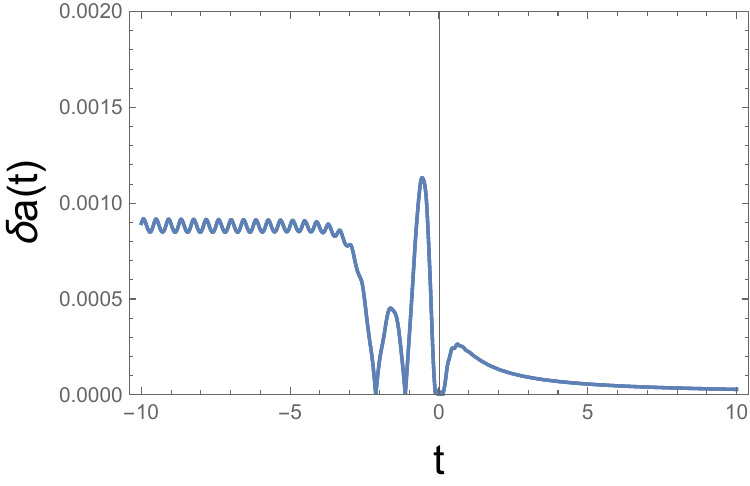}\\
\includegraphics[height=4.cm,width=7cm]{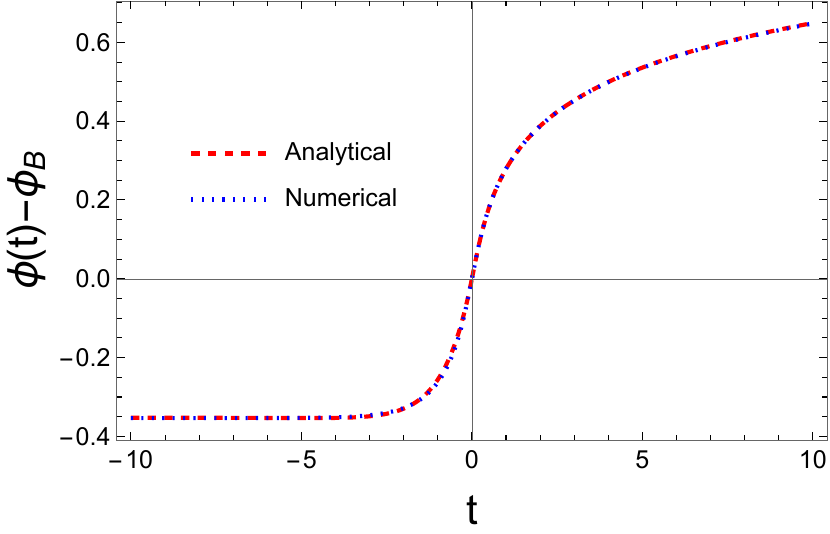}
\includegraphics[height=4.cm,width=7cm]{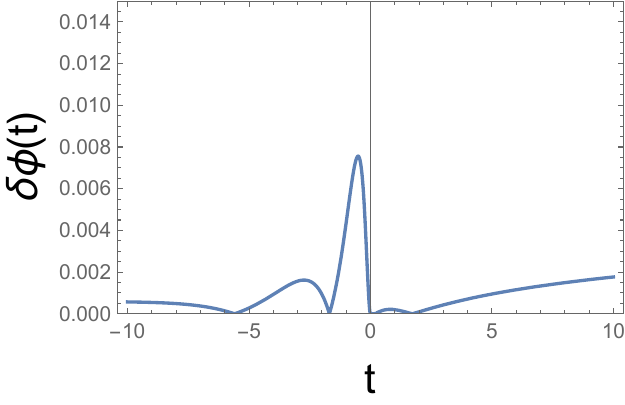}\\
(a)\\
\vspace{.1cm}\\
\includegraphics[height=4.cm,width=7cm]{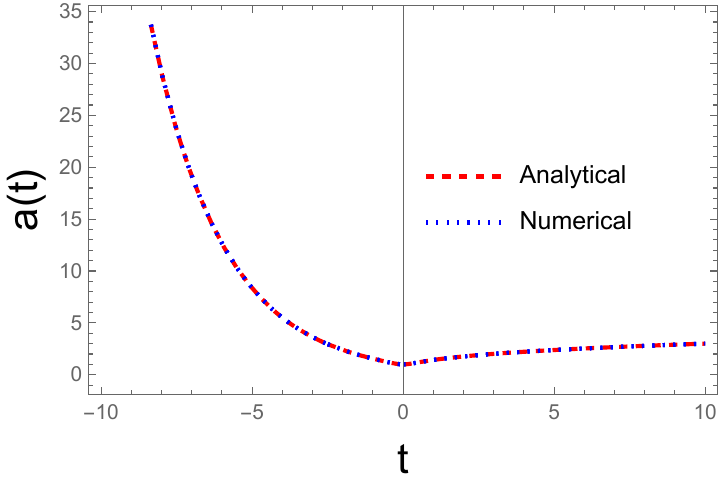}
\includegraphics[height=4.cm,width=7cm]{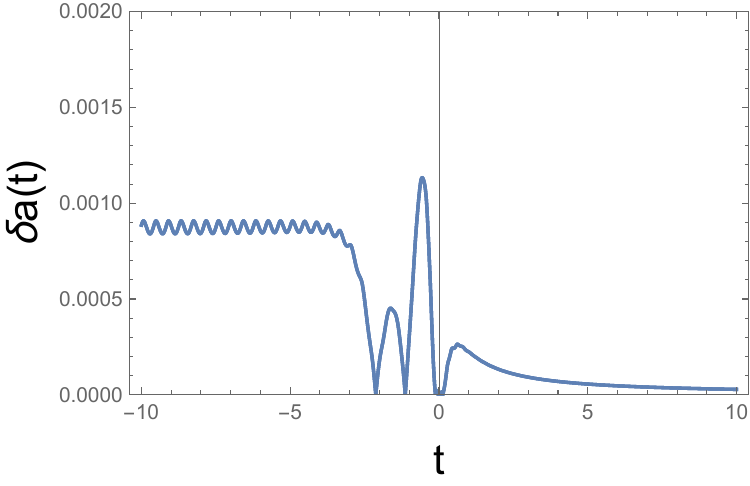}\\
\includegraphics[height=4.cm,width=7cm]{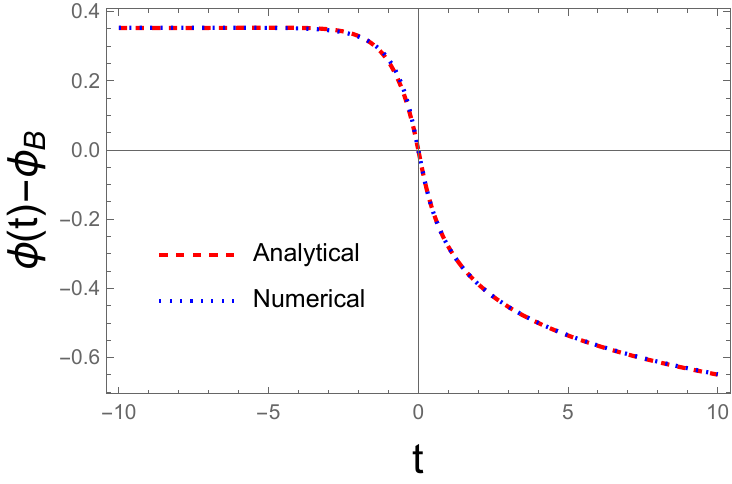}
\includegraphics[height=4.cm,width=7cm]{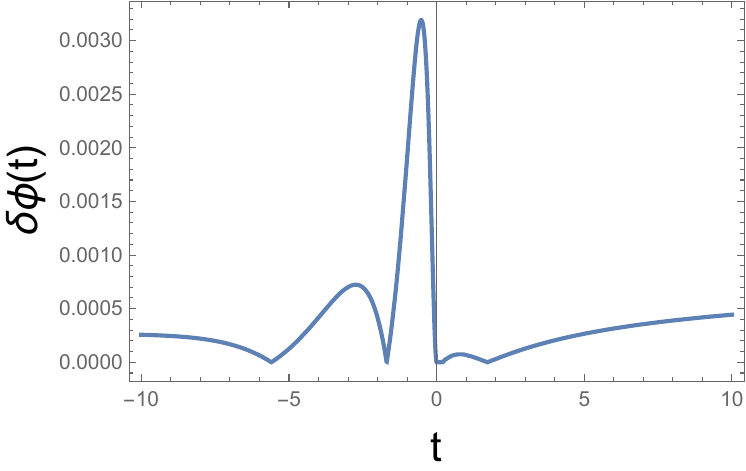}\\
(b)
\end{tabular}}
\caption{The numerical and analytical solutions of $a(t)$ and $\phi(t)$ with various initial values of $\phi_B$ for  the Starobinsky potential (\ref{eq30}), where (a) for $\dot\phi_B > 0$ and (b)
For $\dot\phi_B < 0$. The analytical solutions are given by Eqs.(\ref{eq3.12}) - (\ref{eq3.16}).}
\label{fig32}
\end{figure*}

To analytically model the expansion factor $a(t)$ in the pre-bounce regime, we find that it is convenient to assume that it takes the form
\begin{equation}
\label{eq3.12}
a(t) =   \begin{cases}
\left(1+d_0\rho^I_ct^2  \right)^\frac{1}{6}\sum_{n=2}^4{d_n t^n}, &
t_{m} \leq t \leq t_B, \cr
a(t_{m})\exp\left\{-H_{\Lambda}\left(t-t_{m}\right)\right\}, & t \leq t_m,\cr
\end{cases}
\end{equation}
where $t_m$ is determined by $\dot{H}(t_m) \simeq 0$, and $d_n\; (n = 0, ..., 4)$ are fitting constants, which must satisfy the junction conditions at both the bounce ($t = 0$) and the turning point ($t = t_m$). In particular, we require that {\em $a(t)$ and its first derivative be contiguous across these two points},
where $a(t)$ is given by Eq.(\ref{eq3.3}) in the post-bounce regime ($t \ge 0$). As shown by Fig. \ref{fig31}, the curve of $a(t)$ weakly depends on the initial values of $\phi_B$ and the inflationary potentials, so we can fit Eq.(\ref{eq3.12}) with any given inflationary potential and
initial values of $\phi_B$, as long as the condition (\ref{eq1.1}) is satisfied. In particular, 
for the Starobinsky potential with $\dot\phi_B > 0$ we find 
\bqn
\lb{eq3.14}
d_0 &\simeq& 74.057, \;\;\;
d_2 \simeq 0.104, \nb\\
d_3 &\simeq& 0.001, \;\;\;
d_4 \simeq 0.002,
\eqn
for which the relative errors $\delta a(t)$ are less than $1\%$ at any given moment $t \le t_B$, as shown in Fig. \ref{fig32}, where the relative errors are defined by
\bq
\lb{eq3.15}
\delta{A} \equiv \left|\frac{A_n - A_a}{A_n}\right|,
\eq 
where $A_n$ and $A_a$ denote the numerical and analytical values of $A$ with $A = (a,\; \phi)$.

In addition, to model the scalar field $\phi(t)$, we find that in the whole pre-bounce regime ($t \le t_B$), the scalar field can be cast in the form
\begin{equation}
\lb{eq3.13}
\phi(t) - \phi_B = e_3+\text{sgn}\left(\dot\phi_B\right) (e_0+e_2 t) e^{e_1t},\; (t \leq t_B),
\end{equation}
where $e_m\; (m = 0, ..., 3)$ are the fitting constants. Again, we require that {\em $\phi(t)$ and its first derivative be continuous across the bounce}, where $\phi(t)$ is given by Eq.(\ref{eq3.3}) in the post-bounce regime ($t \ge t_B$).
Then, fitting the above with the numerical data, 
we find that with the choices of fitting constants as
\bqn
\lb{eq3.16}
e_0 &\simeq& 0.3526, \;\;\;
e_1 \simeq 0.9808,\nb\\
e_2 &\simeq& 0.0943, \;\;\;
e_3 \simeq -0.3526,
\eqn
for $\dot\phi_B > 0$, the relative errors $\delta\phi(t)$ are less than $1\%$ at any given moment $t \le t_B$,
as shown in Fig. \ref{fig32}. On the other hand, for $\dot\phi_B < 0$, the coefficients $e_n$ are given by 
\bqn
\lb{eq3.17}
e_0 &\simeq& 0.3526, \;\;\;
e_1 \simeq 0.981,\nb\\
e_2 &\simeq& 0.094, \;\;\;
e_3 \simeq 0.3526.
\eqn
To understanding   the above fitting results,  let us first note that  at the bounce $t = 0$ and at the remote contracting phase $t \rightarrow - \infty$, we have  
\bqn
 \lb{eq3.18}
 && 0 = e_3 + \text{sgn}\left(\dot\phi_B\right) e_0, \nb\\
 && \phi_{-\infty} - \phi_B  = e_3,  
 \eqn
where $\phi_{-\infty} \equiv \phi(t =-\infty)$. On the other hand, from Figs. \ref{fig30} and  \ref{fig31}, 
 we also have $\phi_{-\infty} - \phi_B = -\text{sgn}\left(\dot\phi_B\right) {\cal{A}}_0$, where  ${\cal{A}}_0 \simeq 0.35326$.
 Thus, finally we have
 \bqn
 \lb{eq3.18aa}
 e_0 = {\cal{A}}_0 \simeq 0.3526, \quad e_3 = - \text{sgn}\left(\dot\phi_B\right) {\cal{A}}_0.
 \eqn

\clearpage
\section{Conclusions and Remarks}
\lb{SecV}

In this paper, we have systematically studied the evolution of the Universe before the quantum bounce ($t \le t_B$) in one of the most promising modified theories of loop quantum cosmologies, the so-called mLQC-I \cite{Li:2021mop}, which can be obtained by both bottom-up \cite{Yang:2009fp,Assanioussi:2018hee,Assanioussi:2019iye} and top-down \cite{Dapor:2017rwv,Dapor:2017gdk,Han:2021cwb} approaches, as a complementary to the studies in the post-bounce regime ($t \ge t_B$) carried out recently  in \cite{Li:2018opr,Li:2018fco,Li:2019ipm}, where $t_B$ denotes the moment that the quantum bounce occurs, replacing the classical Big bang singularity.  In this paper, we have mainly focused on the initial conditions for which the kinetic energy of the inflationary field $\phi$ dominates the evolution of the Universe at the bounce $t = t_B$, as given by Eq.(\ref{eq1.1}). This set of initial conditions is important and leads generically to slow-roll inflation, as shown explicitly in \cite{Li:2018opr,Li:2018fco,Li:2019ipm}. Such considerations do not lose their generality, as slow-roll inflation is a generic feature in this model, as shown in
\cite{Li:2019ipm}.

With the above in mind, we have studied several important inflationary potentials tested recently by the Planck 2018 data \cite{Planck:2018jri}. These include chaotic, Starobinsky, generalized Starobinsky, polynomials of the first and second kinds, generalized T-models, and natural inflation. Despite the fact that the chaotic and natural inflationary models have some tensions with current observations \cite{Planck:2018jri}, in this paper, we found that the evolution of the Universe weakly depends on the specific form of the potentials, as long as the initially kinetic-energy dominated condition (\ref{eq1.1}) holds at the quantum bounce.

In particular, we found that the evolution before the bounce can be universally divided into three different epochs, {\em the pre-bouncing, pre-transition, and pre-de Sitter epochs}, as defined in Eq.(\ref{eq3.11}). Although the exact moment of the transition point defined by $\dot{H}(t_m) = 0$ depends on the form of the inflationary potentials, the division of the above three different epochs is sharp and clear, as shown by the equation of state $w(\phi)$ in each of the cases considered in Sec. \ref{SecIII}. In addition, the effective cosmological constant $\rho_{\Lambda}$
defined in Eq.(\ref{eq2.18}) has Planck size and soon dominates the evolution of the Universe as one departs from the bounce to the deep contracting phase ($t \rightarrow - \infty$). Therefore, the pre-bouncing and pre-transition epochs are all very short compared with the pre-de Sitter epoch.

Moreover, the numerical solutions with the initial conditions that satisfy Eq.(\ref{eq1.1}) considered in all the models of this paper are universal, as shown explicitly by Fig. \ref{fig31}, and can be approximated well by analytical solutions cast in the forms (\ref{eq3.12}) for the expansion factor $a(t)$ and (\ref{eq3.13}) for the inflation $\phi(t)$, where the constants $d_n$ and $e_n$ appearing in these expressions cna be obtained by fitting the analytical solutions  with the numerical ones. For the expansion factor $a(t)$ we have found that the fitting constants $d_n$ are given by Eq.(\ref{eq3.14}) for both $\dot\phi_B > 0$ and $\dot\phi_B < 0$. However, for the scalar field, the fitting constant $e_n$ depend on the signs of $\dot\phi_B$, and are given respectively by Eqs.(\ref{eq3.16}) and (\ref{eq3.17}). 

The above remarkable features are also shared in the evolution of the Universe in the post-bounce regime not only in the framework of mLQC-I \cite{Li:2018opr,Li:2018fco,Li:2019ipm,Li:2021mop} but also in that of LQC
\cite{Zhu:2016dkn,Zhu:2017jew,Shahalam:2017wba,Shahalam:2018rby,Sharma:2018vnv,Sharma:2019okc,Shahalam:2019mpw}. We believe that this universal behavior of background evolution will significantly simplify the studies of cosmological perturbations in mLQC-I, as has already been the case in LQC \cite{Zhu:2017jew}.

\begin{acknowledgments}

J.S. and C.B. are supported by the Baylor Physics graduate program, and R.P. and A.W. are partially supported by the US NSF grant: PHY-2308845.

\end{acknowledgments}

\bibliographystyle{apsrev4-1}
\bibliography{main}

\begin{thebibliography}{82}%
\makeatletter
\providecommand \@ifxundefined [1]{%
 \@ifx{#1\undefined}
}%
\providecommand \@ifnum [1]{%
 \ifnum #1\expandafter \@firstoftwo
 \else \expandafter \@secondoftwo
 \fi
}%
\providecommand \@ifx [1]{%
 \ifx #1\expandafter \@firstoftwo
 \else \expandafter \@secondoftwo
 \fi
}%
\providecommand \natexlab [1]{#1}%
\providecommand \enquote  [1]{``#1''}%
\providecommand \bibnamefont  [1]{#1}%
\providecommand \bibfnamefont [1]{#1}%
\providecommand \citenamefont [1]{#1}%
\providecommand \href@noop [0]{\@secondoftwo}%
\providecommand \href [0]{\begingroup \@sanitize@url \@href}%
\providecommand \@href[1]{\@@startlink{#1}\@@href}%
\providecommand \@@href[1]{\endgroup#1\@@endlink}%
\providecommand \@sanitize@url [0]{\catcode `\\12\catcode `\$12\catcode `\&12\catcode `\#12\catcode `\^12\catcode `\_12\catcode `\%12\relax}%
\providecommand \@@startlink[1]{}%
\providecommand \@@endlink[0]{}%
\providecommand \url  [0]{\begingroup\@sanitize@url \@url }%
\providecommand \@url [1]{\endgroup\@href {#1}{\urlprefix }}%
\providecommand \urlprefix  [0]{URL }%
\providecommand \Eprint [0]{\href }%
\providecommand \doibase [0]{http://dx.doi.org/}%
\providecommand \selectlanguage [0]{\@gobble}%
\providecommand \bibinfo  [0]{\@secondoftwo}%
\providecommand \bibfield  [0]{\@secondoftwo}%
\providecommand \translation [1]{[#1]}%
\providecommand \BibitemOpen [0]{}%
\providecommand \bibitemStop [0]{}%
\providecommand \bibitemNoStop [0]{.\EOS\space}%
\providecommand \EOS [0]{\spacefactor3000\relax}%
\providecommand \BibitemShut  [1]{\csname bibitem#1\endcsname}%
\let\auto@bib@innerbib\@empty
\bibitem [{\citenamefont {{Guth}}(1981)}]{1981PhRvD..23..347G}%
  \BibitemOpen
  \bibfield  {author} {\bibinfo {author} {\bibfnamefont {A.~H.}\ \bibnamefont {{Guth}}},\ }\href {\doibase 10.1103/PhysRevD.23.347} {\bibfield  {journal} {\bibinfo  {journal} {\prd}\ }\textbf {\bibinfo {volume} {23}},\ \bibinfo {pages} {347} (\bibinfo {year} {1981})}\BibitemShut {NoStop}%
\bibitem [{\citenamefont {Akrami}\ \emph {et~al.}(2020)\citenamefont {Akrami} \emph {et~al.}}]{Planck:2018jri}%
  \BibitemOpen
  \bibfield  {author} {\bibinfo {author} {\bibfnamefont {Y.}~\bibnamefont {Akrami}} \emph {et~al.} (\bibinfo {collaboration} {Planck}),\ }\href {\doibase 10.1051/0004-6361/201833887} {\bibfield  {journal} {\bibinfo  {journal} {Astron. Astrophys.}\ }\textbf {\bibinfo {volume} {641}},\ \bibinfo {pages} {A10} (\bibinfo {year} {2020})},\ \Eprint {http://arxiv.org/abs/1807.06211} {arXiv:1807.06211 [astro-ph.CO]} \BibitemShut {NoStop}%
\bibitem [{\citenamefont {Borde}\ \emph {et~al.}(2003)\citenamefont {Borde}, \citenamefont {Guth},\ and\ \citenamefont {Vilenkin}}]{Borde:2001nh}%
  \BibitemOpen
  \bibfield  {author} {\bibinfo {author} {\bibfnamefont {A.}~\bibnamefont {Borde}}, \bibinfo {author} {\bibfnamefont {A.~H.}\ \bibnamefont {Guth}}, \ and\ \bibinfo {author} {\bibfnamefont {A.}~\bibnamefont {Vilenkin}},\ }\href {\doibase 10.1103/PhysRevLett.90.151301} {\bibfield  {journal} {\bibinfo  {journal} {Phys. Rev. Lett.}\ }\textbf {\bibinfo {volume} {90}},\ \bibinfo {pages} {151301} (\bibinfo {year} {2003})},\ \Eprint {http://arxiv.org/abs/gr-qc/0110012} {arXiv:gr-qc/0110012} \BibitemShut {NoStop}%
\bibitem [{\citenamefont {Green}\ \emph {et~al.}(2012)\citenamefont {Green}, \citenamefont {Schwarz},\ and\ \citenamefont {Witten}}]{Green_Schwarz_Witten_2012}%
  \BibitemOpen
  \bibfield  {author} {\bibinfo {author} {\bibfnamefont {M.~B.}\ \bibnamefont {Green}}, \bibinfo {author} {\bibfnamefont {J.~H.}\ \bibnamefont {Schwarz}}, \ and\ \bibinfo {author} {\bibfnamefont {E.}~\bibnamefont {Witten}},\ }\href@noop {} {\emph {\bibinfo {title} {Superstring Theory: 25th Anniversary Edition}}},\ Cambridge Monographs on Mathematical Physics\ (\bibinfo  {publisher} {Cambridge University Press},\ \bibinfo {year} {2012})\BibitemShut {NoStop}%
\bibitem [{\citenamefont {Becker}\ \emph {et~al.}(2006)\citenamefont {Becker}, \citenamefont {Becker},\ and\ \citenamefont {Schwarz}}]{Becker:2006dvp}%
  \BibitemOpen
  \bibfield  {author} {\bibinfo {author} {\bibfnamefont {K.}~\bibnamefont {Becker}}, \bibinfo {author} {\bibfnamefont {M.}~\bibnamefont {Becker}}, \ and\ \bibinfo {author} {\bibfnamefont {J.~H.}\ \bibnamefont {Schwarz}},\ }\href {\doibase 10.1017/CBO9780511816086} {\emph {\bibinfo {title} {{String theory and M-theory: A modern introduction}}}}\ (\bibinfo  {publisher} {Cambridge University Press},\ \bibinfo {year} {2006})\BibitemShut {NoStop}%
\bibitem [{\citenamefont {Ashtekar}\ and\ \citenamefont {Lewandowski}(2004)}]{Ashtekar:2004eh}%
  \BibitemOpen
  \bibfield  {author} {\bibinfo {author} {\bibfnamefont {A.}~\bibnamefont {Ashtekar}}\ and\ \bibinfo {author} {\bibfnamefont {J.}~\bibnamefont {Lewandowski}},\ }\href {\doibase 10.1088/0264-9381/21/15/R01} {\bibfield  {journal} {\bibinfo  {journal} {Class. Quant. Grav.}\ }\textbf {\bibinfo {volume} {21}},\ \bibinfo {pages} {R53} (\bibinfo {year} {2004})},\ \Eprint {http://arxiv.org/abs/gr-qc/0404018} {arXiv:gr-qc/0404018} \BibitemShut {NoStop}%
\bibitem [{\citenamefont {Thiemann}(2007)}]{Thiemann_2007}%
  \BibitemOpen
  \bibfield  {author} {\bibinfo {author} {\bibfnamefont {T.}~\bibnamefont {Thiemann}},\ }\href@noop {} {\emph {\bibinfo {title} {Modern Canonical Quantum General Relativity}}},\ Cambridge Monographs on Mathematical Physics\ (\bibinfo  {publisher} {Cambridge University Press},\ \bibinfo {year} {2007})\BibitemShut {NoStop}%
\bibitem [{\citenamefont {Bojowald}(2010)}]{Bojowald_2010}%
  \BibitemOpen
  \bibfield  {author} {\bibinfo {author} {\bibfnamefont {M.}~\bibnamefont {Bojowald}},\ }\href@noop {} {\emph {\bibinfo {title} {Canonical Gravity and Applications: Cosmology, Black Holes, and Quantum Gravity}}}\ (\bibinfo  {publisher} {Cambridge University Press},\ \bibinfo {year} {2010})\BibitemShut {NoStop}%
\bibitem [{\citenamefont {Gambini}\ and\ \citenamefont {Pullin}(2011)}]{Gambini:2011zz}%
  \BibitemOpen
  \bibfield  {author} {\bibinfo {author} {\bibfnamefont {R.}~\bibnamefont {Gambini}}\ and\ \bibinfo {author} {\bibfnamefont {J.}~\bibnamefont {Pullin}},\ }\href@noop {} {\emph {\bibinfo {title} {{A first course in loop quantum gravity}}}}\ (\bibinfo {year} {2011})\BibitemShut {NoStop}%
\bibitem [{\citenamefont {Rovelli}\ and\ \citenamefont {Vidotto}(2014)}]{Rovelli:2014ssa}%
  \BibitemOpen
  \bibfield  {author} {\bibinfo {author} {\bibfnamefont {C.}~\bibnamefont {Rovelli}}\ and\ \bibinfo {author} {\bibfnamefont {F.}~\bibnamefont {Vidotto}},\ }\href@noop {} {\emph {\bibinfo {title} {{Covariant Loop Quantum Gravity}: {An Elementary Introduction to Quantum Gravity and Spinfoam Theory}}}},\ Cambridge Monographs on Mathematical Physics\ (\bibinfo  {publisher} {Cambridge University Press},\ \bibinfo {year} {2014})\BibitemShut {NoStop}%
\bibitem [{\citenamefont {Bojowald}(2005)}]{Bojowald:2005epg}%
  \BibitemOpen
  \bibfield  {author} {\bibinfo {author} {\bibfnamefont {M.}~\bibnamefont {Bojowald}},\ }\href {\doibase 10.12942/lrr-2005-11} {\bibfield  {journal} {\bibinfo  {journal} {Living Rev. Rel.}\ }\textbf {\bibinfo {volume} {8}},\ \bibinfo {pages} {11} (\bibinfo {year} {2005})},\ \Eprint {http://arxiv.org/abs/gr-qc/0601085} {arXiv:gr-qc/0601085} \BibitemShut {NoStop}%
\bibitem [{\citenamefont {Ashtekar}\ and\ \citenamefont {Singh}(2011)}]{Ashtekar:2011ni}%
  \BibitemOpen
  \bibfield  {author} {\bibinfo {author} {\bibfnamefont {A.}~\bibnamefont {Ashtekar}}\ and\ \bibinfo {author} {\bibfnamefont {P.}~\bibnamefont {Singh}},\ }\href {\doibase 10.1088/0264-9381/28/21/213001} {\bibfield  {journal} {\bibinfo  {journal} {Class. Quant. Grav.}\ }\textbf {\bibinfo {volume} {28}},\ \bibinfo {pages} {213001} (\bibinfo {year} {2011})},\ \Eprint {http://arxiv.org/abs/1108.0893} {arXiv:1108.0893 [gr-qc]} \BibitemShut {NoStop}%
\bibitem [{\citenamefont {Ashtekar}\ and\ \citenamefont {Barrau}(2015)}]{Ashtekar:2015dja}%
  \BibitemOpen
  \bibfield  {author} {\bibinfo {author} {\bibfnamefont {A.}~\bibnamefont {Ashtekar}}\ and\ \bibinfo {author} {\bibfnamefont {A.}~\bibnamefont {Barrau}},\ }\href {\doibase 10.1088/0264-9381/32/23/234001} {\bibfield  {journal} {\bibinfo  {journal} {Class. Quant. Grav.}\ }\textbf {\bibinfo {volume} {32}},\ \bibinfo {pages} {234001} (\bibinfo {year} {2015})},\ \Eprint {http://arxiv.org/abs/1504.07559} {arXiv:1504.07559 [gr-qc]} \BibitemShut {NoStop}%
\bibitem [{\citenamefont {Agullo}\ and\ \citenamefont {Singh}(2017)}]{Agullo:2016tjh}%
  \BibitemOpen
  \bibfield  {author} {\bibinfo {author} {\bibfnamefont {I.}~\bibnamefont {Agullo}}\ and\ \bibinfo {author} {\bibfnamefont {P.}~\bibnamefont {Singh}},\ }\enquote {\bibinfo {title} {{Loop Quantum Cosmology}},}\ in\ \href {\doibase 10.1142/9789813220003_0007} {\emph {\bibinfo {booktitle} {{Loop Quantum Gravity}: {The First 30 Years}}}},\ \bibinfo {editor} {edited by\ \bibinfo {editor} {\bibfnamefont {A.}~\bibnamefont {Ashtekar}}\ and\ \bibinfo {editor} {\bibfnamefont {J.}~\bibnamefont {Pullin}}}\ (\bibinfo  {publisher} {WSP},\ \bibinfo {year} {2017})\ pp.\ \bibinfo {pages} {183--240},\ \Eprint {http://arxiv.org/abs/1612.01236} {arXiv:1612.01236 [gr-qc]} \BibitemShut {NoStop}%
\bibitem [{\citenamefont {Wilson-Ewing}(2017)}]{Wilson-Ewing:2016yan}%
  \BibitemOpen
  \bibfield  {author} {\bibinfo {author} {\bibfnamefont {E.}~\bibnamefont {Wilson-Ewing}},\ }\href {\doibase 10.1016/j.crhy.2017.02.004} {\bibfield  {journal} {\bibinfo  {journal} {Comptes Rendus Physique}\ }\textbf {\bibinfo {volume} {18}},\ \bibinfo {pages} {207} (\bibinfo {year} {2017})},\ \Eprint {http://arxiv.org/abs/1612.04551} {arXiv:1612.04551 [gr-qc]} \BibitemShut {NoStop}%
\bibitem [{\citenamefont {Elizaga~Navascu\'es}\ and\ \citenamefont {Marug\'an}(2021)}]{ElizagaNavascues:2020uyf}%
  \BibitemOpen
  \bibfield  {author} {\bibinfo {author} {\bibfnamefont {B.}~\bibnamefont {Elizaga~Navascu\'es}}\ and\ \bibinfo {author} {\bibfnamefont {G.~A.~M.}\ \bibnamefont {Marug\'an}},\ }\href {\doibase 10.3389/fspas.2021.624824} {\bibfield  {journal} {\bibinfo  {journal} {Front. Astron. Space Sci.}\ }\textbf {\bibinfo {volume} {8}},\ \bibinfo {pages} {81} (\bibinfo {year} {2021})},\ \Eprint {http://arxiv.org/abs/2011.04559} {arXiv:2011.04559 [gr-qc]} \BibitemShut {NoStop}%
\bibitem [{\citenamefont {Ashtekar}\ and\ \citenamefont {Bianchi}(2021)}]{Ashtekar:2021kfp}%
  \BibitemOpen
  \bibfield  {author} {\bibinfo {author} {\bibfnamefont {A.}~\bibnamefont {Ashtekar}}\ and\ \bibinfo {author} {\bibfnamefont {E.}~\bibnamefont {Bianchi}},\ }\href {\doibase 10.1088/1361-6633/abed91} {\bibfield  {journal} {\bibinfo  {journal} {Rept. Prog. Phys.}\ }\textbf {\bibinfo {volume} {84}},\ \bibinfo {pages} {042001} (\bibinfo {year} {2021})},\ \Eprint {http://arxiv.org/abs/2104.04394} {arXiv:2104.04394 [gr-qc]} \BibitemShut {NoStop}%
\bibitem [{\citenamefont {Li}\ and\ \citenamefont {Singh}(2023)}]{Li:2023dwy}%
  \BibitemOpen
  \bibfield  {author} {\bibinfo {author} {\bibfnamefont {B.-F.}\ \bibnamefont {Li}}\ and\ \bibinfo {author} {\bibfnamefont {P.}~\bibnamefont {Singh}},\ }\href@noop {} {\  (\bibinfo {year} {2023})},\ \Eprint {http://arxiv.org/abs/2304.05426} {arXiv:2304.05426 [gr-qc]} \BibitemShut {NoStop}%
\bibitem [{\citenamefont {Agull\'o}\ \emph {et~al.}(2023)\citenamefont {Agull\'o}, \citenamefont {Wang},\ and\ \citenamefont {Wilson-Ewing}}]{Agullo:2023rqq}%
  \BibitemOpen
  \bibfield  {author} {\bibinfo {author} {\bibfnamefont {I.}~\bibnamefont {Agull\'o}}, \bibinfo {author} {\bibfnamefont {A.}~\bibnamefont {Wang}}, \ and\ \bibinfo {author} {\bibfnamefont {E.}~\bibnamefont {Wilson-Ewing}},\ }\href@noop {} {\  (\bibinfo {year} {2023})},\ \Eprint {http://arxiv.org/abs/2301.10215} {arXiv:2301.10215 [gr-qc]} \BibitemShut {NoStop}%
\bibitem [{\citenamefont {Gasperini}\ and\ \citenamefont {Veneziano}(1993)}]{gasperini1993pre}%
  \BibitemOpen
  \bibfield  {author} {\bibinfo {author} {\bibfnamefont {M.}~\bibnamefont {Gasperini}}\ and\ \bibinfo {author} {\bibfnamefont {G.}~\bibnamefont {Veneziano}},\ }\href@noop {} {\bibfield  {journal} {\bibinfo  {journal} {Astroparticle Physics}\ }\textbf {\bibinfo {volume} {1}},\ \bibinfo {pages} {317} (\bibinfo {year} {1993})}\BibitemShut {NoStop}%
\bibitem [{\citenamefont {Gasperini}\ \emph {et~al.}(1997)\citenamefont {Gasperini}, \citenamefont {Maggiore},\ and\ \citenamefont {Veneziano}}]{Gasperini:1996fu}%
  \BibitemOpen
  \bibfield  {author} {\bibinfo {author} {\bibfnamefont {M.}~\bibnamefont {Gasperini}}, \bibinfo {author} {\bibfnamefont {M.}~\bibnamefont {Maggiore}}, \ and\ \bibinfo {author} {\bibfnamefont {G.}~\bibnamefont {Veneziano}},\ }\href {\doibase 10.1016/S0550-3213(97)00149-1} {\bibfield  {journal} {\bibinfo  {journal} {Nucl. Phys. B}\ }\textbf {\bibinfo {volume} {494}},\ \bibinfo {pages} {315} (\bibinfo {year} {1997})},\ \Eprint {http://arxiv.org/abs/hep-th/9611039} {arXiv:hep-th/9611039} \BibitemShut {NoStop}%
\bibitem [{\citenamefont {Gasperini}\ and\ \citenamefont {Veneziano}(2003)}]{Gasperini:2002bn}%
  \BibitemOpen
  \bibfield  {author} {\bibinfo {author} {\bibfnamefont {M.}~\bibnamefont {Gasperini}}\ and\ \bibinfo {author} {\bibfnamefont {G.}~\bibnamefont {Veneziano}},\ }\href {\doibase 10.1016/S0370-1573(02)00389-7} {\bibfield  {journal} {\bibinfo  {journal} {Phys. Rept.}\ }\textbf {\bibinfo {volume} {373}},\ \bibinfo {pages} {1} (\bibinfo {year} {2003})},\ \Eprint {http://arxiv.org/abs/hep-th/0207130} {arXiv:hep-th/0207130} \BibitemShut {NoStop}%
\bibitem [{\citenamefont {Haro}(2013)}]{Haro:2013bea}%
  \BibitemOpen
  \bibfield  {author} {\bibinfo {author} {\bibfnamefont {J.}~\bibnamefont {Haro}},\ }\href {\doibase 10.1088/1475-7516/2013/11/068} {\bibfield  {journal} {\bibinfo  {journal} {JCAP}\ }\textbf {\bibinfo {volume} {11}},\ \bibinfo {pages} {068} (\bibinfo {year} {2013})},\ \bibinfo {note} {[Erratum: JCAP 05, E01 (2014)]},\ \Eprint {http://arxiv.org/abs/1309.0352} {arXiv:1309.0352 [gr-qc]} \BibitemShut {NoStop}%
\bibitem [{\citenamefont {Alonso-Serrano}\ \emph {et~al.}(2023)\citenamefont {Alonso-Serrano}, \citenamefont {Liška},\ and\ \citenamefont {Vicente-Becceril}}]{article}%
  \BibitemOpen
  \bibfield  {author} {\bibinfo {author} {\bibfnamefont {A.}~\bibnamefont {Alonso-Serrano}}, \bibinfo {author} {\bibfnamefont {M.}~\bibnamefont {Liška}}, \ and\ \bibinfo {author} {\bibfnamefont {A.}~\bibnamefont {Vicente-Becceril}},\ }\href {\doibase 10.1016/j.physletb.2023.137827} {\bibfield  {journal} {\bibinfo  {journal} {Physics Letters B}\ }\textbf {\bibinfo {volume} {839}},\ \bibinfo {pages} {137827} (\bibinfo {year} {2023})}\BibitemShut {NoStop}%
\bibitem [{\citenamefont {Conzinu}\ \emph {et~al.}(2023)\citenamefont {Conzinu}, \citenamefont {Fanizza}, \citenamefont {Gasperini}, \citenamefont {Pavone}, \citenamefont {Tedesco},\ and\ \citenamefont {Veneziano}}]{Conzinu:2023fth}%
  \BibitemOpen
  \bibfield  {author} {\bibinfo {author} {\bibfnamefont {P.}~\bibnamefont {Conzinu}}, \bibinfo {author} {\bibfnamefont {G.}~\bibnamefont {Fanizza}}, \bibinfo {author} {\bibfnamefont {M.}~\bibnamefont {Gasperini}}, \bibinfo {author} {\bibfnamefont {E.}~\bibnamefont {Pavone}}, \bibinfo {author} {\bibfnamefont {L.}~\bibnamefont {Tedesco}}, \ and\ \bibinfo {author} {\bibfnamefont {G.}~\bibnamefont {Veneziano}},\ }\href {\doibase 10.1088/1475-7516/2023/12/019} {\bibfield  {journal} {\bibinfo  {journal} {JCAP}\ }\textbf {\bibinfo {volume} {12}},\ \bibinfo {pages} {019} (\bibinfo {year} {2023})},\ \Eprint {http://arxiv.org/abs/2308.16076} {arXiv:2308.16076 [hep-th]} \BibitemShut {NoStop}%
\bibitem [{\citenamefont {Khoury}\ \emph {et~al.}(2001)\citenamefont {Khoury}, \citenamefont {Ovrut}, \citenamefont {Steinhardt},\ and\ \citenamefont {Turok}}]{Khoury:2001wf}%
  \BibitemOpen
  \bibfield  {author} {\bibinfo {author} {\bibfnamefont {J.}~\bibnamefont {Khoury}}, \bibinfo {author} {\bibfnamefont {B.~A.}\ \bibnamefont {Ovrut}}, \bibinfo {author} {\bibfnamefont {P.~J.}\ \bibnamefont {Steinhardt}}, \ and\ \bibinfo {author} {\bibfnamefont {N.}~\bibnamefont {Turok}},\ }\href {\doibase 10.1103/PhysRevD.64.123522} {\bibfield  {journal} {\bibinfo  {journal} {Phys. Rev. D}\ }\textbf {\bibinfo {volume} {64}},\ \bibinfo {pages} {123522} (\bibinfo {year} {2001})},\ \Eprint {http://arxiv.org/abs/hep-th/0103239} {arXiv:hep-th/0103239} \BibitemShut {NoStop}%
\bibitem [{\citenamefont {Brown}\ \emph {et~al.}(2008)\citenamefont {Brown}, \citenamefont {Freese},\ and\ \citenamefont {Kinney}}]{Brown:2004cs}%
  \BibitemOpen
  \bibfield  {author} {\bibinfo {author} {\bibfnamefont {M.~G.}\ \bibnamefont {Brown}}, \bibinfo {author} {\bibfnamefont {K.}~\bibnamefont {Freese}}, \ and\ \bibinfo {author} {\bibfnamefont {W.~H.}\ \bibnamefont {Kinney}},\ }\href {\doibase 10.1088/1475-7516/2008/03/002} {\bibfield  {journal} {\bibinfo  {journal} {JCAP}\ }\textbf {\bibinfo {volume} {03}},\ \bibinfo {pages} {002} (\bibinfo {year} {2008})},\ \Eprint {http://arxiv.org/abs/astro-ph/0405353} {arXiv:astro-ph/0405353} \BibitemShut {NoStop}%
\bibitem [{\citenamefont {Battefeld}\ and\ \citenamefont {Peter}(2015)}]{Battefeld:2014uga}%
  \BibitemOpen
  \bibfield  {author} {\bibinfo {author} {\bibfnamefont {D.}~\bibnamefont {Battefeld}}\ and\ \bibinfo {author} {\bibfnamefont {P.}~\bibnamefont {Peter}},\ }\href {\doibase 10.1016/j.physrep.2014.12.004} {\bibfield  {journal} {\bibinfo  {journal} {Phys. Rept.}\ }\textbf {\bibinfo {volume} {571}},\ \bibinfo {pages} {1} (\bibinfo {year} {2015})},\ \Eprint {http://arxiv.org/abs/1406.2790} {arXiv:1406.2790 [astro-ph.CO]} \BibitemShut {NoStop}%
\bibitem [{\citenamefont {Brandenberger}\ and\ \citenamefont {Peter}(2017)}]{Brandenberger:2016vhg}%
  \BibitemOpen
  \bibfield  {author} {\bibinfo {author} {\bibfnamefont {R.}~\bibnamefont {Brandenberger}}\ and\ \bibinfo {author} {\bibfnamefont {P.}~\bibnamefont {Peter}},\ }\href {\doibase 10.1007/s10701-016-0057-0} {\bibfield  {journal} {\bibinfo  {journal} {Found. Phys.}\ }\textbf {\bibinfo {volume} {47}},\ \bibinfo {pages} {797} (\bibinfo {year} {2017})},\ \Eprint {http://arxiv.org/abs/1603.05834} {arXiv:1603.05834 [hep-th]} \BibitemShut {NoStop}%
\bibitem [{\citenamefont {Ijjas}\ and\ \citenamefont {Steinhardt}(2018)}]{Ijjas:2018qbo}%
  \BibitemOpen
  \bibfield  {author} {\bibinfo {author} {\bibfnamefont {A.}~\bibnamefont {Ijjas}}\ and\ \bibinfo {author} {\bibfnamefont {P.~J.}\ \bibnamefont {Steinhardt}},\ }\href {\doibase 10.1088/1361-6382/aac482} {\bibfield  {journal} {\bibinfo  {journal} {Class. Quant. Grav.}\ }\textbf {\bibinfo {volume} {35}},\ \bibinfo {pages} {135004} (\bibinfo {year} {2018})},\ \Eprint {http://arxiv.org/abs/1803.01961} {arXiv:1803.01961 [astro-ph.CO]} \BibitemShut {NoStop}%
\bibitem [{\citenamefont {Chandran}\ and\ \citenamefont {Shankaranarayanan}(2024)}]{Chandran:2024utf}%
  \BibitemOpen
  \bibfield  {author} {\bibinfo {author} {\bibfnamefont {S.~M.}\ \bibnamefont {Chandran}}\ and\ \bibinfo {author} {\bibfnamefont {S.}~\bibnamefont {Shankaranarayanan}},\ }\href@noop {} {\  (\bibinfo {year} {2024})},\ \Eprint {http://arxiv.org/abs/2405.08543} {arXiv:2405.08543 [astro-ph.CO]} \BibitemShut {NoStop}%
\bibitem [{\citenamefont {Taveras}(2008)}]{Taveras:2008ke}%
  \BibitemOpen
  \bibfield  {author} {\bibinfo {author} {\bibfnamefont {V.}~\bibnamefont {Taveras}},\ }\href {\doibase 10.1103/PhysRevD.78.064072} {\bibfield  {journal} {\bibinfo  {journal} {Phys. Rev. D}\ }\textbf {\bibinfo {volume} {78}},\ \bibinfo {pages} {064072} (\bibinfo {year} {2008})},\ \Eprint {http://arxiv.org/abs/0807.3325} {arXiv:0807.3325 [gr-qc]} \BibitemShut {NoStop}%
\bibitem [{\citenamefont {Singh}(2018)}]{Singh:2018rwa}%
  \BibitemOpen
  \bibfield  {author} {\bibinfo {author} {\bibfnamefont {P.}~\bibnamefont {Singh}},\ }\href {\doibase 10.1109/MCSE.2018.042781324} {\bibfield  {journal} {\bibinfo  {journal} {Comput. Sci. Eng.}\ }\textbf {\bibinfo {volume} {20}},\ \bibinfo {pages} {26} (\bibinfo {year} {2018})},\ \Eprint {http://arxiv.org/abs/1809.01747} {arXiv:1809.01747 [physics.comp-ph]} \BibitemShut {NoStop}%
\bibitem [{\citenamefont {Corichi}\ and\ \citenamefont {Singh}(2008)}]{Corichi:2007am}%
  \BibitemOpen
  \bibfield  {author} {\bibinfo {author} {\bibfnamefont {A.}~\bibnamefont {Corichi}}\ and\ \bibinfo {author} {\bibfnamefont {P.}~\bibnamefont {Singh}},\ }\href {\doibase 10.1103/PhysRevLett.100.161302} {\bibfield  {journal} {\bibinfo  {journal} {Phys. Rev. Lett.}\ }\textbf {\bibinfo {volume} {100}},\ \bibinfo {pages} {161302} (\bibinfo {year} {2008})},\ \Eprint {http://arxiv.org/abs/0710.4543} {arXiv:0710.4543 [gr-qc]} \BibitemShut {NoStop}%
\bibitem [{\citenamefont {Kami\'nski}\ \emph {et~al.}(2020)\citenamefont {Kami\'nski}, \citenamefont {Kolanowski},\ and\ \citenamefont {Lewandowski}}]{Kaminski:2019qjn}%
  \BibitemOpen
  \bibfield  {author} {\bibinfo {author} {\bibfnamefont {W.}~\bibnamefont {Kami\'nski}}, \bibinfo {author} {\bibfnamefont {M.}~\bibnamefont {Kolanowski}}, \ and\ \bibinfo {author} {\bibfnamefont {J.}~\bibnamefont {Lewandowski}},\ }\href {\doibase 10.1088/1361-6382/ab7ee0} {\bibfield  {journal} {\bibinfo  {journal} {Class. Quant. Grav.}\ }\textbf {\bibinfo {volume} {37}},\ \bibinfo {pages} {095001} (\bibinfo {year} {2020})},\ \Eprint {http://arxiv.org/abs/1912.02556} {arXiv:1912.02556 [gr-qc]} \BibitemShut {NoStop}%
\bibitem [{\citenamefont {Beetle}\ \emph {et~al.}(2017)\citenamefont {Beetle}, \citenamefont {Engle}, \citenamefont {Hogan},\ and\ \citenamefont {Mendon\c{c}a}}]{Beetle:2017qle}%
  \BibitemOpen
  \bibfield  {author} {\bibinfo {author} {\bibfnamefont {C.}~\bibnamefont {Beetle}}, \bibinfo {author} {\bibfnamefont {J.~S.}\ \bibnamefont {Engle}}, \bibinfo {author} {\bibfnamefont {M.~E.}\ \bibnamefont {Hogan}}, \ and\ \bibinfo {author} {\bibfnamefont {P.}~\bibnamefont {Mendon\c{c}a}},\ }\href {\doibase 10.1088/1361-6382/aa89c6} {\bibfield  {journal} {\bibinfo  {journal} {Class. Quant. Grav.}\ }\textbf {\bibinfo {volume} {34}},\ \bibinfo {pages} {225009} (\bibinfo {year} {2017})},\ \Eprint {http://arxiv.org/abs/1706.02424} {arXiv:1706.02424 [gr-qc]} \BibitemShut {NoStop}%
\bibitem [{\citenamefont {Bojowald}(2021)}]{Bojowald:2021kzv}%
  \BibitemOpen
  \bibfield  {author} {\bibinfo {author} {\bibfnamefont {M.}~\bibnamefont {Bojowald}},\ }\href {\doibase 10.3390/universe7070251} {\bibfield  {journal} {\bibinfo  {journal} {Universe}\ }\textbf {\bibinfo {volume} {7}},\ \bibinfo {pages} {251} (\bibinfo {year} {2021})},\ \Eprint {http://arxiv.org/abs/2108.11936} {arXiv:2108.11936 [gr-qc]} \BibitemShut {NoStop}%
\bibitem [{\citenamefont {Li}\ \emph {et~al.}(2021)\citenamefont {Li}, \citenamefont {Singh},\ and\ \citenamefont {Wang}}]{Li:2021mop}%
  \BibitemOpen
  \bibfield  {author} {\bibinfo {author} {\bibfnamefont {B.-F.}\ \bibnamefont {Li}}, \bibinfo {author} {\bibfnamefont {P.}~\bibnamefont {Singh}}, \ and\ \bibinfo {author} {\bibfnamefont {A.}~\bibnamefont {Wang}},\ }\href {\doibase 10.3389/fspas.2021.701417} {\bibfield  {journal} {\bibinfo  {journal} {Front. Astron. Space Sci.}\ }\textbf {\bibinfo {volume} {8}},\ \bibinfo {pages} {701417} (\bibinfo {year} {2021})},\ \Eprint {http://arxiv.org/abs/2105.14067} {arXiv:2105.14067 [gr-qc]} \BibitemShut {NoStop}%
\bibitem [{\citenamefont {Yang}\ \emph {et~al.}(2009)\citenamefont {Yang}, \citenamefont {Ding},\ and\ \citenamefont {Ma}}]{Yang:2009fp}%
  \BibitemOpen
  \bibfield  {author} {\bibinfo {author} {\bibfnamefont {J.}~\bibnamefont {Yang}}, \bibinfo {author} {\bibfnamefont {Y.}~\bibnamefont {Ding}}, \ and\ \bibinfo {author} {\bibfnamefont {Y.}~\bibnamefont {Ma}},\ }\href {\doibase 10.1016/j.physletb.2009.10.072} {\bibfield  {journal} {\bibinfo  {journal} {Phys. Lett. B}\ }\textbf {\bibinfo {volume} {682}},\ \bibinfo {pages} {1} (\bibinfo {year} {2009})},\ \Eprint {http://arxiv.org/abs/0904.4379} {arXiv:0904.4379 [gr-qc]} \BibitemShut {NoStop}%
\bibitem [{\citenamefont {Thiemann}(1998{\natexlab{a}})}]{Thiemann:1996av}%
  \BibitemOpen
  \bibfield  {author} {\bibinfo {author} {\bibfnamefont {T.}~\bibnamefont {Thiemann}},\ }\href {\doibase 10.1088/0264-9381/15/4/012} {\bibfield  {journal} {\bibinfo  {journal} {Class. Quant. Grav.}\ }\textbf {\bibinfo {volume} {15}},\ \bibinfo {pages} {875} (\bibinfo {year} {1998}{\natexlab{a}})},\ \Eprint {http://arxiv.org/abs/gr-qc/9606090} {arXiv:gr-qc/9606090} \BibitemShut {NoStop}%
\bibitem [{\citenamefont {Thiemann}(1998{\natexlab{b}})}]{Thiemann:1996aw}%
  \BibitemOpen
  \bibfield  {author} {\bibinfo {author} {\bibfnamefont {T.}~\bibnamefont {Thiemann}},\ }\href {\doibase 10.1088/0264-9381/15/4/011} {\bibfield  {journal} {\bibinfo  {journal} {Class. Quant. Grav.}\ }\textbf {\bibinfo {volume} {15}},\ \bibinfo {pages} {839} (\bibinfo {year} {1998}{\natexlab{b}})},\ \Eprint {http://arxiv.org/abs/gr-qc/9606089} {arXiv:gr-qc/9606089} \BibitemShut {NoStop}%
\bibitem [{\citenamefont {Assanioussi}\ \emph {et~al.}(2018)\citenamefont {Assanioussi}, \citenamefont {Dapor}, \citenamefont {Liegener},\ and\ \citenamefont {Paw\l{}owski}}]{Assanioussi:2018hee}%
  \BibitemOpen
  \bibfield  {author} {\bibinfo {author} {\bibfnamefont {M.}~\bibnamefont {Assanioussi}}, \bibinfo {author} {\bibfnamefont {A.}~\bibnamefont {Dapor}}, \bibinfo {author} {\bibfnamefont {K.}~\bibnamefont {Liegener}}, \ and\ \bibinfo {author} {\bibfnamefont {T.}~\bibnamefont {Paw\l{}owski}},\ }\href {\doibase 10.1103/PhysRevLett.121.081303} {\bibfield  {journal} {\bibinfo  {journal} {Phys. Rev. Lett.}\ }\textbf {\bibinfo {volume} {121}},\ \bibinfo {pages} {081303} (\bibinfo {year} {2018})},\ \Eprint {http://arxiv.org/abs/1801.00768} {arXiv:1801.00768 [gr-qc]} \BibitemShut {NoStop}%
\bibitem [{\citenamefont {Assanioussi}\ \emph {et~al.}(2019)\citenamefont {Assanioussi}, \citenamefont {Dapor}, \citenamefont {Liegener},\ and\ \citenamefont {Paw\l{}owski}}]{Assanioussi:2019iye}%
  \BibitemOpen
  \bibfield  {author} {\bibinfo {author} {\bibfnamefont {M.}~\bibnamefont {Assanioussi}}, \bibinfo {author} {\bibfnamefont {A.}~\bibnamefont {Dapor}}, \bibinfo {author} {\bibfnamefont {K.}~\bibnamefont {Liegener}}, \ and\ \bibinfo {author} {\bibfnamefont {T.}~\bibnamefont {Paw\l{}owski}},\ }\href {\doibase 10.1103/PhysRevD.100.084003} {\bibfield  {journal} {\bibinfo  {journal} {Phys. Rev. D}\ }\textbf {\bibinfo {volume} {100}},\ \bibinfo {pages} {084003} (\bibinfo {year} {2019})},\ \Eprint {http://arxiv.org/abs/1906.05315} {arXiv:1906.05315 [gr-qc]} \BibitemShut {NoStop}%
\bibitem [{\citenamefont {Li}\ \emph {et~al.}(2018{\natexlab{a}})\citenamefont {Li}, \citenamefont {Singh},\ and\ \citenamefont {Wang}}]{Li:2018opr}%
  \BibitemOpen
  \bibfield  {author} {\bibinfo {author} {\bibfnamefont {B.-F.}\ \bibnamefont {Li}}, \bibinfo {author} {\bibfnamefont {P.}~\bibnamefont {Singh}}, \ and\ \bibinfo {author} {\bibfnamefont {A.}~\bibnamefont {Wang}},\ }\href {\doibase 10.1103/PhysRevD.97.084029} {\bibfield  {journal} {\bibinfo  {journal} {Phys. Rev. D}\ }\textbf {\bibinfo {volume} {97}},\ \bibinfo {pages} {084029} (\bibinfo {year} {2018}{\natexlab{a}})},\ \Eprint {http://arxiv.org/abs/1801.07313} {arXiv:1801.07313 [gr-qc]} \BibitemShut {NoStop}%
\bibitem [{\citenamefont {Li}\ \emph {et~al.}(2018{\natexlab{b}})\citenamefont {Li}, \citenamefont {Singh},\ and\ \citenamefont {Wang}}]{Li:2018fco}%
  \BibitemOpen
  \bibfield  {author} {\bibinfo {author} {\bibfnamefont {B.-F.}\ \bibnamefont {Li}}, \bibinfo {author} {\bibfnamefont {P.}~\bibnamefont {Singh}}, \ and\ \bibinfo {author} {\bibfnamefont {A.}~\bibnamefont {Wang}},\ }\href {\doibase 10.1103/PhysRevD.98.066016} {\bibfield  {journal} {\bibinfo  {journal} {Phys. Rev. D}\ }\textbf {\bibinfo {volume} {98}},\ \bibinfo {pages} {066016} (\bibinfo {year} {2018}{\natexlab{b}})},\ \Eprint {http://arxiv.org/abs/1807.05236} {arXiv:1807.05236 [gr-qc]} \BibitemShut {NoStop}%
\bibitem [{\citenamefont {Li}\ \emph {et~al.}(2019{\natexlab{a}})\citenamefont {Li}, \citenamefont {Singh},\ and\ \citenamefont {Wang}}]{Li:2019ipm}%
  \BibitemOpen
  \bibfield  {author} {\bibinfo {author} {\bibfnamefont {B.-F.}\ \bibnamefont {Li}}, \bibinfo {author} {\bibfnamefont {P.}~\bibnamefont {Singh}}, \ and\ \bibinfo {author} {\bibfnamefont {A.}~\bibnamefont {Wang}},\ }\href {\doibase 10.1103/PhysRevD.100.063513} {\bibfield  {journal} {\bibinfo  {journal} {Phys. Rev. D}\ }\textbf {\bibinfo {volume} {100}},\ \bibinfo {pages} {063513} (\bibinfo {year} {2019}{\natexlab{a}})},\ \Eprint {http://arxiv.org/abs/1906.01001} {arXiv:1906.01001 [gr-qc]} \BibitemShut {NoStop}%
\bibitem [{\citenamefont {Dapor}\ and\ \citenamefont {Liegener}(2018{\natexlab{a}})}]{Dapor:2017rwv}%
  \BibitemOpen
  \bibfield  {author} {\bibinfo {author} {\bibfnamefont {A.}~\bibnamefont {Dapor}}\ and\ \bibinfo {author} {\bibfnamefont {K.}~\bibnamefont {Liegener}},\ }\href {\doibase 10.1016/j.physletb.2018.09.005} {\bibfield  {journal} {\bibinfo  {journal} {Phys. Lett. B}\ }\textbf {\bibinfo {volume} {785}},\ \bibinfo {pages} {506} (\bibinfo {year} {2018}{\natexlab{a}})},\ \Eprint {http://arxiv.org/abs/1706.09833} {arXiv:1706.09833 [gr-qc]} \BibitemShut {NoStop}%
\bibitem [{\citenamefont {Dapor}\ and\ \citenamefont {Liegener}(2018{\natexlab{b}})}]{Dapor:2017gdk}%
  \BibitemOpen
  \bibfield  {author} {\bibinfo {author} {\bibfnamefont {A.}~\bibnamefont {Dapor}}\ and\ \bibinfo {author} {\bibfnamefont {K.}~\bibnamefont {Liegener}},\ }\href {\doibase 10.1088/1361-6382/aac4ba} {\bibfield  {journal} {\bibinfo  {journal} {Class. Quant. Grav.}\ }\textbf {\bibinfo {volume} {35}},\ \bibinfo {pages} {135011} (\bibinfo {year} {2018}{\natexlab{b}})},\ \Eprint {http://arxiv.org/abs/1710.04015} {arXiv:1710.04015 [gr-qc]} \BibitemShut {NoStop}%
\bibitem [{\citenamefont {Han}\ and\ \citenamefont {Liu}(2021)}]{Han:2021cwb}%
  \BibitemOpen
  \bibfield  {author} {\bibinfo {author} {\bibfnamefont {M.}~\bibnamefont {Han}}\ and\ \bibinfo {author} {\bibfnamefont {H.}~\bibnamefont {Liu}},\ }\href {\doibase 10.1103/PhysRevD.104.024011} {\bibfield  {journal} {\bibinfo  {journal} {Phys. Rev. D}\ }\textbf {\bibinfo {volume} {104}},\ \bibinfo {pages} {024011} (\bibinfo {year} {2021})},\ \Eprint {http://arxiv.org/abs/2101.07659} {arXiv:2101.07659 [gr-qc]} \BibitemShut {NoStop}%
\bibitem [{\citenamefont {de~Haro}(2018)}]{deHaro:2018khb}%
  \BibitemOpen
  \bibfield  {author} {\bibinfo {author} {\bibfnamefont {J.}~\bibnamefont {de~Haro}},\ }\href {\doibase 10.1140/epjc/s10052-018-6402-z} {\bibfield  {journal} {\bibinfo  {journal} {Eur. Phys. J. C}\ }\textbf {\bibinfo {volume} {78}},\ \bibinfo {pages} {926} (\bibinfo {year} {2018})},\ \Eprint {http://arxiv.org/abs/1806.08926} {arXiv:1806.08926 [gr-qc]} \BibitemShut {NoStop}%
\bibitem [{\citenamefont {Saini}\ and\ \citenamefont {Singh}(2019{\natexlab{a}})}]{Saini:2018tto}%
  \BibitemOpen
  \bibfield  {author} {\bibinfo {author} {\bibfnamefont {S.}~\bibnamefont {Saini}}\ and\ \bibinfo {author} {\bibfnamefont {P.}~\bibnamefont {Singh}},\ }\href {\doibase 10.1088/1361-6382/ab1274} {\bibfield  {journal} {\bibinfo  {journal} {Class. Quant. Grav.}\ }\textbf {\bibinfo {volume} {36}},\ \bibinfo {pages} {105014} (\bibinfo {year} {2019}{\natexlab{a}})},\ \Eprint {http://arxiv.org/abs/1812.08937} {arXiv:1812.08937 [gr-qc]} \BibitemShut {NoStop}%
\bibitem [{\citenamefont {Saini}\ and\ \citenamefont {Singh}(2019{\natexlab{b}})}]{Saini:2019tem}%
  \BibitemOpen
  \bibfield  {author} {\bibinfo {author} {\bibfnamefont {S.}~\bibnamefont {Saini}}\ and\ \bibinfo {author} {\bibfnamefont {P.}~\bibnamefont {Singh}},\ }\href {\doibase 10.1088/1361-6382/ab1608} {\bibfield  {journal} {\bibinfo  {journal} {Class. Quant. Grav.}\ }\textbf {\bibinfo {volume} {36}},\ \bibinfo {pages} {105010} (\bibinfo {year} {2019}{\natexlab{b}})},\ \Eprint {http://arxiv.org/abs/1901.01279} {arXiv:1901.01279 [gr-qc]} \BibitemShut {NoStop}%
\bibitem [{\citenamefont {Li}\ and\ \citenamefont {Singh}(2022{\natexlab{a}})}]{Li:2021fmu}%
  \BibitemOpen
  \bibfield  {author} {\bibinfo {author} {\bibfnamefont {B.-F.}\ \bibnamefont {Li}}\ and\ \bibinfo {author} {\bibfnamefont {P.}~\bibnamefont {Singh}},\ }\href {\doibase 10.1103/PhysRevD.105.046013} {\bibfield  {journal} {\bibinfo  {journal} {Phys. Rev. D}\ }\textbf {\bibinfo {volume} {105}},\ \bibinfo {pages} {046013} (\bibinfo {year} {2022}{\natexlab{a}})},\ \Eprint {http://arxiv.org/abs/2108.12553} {arXiv:2108.12553 [gr-qc]} \BibitemShut {NoStop}%
\bibitem [{\citenamefont {Bonga}\ and\ \citenamefont {Gupt}(2016)}]{Bonga:2015xna}%
  \BibitemOpen
  \bibfield  {author} {\bibinfo {author} {\bibfnamefont {B.}~\bibnamefont {Bonga}}\ and\ \bibinfo {author} {\bibfnamefont {B.}~\bibnamefont {Gupt}},\ }\href {\doibase 10.1103/PhysRevD.93.063513} {\bibfield  {journal} {\bibinfo  {journal} {Phys. Rev. D}\ }\textbf {\bibinfo {volume} {93}},\ \bibinfo {pages} {063513} (\bibinfo {year} {2016})},\ \Eprint {http://arxiv.org/abs/1510.04896} {arXiv:1510.04896 [gr-qc]} \BibitemShut {NoStop}%
\bibitem [{\citenamefont {Zhu}\ \emph {et~al.}(2017{\natexlab{a}})\citenamefont {Zhu}, \citenamefont {Wang}, \citenamefont {Kirsten}, \citenamefont {Cleaver},\ and\ \citenamefont {Sheng}}]{Zhu:2016dkn}%
  \BibitemOpen
  \bibfield  {author} {\bibinfo {author} {\bibfnamefont {T.}~\bibnamefont {Zhu}}, \bibinfo {author} {\bibfnamefont {A.}~\bibnamefont {Wang}}, \bibinfo {author} {\bibfnamefont {K.}~\bibnamefont {Kirsten}}, \bibinfo {author} {\bibfnamefont {G.}~\bibnamefont {Cleaver}}, \ and\ \bibinfo {author} {\bibfnamefont {Q.}~\bibnamefont {Sheng}},\ }\href {\doibase 10.1016/j.physletb.2017.08.025} {\bibfield  {journal} {\bibinfo  {journal} {Phys. Lett. B}\ }\textbf {\bibinfo {volume} {773}},\ \bibinfo {pages} {196} (\bibinfo {year} {2017}{\natexlab{a}})},\ \Eprint {http://arxiv.org/abs/1607.06329} {arXiv:1607.06329 [gr-qc]} \BibitemShut {NoStop}%
\bibitem [{\citenamefont {Zhu}\ \emph {et~al.}(2017{\natexlab{b}})\citenamefont {Zhu}, \citenamefont {Wang}, \citenamefont {Cleaver}, \citenamefont {Kirsten},\ and\ \citenamefont {Sheng}}]{Zhu:2017jew}%
  \BibitemOpen
  \bibfield  {author} {\bibinfo {author} {\bibfnamefont {T.}~\bibnamefont {Zhu}}, \bibinfo {author} {\bibfnamefont {A.}~\bibnamefont {Wang}}, \bibinfo {author} {\bibfnamefont {G.}~\bibnamefont {Cleaver}}, \bibinfo {author} {\bibfnamefont {K.}~\bibnamefont {Kirsten}}, \ and\ \bibinfo {author} {\bibfnamefont {Q.}~\bibnamefont {Sheng}},\ }\href {\doibase 10.1103/PhysRevD.96.083520} {\bibfield  {journal} {\bibinfo  {journal} {Phys. Rev. D}\ }\textbf {\bibinfo {volume} {96}},\ \bibinfo {pages} {083520} (\bibinfo {year} {2017}{\natexlab{b}})},\ \Eprint {http://arxiv.org/abs/1705.07544} {arXiv:1705.07544 [gr-qc]} \BibitemShut {NoStop}%
\bibitem [{\citenamefont {Shahalam}\ \emph {et~al.}(2017)\citenamefont {Shahalam}, \citenamefont {Sharma}, \citenamefont {Wu},\ and\ \citenamefont {Wang}}]{Shahalam:2017wba}%
  \BibitemOpen
  \bibfield  {author} {\bibinfo {author} {\bibfnamefont {M.}~\bibnamefont {Shahalam}}, \bibinfo {author} {\bibfnamefont {M.}~\bibnamefont {Sharma}}, \bibinfo {author} {\bibfnamefont {Q.}~\bibnamefont {Wu}}, \ and\ \bibinfo {author} {\bibfnamefont {A.}~\bibnamefont {Wang}},\ }\href {\doibase 10.1103/PhysRevD.96.123533} {\bibfield  {journal} {\bibinfo  {journal} {Phys. Rev. D}\ }\textbf {\bibinfo {volume} {96}},\ \bibinfo {pages} {123533} (\bibinfo {year} {2017})},\ \Eprint {http://arxiv.org/abs/1710.09845} {arXiv:1710.09845 [gr-qc]} \BibitemShut {NoStop}%
\bibitem [{\citenamefont {Shahalam}\ \emph {et~al.}(2018)\citenamefont {Shahalam}, \citenamefont {Sami},\ and\ \citenamefont {Wang}}]{Shahalam:2018rby}%
  \BibitemOpen
  \bibfield  {author} {\bibinfo {author} {\bibfnamefont {M.}~\bibnamefont {Shahalam}}, \bibinfo {author} {\bibfnamefont {M.}~\bibnamefont {Sami}}, \ and\ \bibinfo {author} {\bibfnamefont {A.}~\bibnamefont {Wang}},\ }\href {\doibase 10.1103/PhysRevD.98.043524} {\bibfield  {journal} {\bibinfo  {journal} {Phys. Rev. D}\ }\textbf {\bibinfo {volume} {98}},\ \bibinfo {pages} {043524} (\bibinfo {year} {2018})},\ \Eprint {http://arxiv.org/abs/1806.05815} {arXiv:1806.05815 [astro-ph.CO]} \BibitemShut {NoStop}%
\bibitem [{\citenamefont {Sharma}\ \emph {et~al.}(2018)\citenamefont {Sharma}, \citenamefont {Shahalam}, \citenamefont {Wu},\ and\ \citenamefont {Wang}}]{Sharma:2018vnv}%
  \BibitemOpen
  \bibfield  {author} {\bibinfo {author} {\bibfnamefont {M.}~\bibnamefont {Sharma}}, \bibinfo {author} {\bibfnamefont {M.}~\bibnamefont {Shahalam}}, \bibinfo {author} {\bibfnamefont {Q.}~\bibnamefont {Wu}}, \ and\ \bibinfo {author} {\bibfnamefont {A.}~\bibnamefont {Wang}},\ }\href {\doibase 10.1088/1475-7516/2018/11/003} {\bibfield  {journal} {\bibinfo  {journal} {JCAP}\ }\textbf {\bibinfo {volume} {11}},\ \bibinfo {pages} {003} (\bibinfo {year} {2018})},\ \Eprint {http://arxiv.org/abs/1808.05134} {arXiv:1808.05134 [gr-qc]} \BibitemShut {NoStop}%
\bibitem [{\citenamefont {Sharma}\ \emph {et~al.}(2019)\citenamefont {Sharma}, \citenamefont {Zhu},\ and\ \citenamefont {Wang}}]{Sharma:2019okc}%
  \BibitemOpen
  \bibfield  {author} {\bibinfo {author} {\bibfnamefont {M.}~\bibnamefont {Sharma}}, \bibinfo {author} {\bibfnamefont {T.}~\bibnamefont {Zhu}}, \ and\ \bibinfo {author} {\bibfnamefont {A.}~\bibnamefont {Wang}},\ }\href {\doibase 10.1088/0253-6102/71/10/1205} {\bibfield  {journal} {\bibinfo  {journal} {Commun. Theor. Phys.}\ }\textbf {\bibinfo {volume} {71}},\ \bibinfo {pages} {1205} (\bibinfo {year} {2019})},\ \Eprint {http://arxiv.org/abs/1903.07382} {arXiv:1903.07382 [gr-qc]} \BibitemShut {NoStop}%
\bibitem [{\citenamefont {Shahalam}\ \emph {et~al.}(2020)\citenamefont {Shahalam}, \citenamefont {Al~Ajmi}, \citenamefont {Myrzakulov},\ and\ \citenamefont {Wang}}]{Shahalam:2019mpw}%
  \BibitemOpen
  \bibfield  {author} {\bibinfo {author} {\bibfnamefont {M.}~\bibnamefont {Shahalam}}, \bibinfo {author} {\bibfnamefont {M.}~\bibnamefont {Al~Ajmi}}, \bibinfo {author} {\bibfnamefont {R.}~\bibnamefont {Myrzakulov}}, \ and\ \bibinfo {author} {\bibfnamefont {A.}~\bibnamefont {Wang}},\ }\href {\doibase 10.1088/1361-6382/aba486} {\bibfield  {journal} {\bibinfo  {journal} {Class. Quant. Grav.}\ }\textbf {\bibinfo {volume} {37}},\ \bibinfo {pages} {195026} (\bibinfo {year} {2020})},\ \Eprint {http://arxiv.org/abs/1912.00616} {arXiv:1912.00616 [gr-qc]} \BibitemShut {NoStop}%
\bibitem [{\citenamefont {Zhu}\ \emph {et~al.}(2016)\citenamefont {Zhu}, \citenamefont {Wang}, \citenamefont {Kirsten}, \citenamefont {Cleaver},\ and\ \citenamefont {Sheng}}]{Zhu:2016srz}%
  \BibitemOpen
  \bibfield  {author} {\bibinfo {author} {\bibfnamefont {T.}~\bibnamefont {Zhu}}, \bibinfo {author} {\bibfnamefont {A.}~\bibnamefont {Wang}}, \bibinfo {author} {\bibfnamefont {K.}~\bibnamefont {Kirsten}}, \bibinfo {author} {\bibfnamefont {G.}~\bibnamefont {Cleaver}}, \ and\ \bibinfo {author} {\bibfnamefont {Q.}~\bibnamefont {Sheng}},\ }\href {\doibase 10.1103/PhysRevD.93.123525} {\bibfield  {journal} {\bibinfo  {journal} {Phys. Rev. D}\ }\textbf {\bibinfo {volume} {93}},\ \bibinfo {pages} {123525} (\bibinfo {year} {2016})},\ \Eprint {http://arxiv.org/abs/1604.05739} {arXiv:1604.05739 [gr-qc]} \BibitemShut {NoStop}%
\bibitem [{\citenamefont {Levy}\ and\ \citenamefont {O.~Ramos}(2024)}]{Levy:2024naz}%
  \BibitemOpen
  \bibfield  {author} {\bibinfo {author} {\bibfnamefont {G.~L. L.~W.}\ \bibnamefont {Levy}}\ and\ \bibinfo {author} {\bibfnamefont {R.}~\bibnamefont {O.~Ramos}},\ }\href@noop {} {\  (\bibinfo {year} {2024})},\ \Eprint {http://arxiv.org/abs/2404.10149} {arXiv:2404.10149 [gr-qc]} \BibitemShut {NoStop}%
\bibitem [{\citenamefont {Li}\ \emph {et~al.}(2020{\natexlab{a}})\citenamefont {Li}, \citenamefont {Singh},\ and\ \citenamefont {Wang}}]{Li:2019qzr}%
  \BibitemOpen
  \bibfield  {author} {\bibinfo {author} {\bibfnamefont {B.-F.}\ \bibnamefont {Li}}, \bibinfo {author} {\bibfnamefont {P.}~\bibnamefont {Singh}}, \ and\ \bibinfo {author} {\bibfnamefont {A.}~\bibnamefont {Wang}},\ }\href {\doibase 10.1103/PhysRevD.101.086004} {\bibfield  {journal} {\bibinfo  {journal} {Phys. Rev. D}\ }\textbf {\bibinfo {volume} {101}},\ \bibinfo {pages} {086004} (\bibinfo {year} {2020}{\natexlab{a}})},\ \Eprint {http://arxiv.org/abs/1912.08225} {arXiv:1912.08225 [gr-qc]} \BibitemShut {NoStop}%
\bibitem [{\citenamefont {Li}\ \emph {et~al.}(2020{\natexlab{b}})\citenamefont {Li}, \citenamefont {Olmedo}, \citenamefont {Singh},\ and\ \citenamefont {Wang}}]{Li:2020mfi}%
  \BibitemOpen
  \bibfield  {author} {\bibinfo {author} {\bibfnamefont {B.-F.}\ \bibnamefont {Li}}, \bibinfo {author} {\bibfnamefont {J.}~\bibnamefont {Olmedo}}, \bibinfo {author} {\bibfnamefont {P.}~\bibnamefont {Singh}}, \ and\ \bibinfo {author} {\bibfnamefont {A.}~\bibnamefont {Wang}},\ }\href {\doibase 10.1103/PhysRevD.102.126025} {\bibfield  {journal} {\bibinfo  {journal} {Phys. Rev. D}\ }\textbf {\bibinfo {volume} {102}},\ \bibinfo {pages} {126025} (\bibinfo {year} {2020}{\natexlab{b}})},\ \Eprint {http://arxiv.org/abs/2008.09135} {arXiv:2008.09135 [gr-qc]} \BibitemShut {NoStop}%
\bibitem [{\citenamefont {Li}\ \emph {et~al.}(2024)\citenamefont {Li}, \citenamefont {Motaharfar},\ and\ \citenamefont {Singh}}]{Li:2024xxz}%
  \BibitemOpen
  \bibfield  {author} {\bibinfo {author} {\bibfnamefont {B.-F.}\ \bibnamefont {Li}}, \bibinfo {author} {\bibfnamefont {M.}~\bibnamefont {Motaharfar}}, \ and\ \bibinfo {author} {\bibfnamefont {P.}~\bibnamefont {Singh}},\ }\href@noop {} {\  (\bibinfo {year} {2024})},\ \Eprint {http://arxiv.org/abs/2405.12296} {arXiv:2405.12296 [gr-qc]} \BibitemShut {NoStop}%
\bibitem [{\citenamefont {Li}\ and\ \citenamefont {Singh}(2022{\natexlab{b}})}]{Li:2022evi}%
  \BibitemOpen
  \bibfield  {author} {\bibinfo {author} {\bibfnamefont {B.-F.}\ \bibnamefont {Li}}\ and\ \bibinfo {author} {\bibfnamefont {P.}~\bibnamefont {Singh}},\ }\href {\doibase 10.1103/PhysRevD.106.086015} {\bibfield  {journal} {\bibinfo  {journal} {Phys. Rev. D}\ }\textbf {\bibinfo {volume} {106}},\ \bibinfo {pages} {086015} (\bibinfo {year} {2022}{\natexlab{b}})},\ \Eprint {http://arxiv.org/abs/2206.12434} {arXiv:2206.12434 [gr-qc]} \BibitemShut {NoStop}%
\bibitem [{\citenamefont {Meissner}(2004)}]{Meissner:2004ju}%
  \BibitemOpen
  \bibfield  {author} {\bibinfo {author} {\bibfnamefont {K.~A.}\ \bibnamefont {Meissner}},\ }\href {\doibase 10.1088/0264-9381/21/22/015} {\bibfield  {journal} {\bibinfo  {journal} {Class. Quant. Grav.}\ }\textbf {\bibinfo {volume} {21}},\ \bibinfo {pages} {5245} (\bibinfo {year} {2004})},\ \Eprint {http://arxiv.org/abs/gr-qc/0407052} {arXiv:gr-qc/0407052} \BibitemShut {NoStop}%
\bibitem [{\citenamefont {Rovelli}\ and\ \citenamefont {Smolin}(1995)}]{Rovelli:1994ge}%
  \BibitemOpen
  \bibfield  {author} {\bibinfo {author} {\bibfnamefont {C.}~\bibnamefont {Rovelli}}\ and\ \bibinfo {author} {\bibfnamefont {L.}~\bibnamefont {Smolin}},\ }\href {\doibase 10.1016/0550-3213(95)00150-Q} {\bibfield  {journal} {\bibinfo  {journal} {Nucl. Phys. B}\ }\textbf {\bibinfo {volume} {442}},\ \bibinfo {pages} {593} (\bibinfo {year} {1995})},\ \bibinfo {note} {[Erratum: Nucl.Phys.B 456, 753--754 (1995)]},\ \Eprint {http://arxiv.org/abs/gr-qc/9411005} {arXiv:gr-qc/9411005} \BibitemShut {NoStop}%
\bibitem [{\citenamefont {Ashtekar}\ and\ \citenamefont {Lewandowski}(1997)}]{Ashtekar:1996eg}%
  \BibitemOpen
  \bibfield  {author} {\bibinfo {author} {\bibfnamefont {A.}~\bibnamefont {Ashtekar}}\ and\ \bibinfo {author} {\bibfnamefont {J.}~\bibnamefont {Lewandowski}},\ }\href {\doibase 10.1088/0264-9381/14/1A/006} {\bibfield  {journal} {\bibinfo  {journal} {Class. Quant. Grav.}\ }\textbf {\bibinfo {volume} {14}},\ \bibinfo {pages} {A55} (\bibinfo {year} {1997})},\ \Eprint {http://arxiv.org/abs/gr-qc/9602046} {arXiv:gr-qc/9602046} \BibitemShut {NoStop}%
\bibitem [{\citenamefont {Ashtekar}\ and\ \citenamefont {Sloan}(2011{\natexlab{a}})}]{Ashtekar:2011rm}%
  \BibitemOpen
  \bibfield  {author} {\bibinfo {author} {\bibfnamefont {A.}~\bibnamefont {Ashtekar}}\ and\ \bibinfo {author} {\bibfnamefont {D.}~\bibnamefont {Sloan}},\ }\href {\doibase 10.1007/s10714-011-1246-y} {\bibfield  {journal} {\bibinfo  {journal} {Gen. Rel. Grav.}\ }\textbf {\bibinfo {volume} {43}},\ \bibinfo {pages} {3619} (\bibinfo {year} {2011}{\natexlab{a}})},\ \Eprint {http://arxiv.org/abs/1103.2475} {arXiv:1103.2475 [gr-qc]} \BibitemShut {NoStop}%
\bibitem [{\citenamefont {Ashtekar}\ and\ \citenamefont {Sloan}(2011{\natexlab{b}})}]{Ashtekar:2009mm}%
  \BibitemOpen
  \bibfield  {author} {\bibinfo {author} {\bibfnamefont {A.}~\bibnamefont {Ashtekar}}\ and\ \bibinfo {author} {\bibfnamefont {D.}~\bibnamefont {Sloan}},\ }\href {\doibase 10.1016/j.physletb.2010.09.058} {\bibfield  {journal} {\bibinfo  {journal} {Phys. Lett. B}\ }\textbf {\bibinfo {volume} {694}},\ \bibinfo {pages} {108} (\bibinfo {year} {2011}{\natexlab{b}})},\ \Eprint {http://arxiv.org/abs/0912.4093} {arXiv:0912.4093 [gr-qc]} \BibitemShut {NoStop}%
\bibitem [{\citenamefont {Kallosh}\ and\ \citenamefont {Linde}(2022)}]{Kallosh:2022feu}%
  \BibitemOpen
  \bibfield  {author} {\bibinfo {author} {\bibfnamefont {R.}~\bibnamefont {Kallosh}}\ and\ \bibinfo {author} {\bibfnamefont {A.}~\bibnamefont {Linde}},\ }\href {\doibase 10.1088/1475-7516/2022/04/017} {\bibfield  {journal} {\bibinfo  {journal} {JCAP}\ }\textbf {\bibinfo {volume} {04}},\ \bibinfo {pages} {017} (\bibinfo {year} {2022})},\ \Eprint {http://arxiv.org/abs/2202.06492} {arXiv:2202.06492 [astro-ph.CO]} \BibitemShut {NoStop}%
\bibitem [{\citenamefont {Li}\ \emph {et~al.}(2019{\natexlab{b}})\citenamefont {Li}, \citenamefont {Zhu}, \citenamefont {Wang}, \citenamefont {Kirsten}, \citenamefont {Cleaver},\ and\ \citenamefont {Sheng}}]{Li:2018vzr}%
  \BibitemOpen
  \bibfield  {author} {\bibinfo {author} {\bibfnamefont {B.-F.}\ \bibnamefont {Li}}, \bibinfo {author} {\bibfnamefont {T.}~\bibnamefont {Zhu}}, \bibinfo {author} {\bibfnamefont {A.}~\bibnamefont {Wang}}, \bibinfo {author} {\bibfnamefont {K.}~\bibnamefont {Kirsten}}, \bibinfo {author} {\bibfnamefont {G.}~\bibnamefont {Cleaver}}, \ and\ \bibinfo {author} {\bibfnamefont {Q.}~\bibnamefont {Sheng}},\ }\href {\doibase 10.1103/PhysRevD.99.103536} {\bibfield  {journal} {\bibinfo  {journal} {Phys. Rev. D}\ }\textbf {\bibinfo {volume} {99}},\ \bibinfo {pages} {103536} (\bibinfo {year} {2019}{\natexlab{b}})},\ \Eprint {http://arxiv.org/abs/1812.11191} {arXiv:1812.11191 [gr-qc]} \BibitemShut {NoStop}%
\bibitem [{\citenamefont {Baumann}(2011)}]{Baumann:2009ds}%
  \BibitemOpen
  \bibfield  {author} {\bibinfo {author} {\bibfnamefont {D.}~\bibnamefont {Baumann}},\ }in\ \href {\doibase 10.1142/9789814327183_0010} {\emph {\bibinfo {booktitle} {{Theoretical Advanced Study Institute in Elementary Particle Physics}: {Physics of the Large and the Small}}}}\ (\bibinfo {year} {2011})\ pp.\ \bibinfo {pages} {523--686},\ \Eprint {http://arxiv.org/abs/0907.5424} {arXiv:0907.5424 [hep-th]} \BibitemShut {NoStop}%
\bibitem [{\citenamefont {Linde}(1983)}]{Linde:1983gd}%
  \BibitemOpen
  \bibfield  {author} {\bibinfo {author} {\bibfnamefont {A.~D.}\ \bibnamefont {Linde}},\ }\href {\doibase 10.1016/0370-2693(83)90837-7} {\bibfield  {journal} {\bibinfo  {journal} {Phys. Lett. B}\ }\textbf {\bibinfo {volume} {129}},\ \bibinfo {pages} {177} (\bibinfo {year} {1983})}\BibitemShut {NoStop}%
\bibitem [{\citenamefont {Starobinskii}(1979)}]{starobinskii1979spectrum}%
  \BibitemOpen
  \bibfield  {author} {\bibinfo {author} {\bibfnamefont {A.}~\bibnamefont {Starobinskii}},\ }\href@noop {} {\bibfield  {journal} {\bibinfo  {journal} {JETP Letters}\ }\textbf {\bibinfo {volume} {30}},\ \bibinfo {pages} {682} (\bibinfo {year} {1979})}\BibitemShut {NoStop}%
\bibitem [{\citenamefont {Maeda}(1988)}]{PhysRevD.37.858}%
  \BibitemOpen
  \bibfield  {author} {\bibinfo {author} {\bibfnamefont {K.-i.}\ \bibnamefont {Maeda}},\ }\href {\doibase 10.1103/PhysRevD.37.858} {\bibfield  {journal} {\bibinfo  {journal} {Phys. Rev. D}\ }\textbf {\bibinfo {volume} {37}},\ \bibinfo {pages} {858} (\bibinfo {year} {1988})}\BibitemShut {NoStop}%
\bibitem [{\citenamefont {Ferrara}\ \emph {et~al.}(2013)\citenamefont {Ferrara}, \citenamefont {Kallosh}, \citenamefont {Linde},\ and\ \citenamefont {Porrati}}]{Ferrara:2013rsa}%
  \BibitemOpen
  \bibfield  {author} {\bibinfo {author} {\bibfnamefont {S.}~\bibnamefont {Ferrara}}, \bibinfo {author} {\bibfnamefont {R.}~\bibnamefont {Kallosh}}, \bibinfo {author} {\bibfnamefont {A.}~\bibnamefont {Linde}}, \ and\ \bibinfo {author} {\bibfnamefont {M.}~\bibnamefont {Porrati}},\ }\href {\doibase 10.1103/PhysRevD.88.085038} {\bibfield  {journal} {\bibinfo  {journal} {Phys. Rev. D}\ }\textbf {\bibinfo {volume} {88}},\ \bibinfo {pages} {085038} (\bibinfo {year} {2013})},\ \Eprint {http://arxiv.org/abs/1307.7696} {arXiv:1307.7696 [hep-th]} \BibitemShut {NoStop}%
\bibitem [{\citenamefont {Bhattacharya}\ \emph {et~al.}(2023)\citenamefont {Bhattacharya}, \citenamefont {Dutta}, \citenamefont {Gangopadhyay},\ and\ \citenamefont {Maharana}}]{Bhattacharya:2022akq}%
  \BibitemOpen
  \bibfield  {author} {\bibinfo {author} {\bibfnamefont {S.}~\bibnamefont {Bhattacharya}}, \bibinfo {author} {\bibfnamefont {K.}~\bibnamefont {Dutta}}, \bibinfo {author} {\bibfnamefont {M.~R.}\ \bibnamefont {Gangopadhyay}}, \ and\ \bibinfo {author} {\bibfnamefont {A.}~\bibnamefont {Maharana}},\ }\href {\doibase 10.1103/PhysRevD.107.103530} {\bibfield  {journal} {\bibinfo  {journal} {Phys. Rev. D}\ }\textbf {\bibinfo {volume} {107}},\ \bibinfo {pages} {103530} (\bibinfo {year} {2023})},\ \Eprint {http://arxiv.org/abs/2212.13363} {arXiv:2212.13363 [astro-ph.CO]} \BibitemShut {NoStop}%
\bibitem [{\citenamefont {Kallosh}\ \emph {et~al.}(2013)\citenamefont {Kallosh}, \citenamefont {Linde},\ and\ \citenamefont {Roest}}]{Kallosh:2013yoa}%
  \BibitemOpen
  \bibfield  {author} {\bibinfo {author} {\bibfnamefont {R.}~\bibnamefont {Kallosh}}, \bibinfo {author} {\bibfnamefont {A.}~\bibnamefont {Linde}}, \ and\ \bibinfo {author} {\bibfnamefont {D.}~\bibnamefont {Roest}},\ }\href {\doibase 10.1007/JHEP11(2013)198} {\bibfield  {journal} {\bibinfo  {journal} {JHEP}\ }\textbf {\bibinfo {volume} {11}},\ \bibinfo {pages} {198} (\bibinfo {year} {2013})},\ \Eprint {http://arxiv.org/abs/1311.0472} {arXiv:1311.0472 [hep-th]} \BibitemShut {NoStop}%
\bibitem [{\citenamefont {Adams}\ \emph {et~al.}(1993)\citenamefont {Adams}, \citenamefont {Bond}, \citenamefont {Freese}, \citenamefont {Frieman},\ and\ \citenamefont {Olinto}}]{Adams:1992bn}%
  \BibitemOpen
  \bibfield  {author} {\bibinfo {author} {\bibfnamefont {F.~C.}\ \bibnamefont {Adams}}, \bibinfo {author} {\bibfnamefont {J.~R.}\ \bibnamefont {Bond}}, \bibinfo {author} {\bibfnamefont {K.}~\bibnamefont {Freese}}, \bibinfo {author} {\bibfnamefont {J.~A.}\ \bibnamefont {Frieman}}, \ and\ \bibinfo {author} {\bibfnamefont {A.~V.}\ \bibnamefont {Olinto}},\ }\href {\doibase 10.1103/PhysRevD.47.426} {\bibfield  {journal} {\bibinfo  {journal} {Phys. Rev. D}\ }\textbf {\bibinfo {volume} {47}},\ \bibinfo {pages} {426} (\bibinfo {year} {1993})},\ \Eprint {http://arxiv.org/abs/hep-ph/9207245} {arXiv:hep-ph/9207245} \BibitemShut {NoStop}%
\end{thebibliography}%

\end{document}